\documentclass[
 reprint,
 onecolumn,
 aps,superscriptaddress,
 amsmath,amssymb,
 aps,english,
]{revtex4-2}
\usepackage{geometry}
\usepackage[utf8]{inputenc}
\usepackage{amssymb}
\usepackage[caption=false]{subfig}
\usepackage{quantikz}
\usepackage{dcolumn}
\usepackage{bm}
\usepackage{calrsfs}
\usepackage{nameref}
\usepackage{mathtools}
\usepackage{floatrow}   
\usepackage{float}
\usepackage{graphicx,xcolor} 
\usepackage{mathtools}
\usepackage{physics}
\usepackage{microtype}
\usepackage[english]{babel}
\usepackage{hyperref}
\usepackage{cleveref}
\usepackage{natbib}

\usepackage{amsmath}

\DeclarePairedDelimiter{\altket}{\lvert}{\rangle}
\tikzset{
  customgate/.pic={
    \draw (0,0) rectangle (1,1);
    \node at (0.5,0.5) {$G$};
  }
}

\DeclareMathAlphabet{\mathcal}{OMS}{cmsy}{m}{n}

\DeclareMathOperator*{\argmin}{arg\, min}

\usepackage{setspace}

\usepackage{amsmath}

\newtheorem{lemma}{Lemma}
\newtheorem{corollary}{Corollary}
\newtheorem{assumption}{Assumption}
\newtheorem{definition}{Definition}
\newtheorem{result}{Result}
\newtheorem{fact}{Fact}

\Crefname{assumption}{Assumption}{Assumptions}
\Crefname{result}{Result}{Results}
\Crefname{fact}{Fact}{Facts}
\Crefname{definition}{Definition}{Definitions}
\setcitestyle{square}
\usepackage{chngcntr}
\allowdisplaybreaks
\begin{document}

\title{Deep learning-based quantum algorithms for solving nonlinear partial differential equations}
\author{Lukas Mouton}
\thanks{\tt lukasjmouton@gmail.com}
\affiliation{Centre for Quantum Technologies, National University of Singapore, Singapore}
\affiliation{Institute for Quantum Electronics, ETH Zürich, 8093 Zürich, Switzerland}
\author{Florentin Reiter}
\affiliation{Institute for Quantum Electronics, ETH Zürich, 8093 Zürich, Switzerland}
\author{Ying Chen}
\affiliation{Department of Mathematics, Asian Institute of Digital Finance, and Risk Management Institute, National University of Singapore, Singapore}
\author{Patrick Rebentrost}
\thanks{\tt cqtfpr@nus.edu.sg}
\affiliation{Centre for Quantum Technologies, National University of Singapore, Singapore}

\date{\today}

\begin{abstract}
Partial differential equations frequently appear in the natural sciences and related disciplines. Solving them is often challenging, particularly in high dimensions, due to the ``curse of dimensionality".
In this work, we explore the potential for enhancing a classical deep learning-based method for solving high-dimensional nonlinear partial differential equations with suitable quantum subroutines. First, with near-term noisy intermediate-scale quantum computers in mind, we construct architectures employing variational quantum circuits and classical neural networks in conjunction. 
While the hybrid architectures show equal or worse performance than their fully classical counterparts in simulations, they may still be of use in very high-dimensional cases or if the problem is of a quantum mechanical nature. 
Next, we identify the bottlenecks imposed by Monte Carlo sampling and the training of the neural networks. We find that quantum-accelerated Monte Carlo methods offer the potential to speed up the estimation of the loss function. In addition, we identify and analyse the trade-offs when using quantum-accelerated Monte Carlo methods to estimate the gradients with different methods, including a recently developed backpropagation-free forward gradient method. Finally, we discuss the usage of a suitable quantum algorithm for accelerating the training of feed-forward neural networks. 
Hence, this work provides different avenues with the potential for polynomial speedups for deep learning-based methods for nonlinear partial differential equations.
\end{abstract}

\maketitle

\section{Introduction}
\label{sct:intro}
\label{subsct:motivation}

Differential equations naturally appear in many disciplines and play a fundamental role in the sciences and engineering \cite{braun1983differential,pde_book} by mathematically modelling processes in fields such as physics and biology, as well as finance and sociology. In particular, nonlinear parabolic partial differential equations (PDEs) can model, e.g., the pricing of financial derivatives \cite{finance_pricing}, intelligent decision making in game theory \cite{dynamic_programming} and reaction-diffusion processes in physics \cite{physics_pde}. 
Solving high-dimensional PDEs is particularly challenging, as numerical methods typically rely on high-dimensional discretisations
of continuous functions. This procedure leads to what is known as the ``curse of dimensionality", i.e., the computational cost scaling exponentially with the dimensionality \cite{dim_curse}. If the PDE is nonlinear, approximating nonlinear terms in high-dimensional PDEs with polynomials or other basis functions further contributes to this problem.

Quantum computers have the potential to provide speedups for problems that may otherwise be intractable for classical computers, e.g., in machine learning \cite{qml}, linear algebra \cite{zhao2021compiling}, optimisation \cite{farhi2001quantum} or chemistry \cite{chemistry}.  Feynman famously envisioned the possibility of simulating quantum physics by using a quantum mechanical device \cite{feynman2018simulating}. Subsequent early breakthrough algorithms were Shor's algorithm \cite{shor} for factoring integers and Grover's algorithm \cite{grover} for searching an unstructured database. These algorithms motivated the development of increasingly better hardware, as well as a broader set of quantum algorithms and software \cite{Preskill_algorithms, Montanaro_algorithms, Krinner_2022, google_supremacy, google2023suppressing}. The seminal algorithm by Harrow, Hassidim and Lloyd (HHL) shows the potential for exponential quantum speedup for solving linear systems of equations \cite{HHL}. It has subsequently been applied to solving ordinary differential equations (ODEs) \cite{Berry_2014}. 
Many of these algorithms require fault-tolerance, meaning that a certain amount of inaccuracy on the hardware level is tolerable thanks to error-correction codes \cite{shor_faulttolerant, devitt2013quantum}. These quantum algorithms are commonly referred to as fault-tolerant quantum algorithms.
In addition to the work on fault-tolerant quantum algorithms, there is also active research on quantum algorithms which aim to make use of the currently available noisy intermediate-scale quantum (NISQ) computers to solve problems of practical relevance \cite{nisq,mcclean2016theory}. 

Recently, an effective classical algorithm has been proposed for solving high-dimensional semilinear parabolic PDEs (see \Cref{subsct:dl_architecture} and \cite{Han_2018}).
This algorithm reformulates the nonlinear PDE in terms of a stochastic differential equation (SDE), exploiting a link that has been extensively investigated \cite{pardoux1992backward,pardoux1999forward,karoui1997backward,gobet2016monte}. The authors of \cite{Han_2018} then use deep learning methods to approximate the spatial gradient of the sought after function. It is well known that neural networks (NNs) can approximate a wide range of functions \cite{hornik1989multilayer}. By employing NNs instead of polynomial or other basis functions, the authors in \cite{Han_2018} avoid the curse of dimensionality. The gradient is then used in a numerical scheme to solve the nonlinear PDE over a given time interval. 
While the nature of the algorithm from \cite{Han_2018}, namely, being based on deep learning techniques, does not allow for provable guarantees of finding a solution, it has several benefits in practice. The algorithm from \cite{Han_2018} is able to solve nonlinear parabolic PDEs, which, as outlined above, are relevant in a variety of fields, but are often hard to solve, particularly in high dimensions. Furthermore, the kinds of PDEs that can be solved are very general, more so than the kinds of PDEs for which quantum algorithms have already been proposed, see \Cref{subsct:quantum_diffeq}.
The introduction of NNs improves upon using polynomials or other basis functions to approximate nonlinear unknowns. However, some computational bottlenecks remain in this algorithm. By reformulating the PDE in terms of SDEs, the authors introduce stochasticity into the architecture. Therefore, as described by Chebyshev's inequality, one requires a certain number of samples to reliable estimate sought after quantities (such as the loss function or the gradient of the NNs) with a certain error tolerance. Furthermore, the runtime of evaluating and training the NNs scales approximately as the input dimension squared, where the input dimension is the spatial dimension.

By finding ways in which we may speed up the algorithm from \cite{Han_2018} with quantum subroutines, we thus hope to find a quantum algorithm that can achieve a speedup when solving a relatively wide range of nonlinear PDEs, as well as introducing a new paradigm for solving differential equations to the field of quantum algorithms, since, to the best of our knowledge, other quantum algorithms for solving differential equations have not attempted to make use of such a deep learning approach.

In this work, we investigate quantum algorithms for the solving of PDEs by exploring ways in which the deep learning architecture from \cite{Han_2018} can be sped up with quantum subroutines.  
As the deep learning architecture considers a sequence of NNs, we develop a hybridised classical-quantum architecture for solving PDEs by employing variational circuits as feature maps in the NNs. We make use of an existing scheme to avoid the barren plateau issue and carry out simulations to assess the effectiveness of this hybrid approach. 
Stepping away from the variational method, we investigate fault-tolerant approaches. To overcome the limitation imposed by Chebyshev's inequality on Monte Carlo (MC) sampling, we employ the quantum-accelerated MC (QAMC) method in a separate approach. To this end, we investigate methods, such as the recently introduced forward gradient method, for combining QAMC with NNs in order to speed up the estimation of the loss function and of the gradient of the NN. We carry out error and query complexity analyses to quantify the possible speedup.
Moreover, to address the bottleneck of training the NNs, we incorporate a quantum algorithm for accelerating the training and evaluation of classical NNs in a separate approach. We outline the advantages and limitations of this approach, in particular its lack of suitability for being combined with the previous QAMC approach.

In \Cref{sct:variational} we give an introduction to variational quantum algorithms as well as the challenges associated with them. Furthermore, we investigate ways in which we can introduce variational quantum algorithms into the algorithm from \cite{Han_2018} in order to possibly obtain a quantum advantage.
\Cref{sct:training_qmc_nn} outlines the idea of using QAMC (more details on which can be found in \Cref{subsubsubsct:qmc}) in combination with NNs. We highlight hurdles that arise when combining these two methods and examine different avenues to address these, among which is the recently introduced forward gradient method. We then use this tool in the following section.
In \Cref{sct:qmc_pde} we return to our goal of accelerating the algorithm from \cite{Han_2018} and make use of the discussion in \Cref{sct:training_qmc_nn}. Using our results from the preceding section, we incorporate QAMC into the algorithm from \cite{Han_2018}. Finally, we carry out error analyses and calculate query complexities to quantify the performance.
In \Cref{sct:quantum_feedforward} we summarise a fault-tolerant quantum algorithm that accelerates the evaluation and training of classical feedforward NNs. We proceed to outline how one can introduce this algorithm into the architecture from \cite{Han_2018}, and what advantages and drawbacks this entails. 
Finally, in \Cref{sct:discussion} we review our work and results, draw conclusions and outline possible future research directions. 

In \Cref{sct:comp_model}, we present the computational model we will work in. 
In \Cref{sct:classical_quantum_methods} we give an introduction to NNs and automatic differentiation (AD) and present the algorithm from \cite{Han_2018} in mathematical detail. This classical algorithm for solving nonlinear PDEs will be the starting point of our work, i.e., the algorithm which we want to speed up using quantum subroutines. Furthermore, we introduce MC methods and their quantum-accelerated version as well as other quantum algorithms and subroutines. 

\subsection{Notation}
\label{subsct:notation}

As in \cite{rubinstein2018big} in Definition 8.1, we define big $O$ notation as follows. Let $f$ and $g$ be two functions $f,g: X \rightarrow X$. We say that $f = O(g)$ if there is a constant $C\in\mathbb{R}^+$ such that for all $x\in X$ $|f(x)| \leq C\cdot |g(x)|$. We also use $\Tilde{O}$, which hides polylogarithmic factors.

We define Lipschitz continuity as in \cite{searcoid2006metric} (Definition 9.4.1). Suppose $(X,d_X)$ and $(Y,d_Y)$ are metric spaces (where $d_X$ and $d_Y$ denote metrics on $X$ and $Y$, respectively) and $f: X \rightarrow Y$. If there exists $L\in\mathbb{R}^+$ such that $d_Y (f(a), f(b)) \leq L\cdot d_X (a,b)$ for all $a,b\in X$, then $f$ is called Lipschitz continuous on $X$ with Lipschitz constant $L$. In this work we will use the squared $l_2$ norm, denoted by $\lVert \cdot \rVert_2^2$, when dealing with Lipschitz continuity.

In the relevant literature, one often encounters the additive and the mean-squared error. When estimating a quantity $a$ with an estimator $\Tilde{a}$, the additive error is given by $|a - \Tilde{a}|$ and the mean-squared error by $\mathbb{E}[(a - \Tilde{a})^2]$. In Appendix A of \cite{an2021quantum} the authors point out that these two definitions are almost equivalent. Concretely, up to a logarithmic overhead, the mean-squared error being $\epsilon^2$ indicates that the additive error is $\epsilon$ with probability at least $0.99$ and vice versa, using Chebyshev's inequality. Unless stated otherwise we shall use, as is common in the literature, the mean-squared error for classical results and the additive error for quantum results.

In the following section, we provide a detailed description of the algorithm from \cite{Han_2018} in \Cref{subsct:dl_architecture}. As mentioned above, the authors of \cite{Han_2018} present a deep learning-based algorithm for solving nonlinear PDEs which will serve as a starting point for this work, from where we will proceed to develop quantum-enhanced methods for solving nonlinear PDEs in later sections.

\subsection{Solving PDEs with Deep Learning}
\label{subsct:dl_architecture}
In this section we outline the deep learning-based algorithm for solving partial differential equations (PDEs) from \cite{Han_2018}, which will serve as a starting point for much of our work. In \cite{Han_2018}, the authors put forward a method to solve so-called semilinear parabolic PDEs. Semilinear refers to a sum of linear terms (in the variable of the PDE) and one nonlinear one. To obtain an approximation of the solution of the PDE, the authors reformulate the PDE as a backward stochastic differential equation (SDE). They then discretise the SDE temporally and approximate the spatial gradient of the unknown solution using NNs at each time step in the approximation. Let $x\in\mathbb{R}^d$ be the spatial variable, $t\in\mathbb{R}$ the temporal variable, and $u(t,x)\in\mathbb{R}$ the unknown solution. Furthermore, let $\sigma(t,x)\in\mathbb{R}^{d\times d}$, $\mu(t,x)\in\mathbb{R}^d$, and $f(t, x, u(t,x)\sigma^\top(t,x)\nabla u(t,x))\in\mathbb{R}$ be known functions, where $f$ is the nonlinearity.  
The main task is to solve PDEs of the following form \cite{Han_2018},
\begin{equation}
    \frac{\partial u}{\partial t}(t,x) + \frac{1}{2}\mathrm{Tr}(\sigma\sigma^{\top}(t,x)(\mathrm{Hess}_x u)(t,x)) + \nabla u(t,x)\cdot \mu(t,x) + f(t, x, u(t,x), \sigma^\top(t,x)\nabla u(t,x)) = 0,
    \label{eqn:pde}
\end{equation}
on some interval $[t_0, T]$. We assume to have a known terminal condition $u(T, x) = g(x) \in \mathbb{R}$. The quantities $\nabla u$ and $\mathrm{Hess}_x u$ refer to the gradient and Hessian, respectively, of $u$ with respect to the spatial variable $x$. 

\begin{figure}[t]
    \centering
    \includegraphics[width=\textwidth]{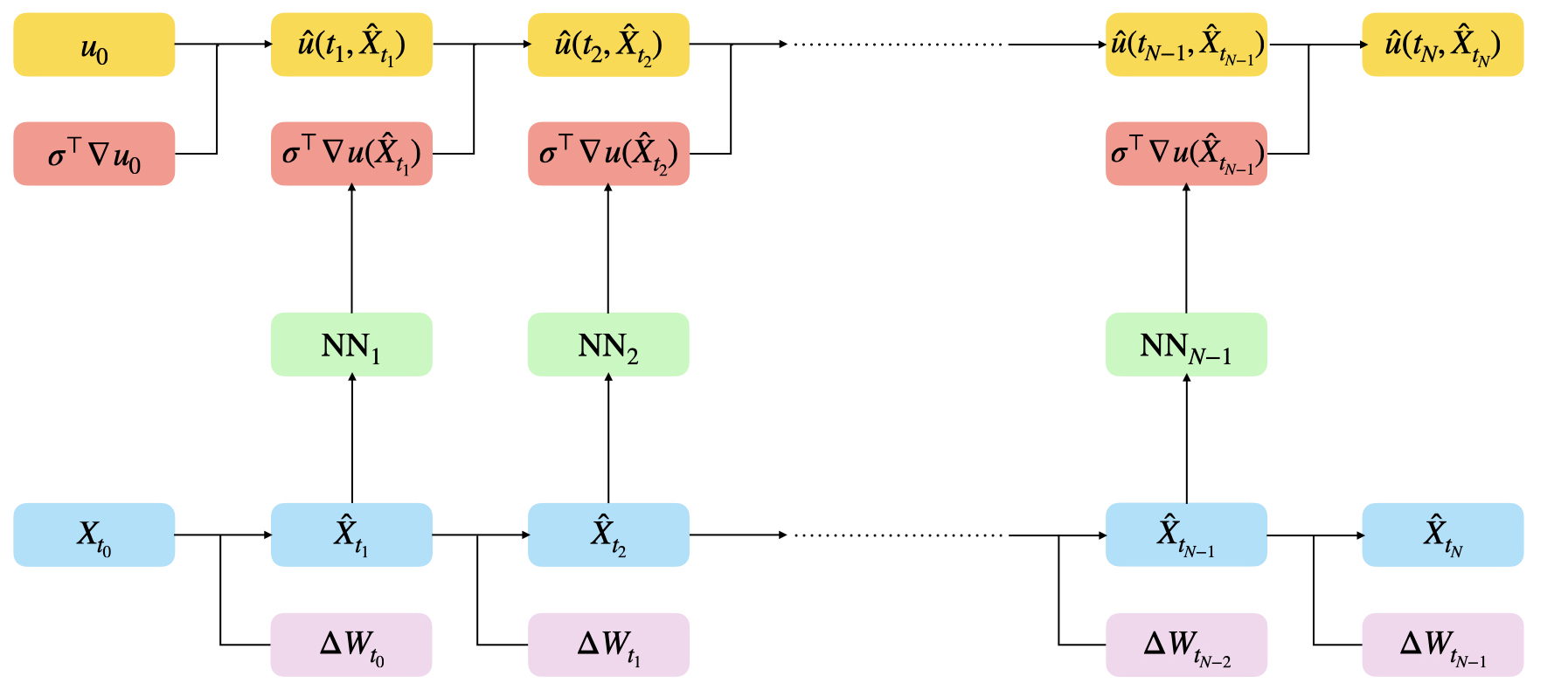}
    \caption{Architecture used for solving semilinear parabolic partial differential equations via a stochastic differential equation approach and the usage of neural networks for approximating spatial gradient information. The approach is a temporally discretised $N$-step procedure ($i\in [N]$) which is driven by stochastic Brownian increments $\Delta W_{t_i}$ determining the randomness of the state variables $X_{t_i}$. The neural networks represent the computation of the spatial gradient information from the state variables, which is then used to propagate the solution $u$. The arrows indicate which quantities are used to approximate which other quantities. Figure adapted from \cite{Han_2018}.
    }
    \label{fig:dl_arch}
\end{figure}

Let $W_t$ be a Brownian motion, meaning that $W_0 = 0$, $W_t$ is continuous in $t$, and for the increments $W_{t + t_1} - W_{t}$ it holds that $W_{t + t_1} - W_{t}$ is distributed according to $\mathcal{N}(0, t_1)$ and independent of past values $W_s$ for $s<t$ \cite{durrett2019brownian}. In the case of a multivariate Brownian motion, the above holds for each entry, which are independently and identically distributed (iid) according to $\mathcal{N}(0, t_1)$. For a $d$-dimensional Brownian motion $\{W_{t}\}_{t\in [0,T]}$ and a $d$-dimensional stochastic process $\{X_{t}\}_{t\in [0,T]}$ satisfying,
\begin{equation}
    X_t = \xi + \int_{0}^{T} \mu (s, X_s) \,ds + \int_{0}^{T} \sigma (s, X_s) \,dW_s,
    \label{eqn:stoch_proc}
\end{equation}
the solution of \Cref{eqn:pde} satisfies the following backward SDE in integral form \cite{pardoux1999forward, pardoux1992backward},
\begin{equation}
    u(t, X_t) = u(t, X_{t_0}) - \int_{0}^{t} f(s, X_s, u(s,X_s), \sigma^\top(s,X_s)\nabla u(s,X_s)) \,ds + \int_{0}^{t} (\nabla u(s, X_s))^\top\sigma (s, X_s) \,d W_s.
    \label{eqn:bsde}
\end{equation}
One can approximate \Cref{eqn:stoch_proc,eqn:bsde} by temporally discretising them, using e.g. Euler's method, with $t\in\{t_0 = 0, t_1,\ldots, t_N = T\}$. We then arrive at
\begin{equation}
    X_{t_{n+1}} - X_{t_{n}} \approx \mu(t_{n}, X_{t_{n}})\Delta t_{n} + \sigma(t_{n}, X_{t_{n}})\Delta W_{t_n},
    \label{eqn:pde_disc_1}
\end{equation}
and 
\begin{equation}
    \begin{aligned}
    u(t_{n+1}, X_{t_{n+1}}) - u(t_{n}, X_{t_{n}}) \approx {} & -f(t_n, X_{t_{n}}, u(t_n, X_{t_{n}}), \sigma^\top(t_n, X_{t_{n}})\nabla u(t_n, X_{t_{n}}))\Delta t_n \\ & + (\nabla u(t_n, X_{t_{n}}))^\top\sigma (t_n, X_{t_{n}})\Delta W_{t_n},
    \end{aligned}
    \label{eqn:pde_disc_2}
\end{equation}
where
\begin{equation}
    \Delta t_n = t_{n+1} - t_n,\hspace{1em} \Delta W_{t_n} = W_{t_{n+1}} - W_{t_n},
    \label{eqn:delta}
\end{equation}
where the entries of $\Delta W_{t_n}$ are iid according to $\mathcal{N}(0,\,\Delta t_n)$. The discretisation in \Cref{eqn:pde_disc_1} allows us to sample paths $\{\hat{X}_{t_n}\}_{n\in [0,N]}$, where we use the hat (i.e., $\hat{X}_t$) to indicate the discretised estimation of $X_t$. In a next step, the authors in \cite{Han_2018} approximate the function $x \rightarrow \sigma^\top(t, X_t)\nabla u(t, X_t)$ at each time step $t = t_n$ using a feedforward NN,
\begin{equation}
    \sigma^\top(t, X_t)\nabla u(t, X_t) = (\sigma^\top\nabla u)(t, X_t) \approx (\sigma^\top\nabla {u})(\hat{X}_t|\theta_n),
    \label{eqn:ffnn}
\end{equation}
where $\theta_n$ denotes the parameters of the $n$th NN. The reason for assuming this would work is that feedforward NNs may serve as universal function approximators \cite{hornik1989multilayer}. The authors in \cite{Han_2018} stack together the NNs together with \Cref{eqn:stoch_proc,eqn:bsde} to step by step calculate $\hat{u}$. \Cref{fig:dl_arch} illustrates this process. As seen in \Cref{fig:dl_arch}, no discretisation of the whole spatial domain takes place in this deep learning-based algorithm. The discretisation of the spatial domain is what typically leads to the curse of dimensionality when solving PDEs, as the cost of solving the problem then grows exponentially in its size (the spatial dimension, in this case).

In order to train the NNs, the authors use the expected difference between the terminal condition $g(\hat{X}_{t_N})$, and the computed $\hat{u}_{t_N}$ after the final step, as a loss function,
\begin{equation}
    l(\theta) = \mathbb{E}(\lvert g(\hat{X}_{t_N}) - \hat{u}_{t_N}(\{\hat{X}_{t_n}\}_{0\leq n\leq N}, \{\Delta W_{t_n}\}_{1\leq n\leq N})\rvert^2).
    \label{eqn:loss}
\end{equation}
We refer to the function inside the expectation value as the payoff function $f_p$,
\begin{equation}
    \label{eqn:payoff}
    f_p(\{\hat{X}_{t_n}\}_{0\leq n\leq N}, \{\Delta W_{t_n}\}_{1\leq n\leq N}) = \lvert g(\hat{X}_{t_N}) - \hat{u}_{t_N}(\{\hat{X}_{t_n}\}_{0\leq n\leq N}, \{\Delta W_{t_n}\}_{1\leq n\leq N})\rvert^2.
\end{equation}
The total set of trainable parameters consists of the initial conditions $u(0, X_{t_0})$ and $\nabla u(0, X_{t_0})$ as well as the parameters for each of the NNs, $\{\theta_n\}_{0\leq n\leq N-1}$. The total number of trainable parameters we call $n_\theta$. The focus of this work is to investigate methods for accelerating the deep learning method introduced in this section.

\subsection{Related Work: Quantum Algorithms for Differential Equations}
\label{subsct:quantum_diffeq}
There exist several proposed quantum algorithms for solving differential equations (ODEs as well as PDEs). In \cite{Berry_2014} the author puts forward a fault-tolerant quantum algorithm which solves sparse systems of linear ODEs by discretising the system of ODEs and subsequently employing the HHL algorithm \cite{HHL} to solve the resulting system of linear equations. In \cite{childs2020quantum_spectral}, the authors also propose a quantum algorithm for solving linear ODEs, which, however, relies on spectral methods. Spectral methods use linear combinations of basis functions (e.g., a Fourier basis) to approximate the solution. This approach also ends with solving a linear system of equations on a quantum computer. A quantum algorithm to solve quadratically nonlinear ODEs under certain conditions is described in \cite{Liu_2021}. The authors make use the Carleman linearisation \cite{carleman1932application,forets2017explicit_carleman,kowalski1991nonlinear_carleman} to approximate the nonlinear part. The Carleman linearisation represents a finite-dimensional polynomially nonlinear system by an infinite-dimensional linear system. To make use of the Carleman linearisation, the infinite-dimensional linear system is truncated at a certain point. Subsequently, the authors from \cite{Liu_2021} again discretise the resulting system and solve the linear system with HHL \cite{HHL}. The algorithm presented in \cite{Liu_2021} may also be applied to solving certain PDEs (with a restricted kind of nonlinearity), as the discretisation of a PDE in all but one dimension generally results in a system of ODEs. In \cite{krovi2023improved}, the author improves on the algorithm from \cite{Liu_2021} and in \cite{liu2023towards} it is modified to address problems in machine learning. The authors in \cite{an2022efficient} present an algorithm for solving (nonlinear) reaction-diffusion equations. Using Euler's method, as well as the Carleman linearisation for the nonlinearity, they discretise the PDE and solve the resulting system with HHL. The authors of \cite{childs_pde} present quantum algorithms for solving linear PDEs by making use of the finite difference method (FDM) and spectral methods separately. In the FDM approach, they discretise the PDE on a grid. Both cases result in a linear system of equations which needs to be solved. The authors of \cite{arrazola2019quantum_linearpde} also present a quantum algorithm for solving linear PDEs, which relies on Hamiltonian simulation of a cleverly chosen Hamiltonian, which encodes certain properties of the PDE. In \cite{lloyd2020quantum} a quantum algorithm to solve nonlinear ODEs by mapping the ODE to the nonlinear Schr\"odinger equation, which is then solved using Trotterisation, is outlined. Another numerical scheme, the finite element method (FEM, which approximates the solution by using interpolating functions within each discretised element) and HHL are used to solve elliptic PDEs in \cite{montanaro2016quantum_fem}. In \cite{jin2022time,jin2022quantum}, the authors derive quantum algorithms for solving nonlinear ODEs as well as the nonlinear Hamilton-Jacobi (HJ) equation (which is a special case of the nonlinear Hamilton-Jacobi-Bellman PDE). They do so, by mapping the nonlinear ODEs and the nonlinear HJ equation to linear ODEs and HJ equations using linear representation methods, such as the level set method, and then use HHL to solve the linear system. For a detailed discussion on HHL-based quantum algorithms for solving differential equations, we refer to \cite{an2022theory}. 

A comment on the form of the solution of HHL is in order. In a classical setting, the vector describing the solution to a linear system of equations can be copied and its entries can be accessed at will. In the quantum setting, however, the solution is encoded in the amplitudes of a quantum superposition (amplitude encoding). The no-cloning theorem \cite{wootters1982single} states that quantum states cannot be cloned (duplicated), and a measurement would collapse the quantum state to the eigenstate corresponding to the measurement outcome. One therefore resorts to multiple computations and subsequent measurements in order to generate reliable statistics to infer the sought results. The procedure of producing a quantum state proportional to the solution (of a system of linear equations, or differential equations) may in some applications not be enough, as a full quantum state cannot be read out efficiently \cite{Liu_2021}. In some scenarios it may suffice to extract relevant information from sampling an observable. However, in other cases, more intricate post-processing may be needed to gain the desired insights.

An example of a variational quantum algorithm for solving differential equations can be found in \cite{kyriienko2021solving}. The authors define a loss function which measures how well a certain candidate function solves the differential equation. The differential equation is then encoded via its key properties such that it is mapped to a high-dimensional feature space, which is explored by a classical solver in order to find a solution. The variational quantum algorithm put forth in \cite{lubasch2020variational} aims to solve nonlinear differential equations by using a so-called quantum nonlinear processing unit. The latter computes polynomially nonlinear terms appearing in the differential equations. 

In addition, there exist multiple proposed quantum algorithms to solve specific differential equations, such as the Poisson equation \cite{cao2013quantum}, the wave equation \cite{costa2019quantum}, the Vlasov equation \cite{engel2019quantum} and the heat equation \cite{linden2022quantum}. 

\section{Variational Approach for Deep Learning Method for PDEs}
\label{sct:variational}

In this section we outline our first approach for enhancing the deep learning architecture from \cite{Han_2018} using quantum computation. We consider employing variational quantum methods in order to enhance the algorithm from \cite{Han_2018} by introducing parametrised quantum circuits (PQCs) into the neural networks (NNs). 

In the context of machine learning, a motivation for making use of variational quantum circuits is the exploitation of the exponentially large Hilbert space via controllable entanglement and interference \cite{schuld2019quantum,quantum_feature,farhi2018classification,goto2020universal,lloyd2020quantum_embeddings}. In this sense, the quantum circuit is employed as a feature map. As the circuit depth may be kept relatively short compared to fault-tolerant quantum algorithms, this approach is more suitable for the noisy intermediate-scale quantum (NISQ) era. If the classical starting point is a classical NN (as in the architecture from \cite{Han_2018}), a straightforward place to introduce PQCs is within the NNs, as described in \cite{qiskit_textbook}. The resulting network is termed a hybrid quantum neural network (hybrid QNN). 

In a hybrid QNN, a PQC is inserted into the classical NN such that the outputs (post-activation) of a given layer are fed as arguments (e.g, as the rotation angles of gates in the PQC) into the PQC. The measurement outcomes of the PQC, measured in the computational basis, are then again passed forward as inputs to the succeeding classical layer. In order to update the weights of a hybrid QNN, one generally makes use of backpropagation, see \Cref{subsct:neuralnetworks}. In order to evaluate the gradients of the parameters in the PQC, we make use of the parameter shift rule, see \Cref{lemma:parameter_shift}.

While the hybrid QNN architecture outlines a possibility for introducing PQCs into classical NNs, it does not specify what kind of PQCs are inserted into the NNs. A commonly used ansatz circuit is the so-called hardware efficient ansatz (HEA), because it makes use of the native gates of a given quantum hardware platform, thus avoiding the transpilation overhead \cite{kandala2017hardwareefficient}. By doing so, it can keep the circuit depth very low while still allowing for entanglement among all qubits and introducing trainable parameters for each qubit.
The HEA is of the form,
\begin{equation}
    \label{eqn:hea}
    \mathcal{U}(\theta) = \prod_k \mathcal{U}_k(\theta_k)\mathcal{W}_k,
\end{equation}
where $\mathcal{U}_k(\theta_k)$ consists of single qubit rotations on each qubit, the angles of which are the variational parameters to be optimised. Furthermore, $\mathcal{W}_k$ are (entangling) two-qubit gates. Note that the form of the HEA is similar to the form outlined in \Cref{eqn:var_ansatz}. The HEA is known to suffer from barren plateaus, especially for larger circuit depths \cite{mcclean2018barren}. We proceed to outline a technique from the authors of \cite{leone2022practical} for preventing the barren plateau issue from occurring in the HEA for short circuit depths. Their finding suggests that using short PQCs of the form of the HEA may still be a viable option for incorporating quantum circuits within hybrid QNNs. However, one can only expect a quantum speedup when using the HEA if the circuit is sufficiently shallow and if the quantum state in which the circuit is initialised satisfies certain entanglement laws. The authors propose an initial state which satisfies the entanglement criteria,
\begin{equation}
    \label{eqn:hea_initial_state}
    |\psi_t\rangle = e^{-i H t} |\psi_0\rangle,
\end{equation}
where $|\psi_0\rangle$ is a product state, i.e., not entangled. For $t$ above a certain threshold, the qubits are suitably entangled. The Hamiltonian $H$ is of the form,
\begin{equation}
    \label{eqn:hea_hamiltonian}
    H = \sum_{i=1}^n X_i X_{i+1} + Y_i Y_{i+1} + 2 Z_i Z_{i+1} + X_i,
\end{equation}
where $n$ is the number of qubits in the circuit and $X_i X_{i+1}$, $Y_i Y_{i+1}$ and $Z_i Z_{i+1}$ refer to the two-qubit Pauli gates acting on the $i$th and $i+1$th qubit. The index $n+1$ refers back to $1$. Next, we discuss how the methods introduced above may be applied to the deep learning architecture.

\subsection{Applying Variational Algorithms to the Deep Learning Approach}
\label{subsct:var_dl}
We proceed to discuss the application of variational quantum methods in the deep learning architecture from \cite{Han_2018} and its consequences. In the deep learning architecture, a place where PQCs can straightforwardly be employed is in (or instead of) the NNs. When replacing the NNs in the architecture from \cite{Han_2018} with (generic) PQCs and calculating the gradient of the loss function $l(\theta)$ (see \Cref{eqn:loss}) with respect to a parameter $\theta_k$ in the $n$th variational circuit $\mathcal{V}_n$ (i.e., at the $n$th step of the temporal discretisation) we have,
\begin{equation}
    \begin{aligned}
    \label{eqn:variational_loss_gradient}
    & \partial_{\theta_k} l(\theta) = \partial_{\theta_k} \mathbb{E}[\lvert g - \hat{u}_{t_N}  \rvert^2] = 2 \mathbb{E}[(g - \hat{u}_{t_N}) \partial_{\theta_k} \hat{u}_{t_N}]\\ &= 2 \mathbb{E}\left[(g - \hat{u}_{t_N}) \left(\prod_{i = N-1}^{n+1} \partial_{\hat{u}_i} f(u_i,\,\sigma^\top \nabla u_i) \Delta t_i \right) \left( (\partial_{\sigma^\top \nabla u_n}f(u_n,\, \sigma^\top \nabla u_n)\Delta t_n + \Delta W_{t_n}) \partial_{\theta_k} \mathcal{V}_n\right)\right],
    \end{aligned}
\end{equation}
where we omitted the arguments of functions that are not of importance for this calculation. The key observation is that the term $\partial_{\theta_k} \mathcal{V}_n$ appears as a factor in the derivative. Therefore, we expect the barren plateau issue to show up in our context as well, provided that the PQCs we employ to replace the classical NNs exhibit barren plateaus in the first place. In other words, embedding a PQC suffering from the barren plateau problem in the architecture from \cite{Han_2018} will result in the same problem showing up there as well. Thus, we will still have to make use of techniques to prevent the barren plateau issue from occurring. Next, we describe how we insert PQCs into the deep learning architecture, using the methods introduced above, followed by simulations to evaluate the performance of the hybrid architectures.

\subsection{Simulations}
\label{subsct:simulations}

We now proceed to apply the methods introduced above to the deep learning architecture. We employ hybrid QNNs in order to allow for PQCs to serve as feature maps and we make use of the HEA with the considerations from \cite{leone2022practical} to address the barren plateau issue. Having done so, we carry out numerical experiments to compare the classical case to the case with PQCs.

\begin{figure}[ht]
    \centering
\begin{quantikz}
\lstick[wires=5]{$|\psi_t\rangle$} & \gate{R_X(z^{(1)})} & \gate{R_X(\theta_1)}\gategroup[wires=5,steps
=6,style={inner sep=3pt}]{Repeat $r$ times} & \ctrl{1} & \qw & \qw & \qw & \targ{} & \qw 
\\
& \gate{R_X(z^{(2)})} &  \gate{R_X(\theta_2)} &  \targ{} & \ctrl{1} & \qw & \qw & \qw & \qw 
\\
& \gate{R_X(z^{(3)})} &  \gate{R_X(\theta_3)} & \qw & \targ{} & \ctrl{1} & \qw & \qw & \qw  
\\
& \gate{R_X(z^{(4)})} &  \gate{R_X(\theta_4)} & \qw & \qw & \targ{} & \ctrl{1} & \qw & \qw 
\\
& \gate{R_X(z^{(5)})} &  \gate{R_X(\theta_5)} & \qw & \qw & \qw & \targ{} & \ctrl{-4} & \qw 
\end{quantikz}
\caption{Example PQC with 5 qubits with $|\psi_t\rangle$ from \Cref{eqn:hea_initial_state}. The $z$ values are inputs to the circuit (which are given by the outputs of the preceding classical layer, in a hybrid QNN) and the $\theta$ values are trainable parameters. The part in the box is repeated $r$ times, with different parameters $\theta$ for each repetition.}
\label{fig:pqc_hybrid}
\end{figure}
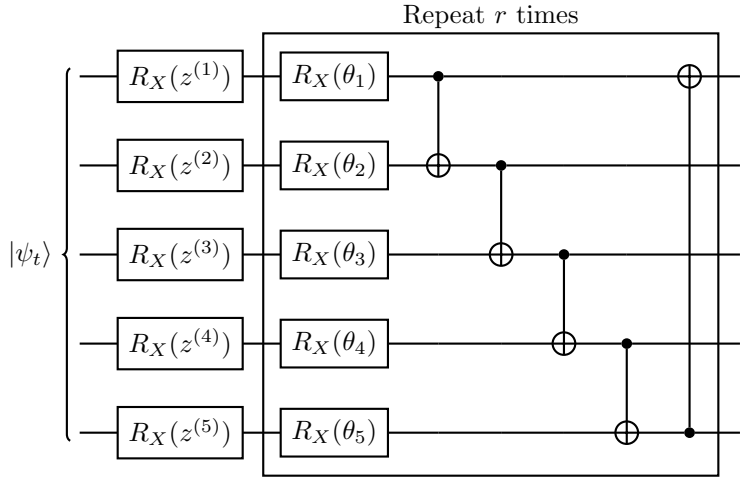

In order to evaluate the performance of variational circuits in the architecture from \cite{Han_2018}, we carry out simulations with small-sized quantum circuits, such that we can simulate them within reasonable time. We investigate whether including a PQC in the NNs in the architecture from \cite{Han_2018} can improve the performance, measured by the loss function. The PQC we employ, for a variable number of qubits, is displayed in \Cref{fig:pqc_hybrid}.

\noindent The circuit in \Cref{fig:pqc_hybrid} is initialised in the state displayed in \Cref{eqn:hea_initial_state} (to avoid the barren plateau from occurring), the inputs are loaded via angle embedding (using $R_X$ rotations), followed by a HEA with $R_X$ rotations and circular entanglement via CNOT gates. On actual quantum hardware, the single qubit rotation gates and the two qubit gates would be the native gates of the hardware platform to reduce compilation overhead. In order to compare the classical base case to hybrid scenarios, we solve the Hamilton-Jacobi-Bellman (HJB) equation, which is a special case of the PDE from \Cref{eqn:pde} with $\sigma(t,X)=2\mathbb{I}$, $\mu(t,X)=0$ and $f(t,X,u(t,X),\sigma^\top(t,X)\nabla u(t,X)) = \lVert \nabla u(t,X)\rVert_2^2$. To evaluate the loss function, we have $g(X) = \log((1+\lVert X \rVert_2^2)/2)$ for the HJB partial differential equation (PDE). We choose to simulate the classical and the quantum case for the HJB PDE, because the output of the NNs plays a more important role compared to the other PDEs, and the performance of the NNs (or hybrid QNNs) is under investigation. For our implementations we use PyTorch \cite{paszke2019pytorch} and PennyLane \cite{bergholm2018pennylane}.

\begin{figure}[t]
\centering
\subfloat[\label{fig:hybrid_arch}]{
  \includegraphics[width=0.3\textwidth]{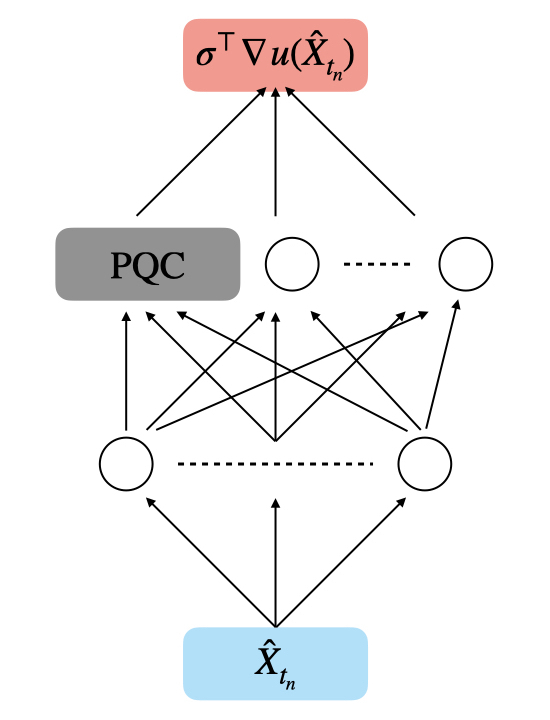}
}\hspace{0.1\textwidth}
\subfloat[\label{fig:pqc_arch}]{
  \includegraphics[width=0.3\textwidth]{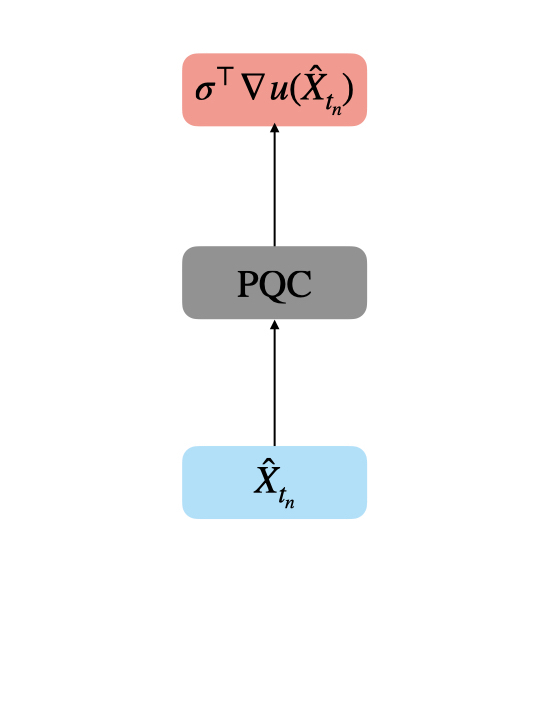}
}
\caption{(a) Hybrid QNN used for the first experiment, where the circles represent classical neurons, the grey box a PQC, and the input and output are labelled according to \Cref{fig:dl_arch} (b) PQC employed in the deep learning architecture as in \Cref{fig:dl_arch} without the assistance of classical neurons, as used in the second experiment}
\end{figure}

In our first experiment we compare the baseline classical architectures with hybrid QNNs with the same total number of trainable parameters in order to investigate whether one can observe improvements of the hybrid case over the classical case, by using the PQC as a feature map. The classical NNs have an input layer, two hidden layers, and an output layer (as in \cite{Han_2018}), which are all fully connected. In the hybrid QNN, we replace nodes in the second hidden layer with a PQC of the form from \Cref{fig:pqc_hybrid}, as seen in \Cref{fig:hybrid_arch}. 

\begin{figure}[t]
    \centering
    \includegraphics[width=0.8\textwidth]{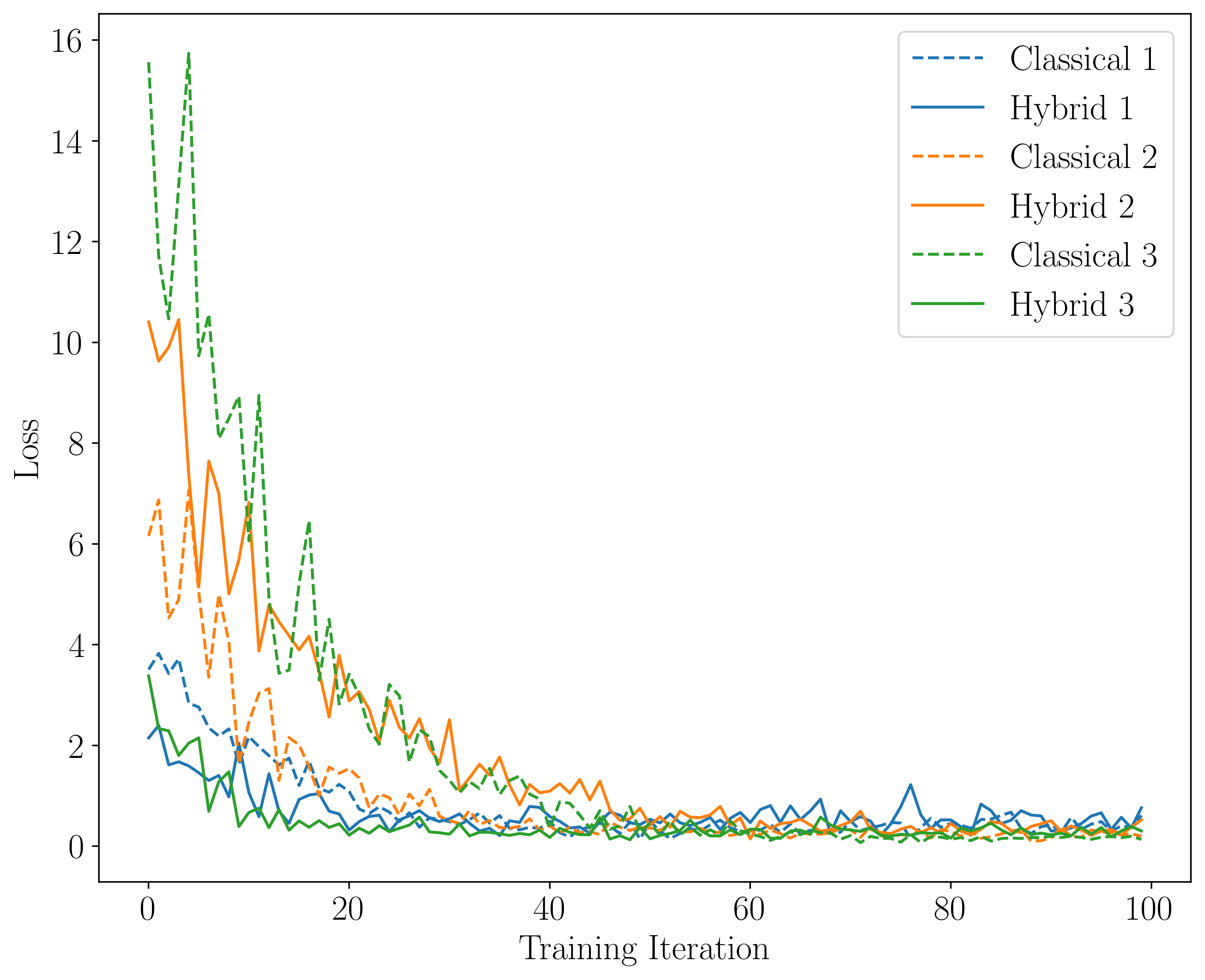}
    \caption{Loss from \Cref{eqn:loss} for the HJB PDE against the number of training iterations with 20 samples per iteration for three different scenarios (225, 565 and 1260 total trainable parameters in the classical and hybrid case where the spatial dimension is 5, 10 and 20, respectively). The learning rate in all cases is 0.05. In each of the hybrid cases, 16 of the trainable parameters are variational parameters in a PQC in a hybrid QNN, as shown in \Cref{fig:hybrid_arch}, where the PQC is of the form of \Cref{fig:pqc_hybrid} with $r=2$.}
    \label{fig:loss_p}
\end{figure}

We introduce the PQC in a way such that the total number of trainable parameters does not change, by tuning the number of classical neurons in the hidden layers. In the hybrid case, some of the trainable parameters are parameters in the PQC (the values of $\theta$ in \Cref{fig:pqc_hybrid}), as opposed to trainable weights in the classical part. We compare three cases, each having a different number of trainable parameters. We compare the performance for the classical base case and the hybrid case, allowing us to estimate if a greater percentage of the trainable parameters in the PQC offers any advantage. Note that while we can easily increase the number of classical parameters, increasing the number of variational quantum parameters is computationally expensive to simulate. Consequently, in the hybrid scenarios of the different cases, a different percentage of the total number of trainable parameters are variational quantum parameters. The cases 1, 2 and 3 we compare in \Cref{fig:loss_p} solve the HJB PDE using the architecture from \cite{Han_2018} in 5, 10 and 20 dimensions with 225, 565 and 1260 total trainable parameters, respectively. In the hybrid cases, we employ a 8-dimensional PQC of the form from \Cref{fig:pqc_hybrid} with $r=2$ in the second hidden layer, introducing 16 trainable variational quantum parameters. Thus, the hybrid model in case 1 has the greatest percentage of variational quantum parameters, and the hybrid model in case 3 the smallest.

As one can see in \Cref{fig:loss_p}, the classical models and the hybrid models in each of the three cases perform comparably well after a sufficient number of training iterations, with the models with a greater number of trainable parameters taking longer to converge. \Cref{fig:loss_p} suggest that introducing the PQC does not provide an improvement over the classical base case. Furthermore, we cannot observe a difference in performance for a different percentage of variational quantum parameters in the hybrid models. Possible reasons for this include that the percentage of variational quantum parameters is too small to make a difference, or that the classical models are just as suited (if not more so) for the problem at hand. To understand which of these possibilities is more plausible, we carry out a second experiment.

In the next experiment, we directly compare PCQs (as opposed to hybrid QNNs) against simple NNs with the same number of trainable parameters,  as shown in \Cref{fig:pqc_arch}.

\begin{figure}[t]
    \centering
    \includegraphics[width=\textwidth]{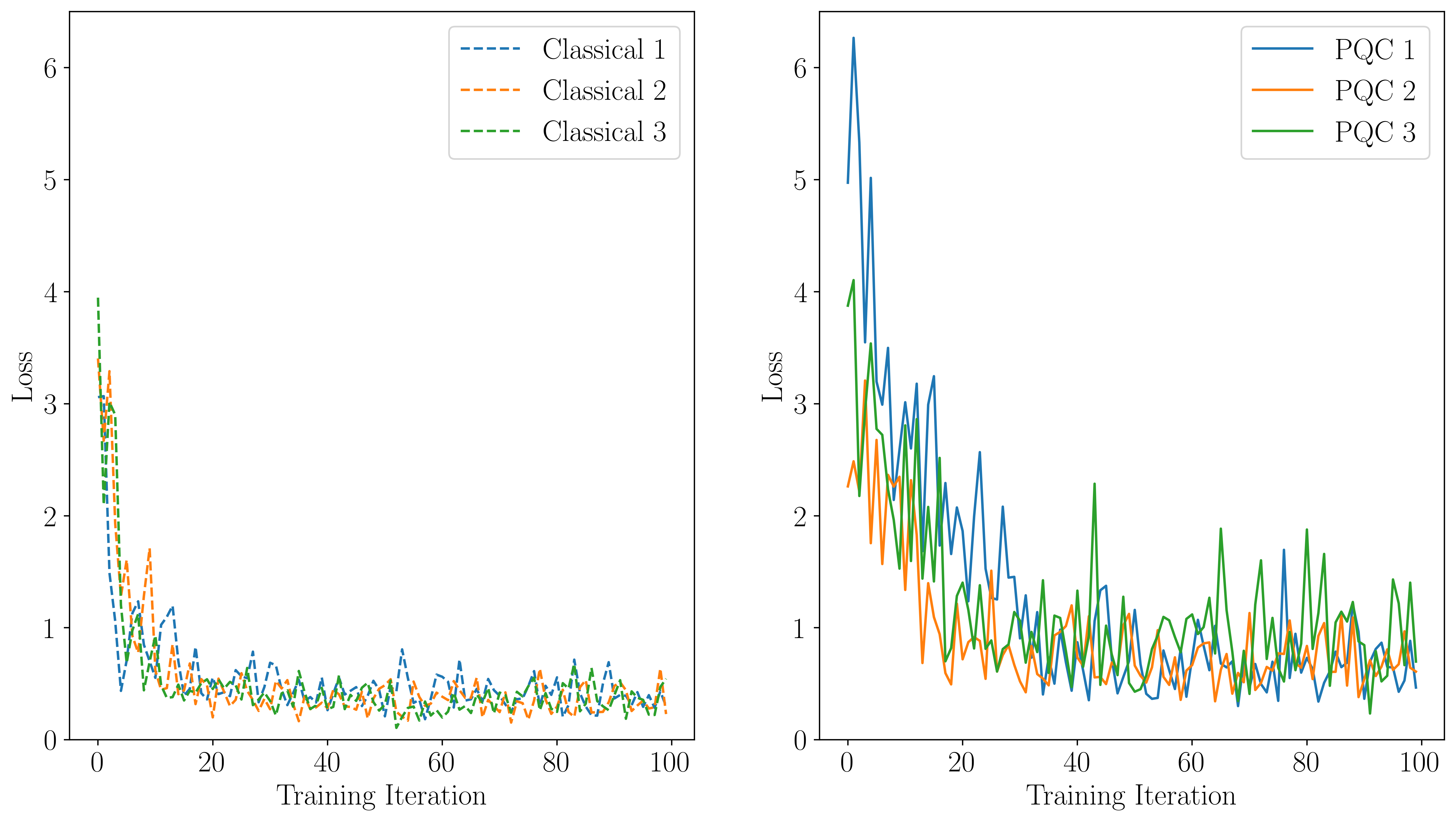}
    \caption{Loss from \Cref{eqn:loss} for the HJB PDE against the number of training iterations with 20 samples per iteration for three different scenarios (20, 30 and 42 trainable parameters, respectively) for classical NNs and PQCs with the same number of trainable parameters. The learning rate in all cases is 0.05. In each of the quantum cases, all of the trainable parameters are variational parameters in a PQC, as shown in \Cref{fig:pqc_arch}, where the PQC is of the form of \Cref{fig:pqc_hybrid}.}
    \label{fig:loss_sep}
\end{figure}

Again, an argument for introducing the PQC is the hope that it may find solutions in the Hilbert space serving as a feature space. We want to see whether the hybrid case still performs similarly well as the classical case. Based on the outcome, we want to learn whether in the first experiment, introducing the PQC did not improve the performance because the PQCs and classical NNs perform similarly well, or because the NNs were ``carrying'' the hybrid models. We again compare three cases where we solve the HJB PDE using the deep learning architecture for the 4, 5, and 6-dimensional case. The classical models are NNs with just one one input layer and one output layer and the PQCs are again of the form from \Cref{fig:pqc_hybrid} with the number of qubits corresponding to the dimension $d$ of the input, and the number of repetitions $r=d+1$ such that the total number of trainable parameters is the same in the classical and the hybrid case. The loss for each of these cases is shown in \Cref{fig:loss_sep}. We separate the classical and the hybrid models for better visibility.

In \Cref{fig:loss_sep} it is clearly visible that the classical models outperform the hybrid models. This appears to refute the hypothesis that the classical models and PQCs are similarly suited in this context, suggesting that in the first experiment, the classical parts of the hybrid QNNs were to some extent carrying the PQC, allowing the hybrid model to perform as well as the classical base model.

To conclude, our evidence suggests that, in the deep learning architecture from \cite{Han_2018}, hybrid models perform similarly well as classical models, but completely replacing the NNs with PQCs worsens the performance. However, we cannot rule out the possibility, that in very high-dimensional scenarios, which are beyond the reach of classical simulations, hybrid architectures may provide better performance. Since the problem of learning the quantity $\sigma^\top(t,X)\nabla u(t,X)$ from samples of $X_t$ is of classical nature, the motivation for including PQCs into the NNs was to use the PQCs as feature map, as done elsewhere. A hypothetical scenario in which the introduction of PQCs into the architecture would be further motivated by the nature of the problem at hand being quantum mechanical, might be if the dependence of the spatial gradient on the stochastic process $X_t$ was given according to some unitary evolution. 

\section{Training Neural Networks with Quantum-Accelerated Monte Carlo Methods}
\label{sct:training_qmc_nn}
In this section, we explore the combination of classical neural networks (NNs) trained on data sampled from known distributions with a suitable quantum subroutine which may offer a speedup in the error tolerance, namely the quantum-accelerated Monte Carlo (QAMC) method, as reviewed in \Cref{subsubsubsct:qmc}. We analyse different methods for training classical NNs in quantum circuits on data sampled from known distributions and compare them to each other, as well as to classical methods.

We intend to apply the methods discussed in this section to the algorithm from \cite{Han_2018}, where NNs are trained on data sampled from known distributions. We will proceed to do so in \Cref{sct:qmc_pde}. Training and evaluating NNs with data obtained with MC sampling is, however, also used elsewhere, e.g., in quantitative finance \cite{de2000sequential,pagnottoni2019neural}, physics and chemistry simulations \cite{dirisio2021gpu,hermann2022ab}, image and video processing \cite{liu2023monte,tsui2018modelling}, drug discovery \cite{xie2021mars}, and robotics \cite{riccio2017using}. We thus see potential for this discussion being relevant beyond the scope of this work.

A key direction we pursue is to apply QAMC to estimating loss functions of NNs, which we do in \Cref{subsct:nn_output_estimate}. However, the NNs may not be pretrained, their parameters thus requiring fine-tuning in order to be useful. In the classical scenario, backpropagation (see \Cref{subsct:neuralnetworks}) is a widely used algorithm to train the weights in NNs, in spite of its excessive memory usage \cite{gruslys2016memory,griewank2000memory_reverse}, which is difficult to predict in advance \cite{margossian2019review}. Due to memory in quantum circuits (i.e., qubits) being especially expensive for the foreseeable future, this concern is further amplified in the quantum case. It is worth pointing out, that when backpropagation is applied to a classical NN implemented in a quantum circuits, where the inputs may be in superposition, the forward and the backward pass occur in the same circuit, and measurement crucially only takes place after both passes have been carried out, lest the superpositions be collapsed.

In the classical machine learning literature, a novel method for training the weights of NNs, termed the forward gradient \cite{baydin2022gradients}, has been proposed, and we will introduce it in \Cref{subsct:forward_gradient}. This method relies on forward mode automatic differentiation (AD) as opposed to reverse mode AD, on which backpropagation is based (see \Cref{subsct:ad}). This difference results in the forward gradient requiring less memory, at the cost of introducing stochasticity into the gradient. Furthermore, being based on AD, it does not suffer from numerical issues as numerical differentiation does. Since MC methods (classical or quantum) leverage stochasticity, as the forward gradient method does, we investigate how the forward gradient method interacts with MC methods. We then compare different approaches to incorporating the forward gradient with other methods for training NNs in the classical and quantum scenario in \Cref{subsct:qamc_nn}. 

In our model, we consider the NNs to be classical circuits which are implemented via a unitary (see \Cref{def:unitary}) performing the equivalent quantum circuit.
We briefly comment on how a classical feed-forward NN may be implemented in a quantum circuit. As mentioned in \Cref{sct:comp_model}, any classical circuit can be implemented on a quantum circuit with only a minimal overhead. For more on how classical feedforward NNs may be implemented in a quantum circuit, we refer to  \cite{wan2017quantum_generalisation}. In this reference the authors only outline how to implement a certain activation function which is not Lipschitz continuous. Common activation functions such as the sigmoid function or the rectified linear unit (ReLU) are, however, Lipschitz continuous. 

\subsection{Estimating Output of Neural Network}
\label{subsct:nn_output_estimate}
 
We first quantify the potential speedup obtained when using QAMC to estimate the loss function of a NN, where the stochasticity comes from the input of the neural network.
Note \Cref{def:dist_loading} on preparing probability distributions in in superposition.
We proceed to outline the setting with a quantum circuit. In this section, a sample or a query refers to a query to the unitary we introduce below (or its modified forms, which we introduce later on in \Cref{subsct:qamc_nn}). Note that the number of wires is not representative of how many qubits are used to represent a certain state.

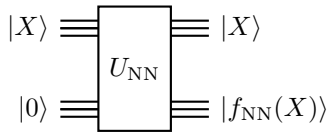
\begin{figure}[H]
\centering
\begin{quantikz}
     \lstick{$|{X}\rangle$} & \gate[2]{U_{\mathrm{NN}}}\qwbundle[alternate]{} &   \rstick{$|{X}\rangle$}\qwbundle[alternate]{} 
     \\
     \lstick{$\ket{0}$} & \qwbundle[alternate]{} & \rstick{$|f_\mathrm{NN}(X)\rangle$}\qwbundle[alternate]{} 
\end{quantikz}
\caption{Unitary evaluating NN}
\label{fig:circ_nn_eval1}
\end{figure}

Consider a classical NN including a scalar loss function represented by the function $f_\mathrm{NN}: \mathbb{R}^{d} \mapsto \mathbb{R}$ and a random variable (RV) $X$, distributed according to $p_X$, i.e., $X_i$ occurs with probability $p_{X_i}$. Let the NN be implemented in a unitary in a quantum circuit, as displayed in \Cref{fig:circ_nn_eval1}. Since the trainable parameters are classical parameters (i.e., they are not encoded in quantum states) we do not make them explicit in this definition.

\noindent For the circuit shown in \Cref{fig:circ_nn_eval1}, when we query the unitary $U_\mathrm{NN}$ with an input state as seen in \Cref{def:dist_loading}, we have,
\begin{equation}
    \label{eqn:nn_oracle_basic_input}
    \sum_{i} \sqrt{p_{X,i}} |X_i\rangle |0\rangle.
\end{equation}
We then get the following output state,
\begin{equation}
    \label{eqn:nn_oracle_basic_output}
    \sum_{i} \sqrt{p_{X,i}} |X_i\rangle |f_\mathrm{NN}(X_i)\rangle.
\end{equation}
Applying QAMC to estimate the mean of $f_\mathrm{NN}(X)$, weighted by the distribution $p_X$, as described in \Cref{subsubsubsct:qmc}, we need to bound the variance of $f_\mathrm{NN}(X)$ in order to determine the required sample complexity and quantify the potential quantum speedup. 

At this point, we comment on a property of NNs, namely that NNs are in practice Lipschitz continuous. As shown in \cite{szegedy2013intriguing}, the Lipschitz constant of a (feedforward) NN can be upper bounded by the product of the Lipschitz constants of its constituent layers. The Lipschitz constant of a particular layer can, in turn, be upper bounded by the largest singular value of the matrix describing the weights of that particular layer, $\sigma_\mathrm{max}(W_l)$, multiplied by the Lipschitz constant $L_{f_{\mathrm{nl},l}}$ of the activation function of the same layer, $f_{\mathrm{nl},l}$,
\begin{equation}
    \label{eqn:lipschitz_nn}
    L_{f_\mathrm{NN}} \leq \prod_{l=1}^L \sigma_\mathrm{max}(W_l)L_{f_{\mathrm{nl},l}},
\end{equation}
where there are $L$ layers in the NN. The individual layers are Lipschitz continuous if the nonlinear activation functions $f_{\mathrm{nl},l}$ are so, and common activation functions, such as ReLU or the sigmoid function, are Lipschitz continuous. Furthermore, we assume the loss function is Lipschitz continuous, at least on the bounded domain of interest. Thus, in practice, NNs are typically Lipschitz continuous. There exist regularisation techniques, termed spectral normalisation, that control the Lipschitz constant of a NN by normalising the singular values of the weight matrices for each layer to a desired value \cite{miyato2018spectral,anil2019sorting_lipschitz}. 

Given the Lipschitz constant for a NN, $L_{f_\mathrm{NN}}$, we can also derive the following growth bound. By the Lipschitz continuity of $f_\mathrm{NN}$ we have,
\begin{equation}
    \label{eqn:nn_growth_bound}
    \lvert f_\mathrm{NN}(X) - f_\mathrm{NN}(0)\lvert^2 \leq L_{f_\mathrm{NN}} \lVert X - 0 \rVert_2^2 = L_{f_\mathrm{NN}} \lVert X  \rVert_2 ^2.
\end{equation}
Therefore, we have that,
\begin{equation}
    \label{eqn:nn_growth_bound2}
    \begin{aligned}
    \lvert f_\mathrm{NN}(X)\rvert^2 &= \lvert f_\mathrm{NN}(X) - f_\mathrm{NN}(0) + f_\mathrm{NN}(0) \rvert^2 \\ 
    &\leq \lvert f_\mathrm{NN}(X) - f_\mathrm{NN}(0)\rvert^2 + \lvert f_\mathrm{NN}(0) \rvert^2 \\
    & \leq L_{f_\mathrm{NN}} \lVert X \rVert_2^2 + \lvert f_\mathrm{NN}(0) \rvert^2 \\
    & \leq (L_{f_\mathrm{NN}} + \lvert f_\mathrm{NN}(0) \rvert^2) (1 + \lVert X \rVert_2^2).
    \end{aligned}
\end{equation}

\noindent We now proceed to find an upper bound on the variance of $f_\mathrm{NN}(X)$. We have,
\begin{equation}
    \label{eqn:x_var_bound}
    \begin{aligned}
        \mathbb{V}\left[ f_\mathrm{NN}(X) \right] & = \mathbb{E}\left[ f_\mathrm{NN}(X)^2 \right] -\mathbb{E}\left[ f_\mathrm{NN}(X) \right]^2 \\
        & \leq \mathbb{E}\left[ f_\mathrm{NN}(X)^2 \right] \\
        & \leq \mathbb{E}\left[ (L_{f_\mathrm{NN}} + \lvert f_\mathrm{NN}(0) \rvert^2) (1 + \lVert X \rVert_2^2) \right] \\
        & = (L_{f_\mathrm{NN}} + \lvert f_\mathrm{NN}(0) \rvert^2) (1 + \mathbb{E}\left[\lVert X \rVert_2^2\right] ),
    \end{aligned}
\end{equation}
which allows us the formulate the following result, by applying \Cref{lemma:qmc}.

\begin{result}[Quantum-Accelerated Estimation of Neural Network Loss Function]
\label{thm:nn_qamc_eval}
 Consider a classical NN including its scalar loss function, $f_\mathrm{NN}: \mathbb{R}^d \mapsto \mathbb{R}$, with Lipschitz constant $L_{f_\mathrm{NN}}$. Let the NN be implemented in a quantum circuit in the form of a unitary $U_\mathrm{NN}$. Let the input to $U_\mathrm{NN}$ be given by a RV $X$, governed by a distribution $p_X$. Then we can estimate the mean output of $f_\mathrm{NN}$ using QAMC with a sample complexity, with respect to $U_\mathrm{NN}$, of,
\begin{equation}
    \Tilde{O}((L_{f_\mathrm{NN}} + \lvert f_\mathrm{NN}(0) \rvert^2) (1 + \mathbb{E}\left[\lVert X \rVert_2^2\right] ) / \epsilon),
\end{equation}
up to additive error $\epsilon$ with probability 0.99, thus offering a quadratic speedup over the classical case.
\end{result}
We proceed to discuss the nature of the speedup, given the upper bound on $\mathbb{V}\left[ f_\mathrm{NN}(X) \right]$. The first factor, $(L_{f_\mathrm{NN}} + \lvert f_\mathrm{NN}(0) \rvert^2)$, is hard to upper bound in practice, as hinted at in \Cref{eqn:lipschitz_nn}. As mentioned above, there exist regularisation techniques to normalise the Lipschitz constant of a NN. The quantity $\lvert f_\mathrm{NN}(0) \rvert^2$ is also hard to estimate in a general setting, and we offer no general bound on it. The quantity $\mathbb{E}\left[\lVert X \rVert_2^2\right]$ depends naturally on the distribution $p_X$, but we can expect it to grow with the dimension $d$ of $X$. A key variable in which we get a speedup is the error $\epsilon$, allowing us to achieve the same error tolerance with quadratically fewer samples.

Next, we discuss the problem of training the classical NNs implemented in quantum circuits. 
 
\subsection{Forward Gradient and Other Training Methods}
\label{subsct:forward_gradient}
We now discuss methods for training the NNs in a quantum circuit in the setting outlined above, when the input is distributed according to a known distribution. We first discuss numerical differentiation, then proceed to methods based on AD. In particular, we introduce a method from the classical machine learning literature, namely the so-called forward gradient from \cite{baydin2022gradients}. 

When updating the weights in NNs, a straightforward method is to resort to numerical differentiation. Let $\{\theta_k\}_k$ denote the set of trainable parameters in the NN, where $k\in\{1,\ldots,n_\theta\}$, $n_\theta$ being the number of trainable parameters. Numerical differentiation then approximates the derivative with respect to an individual weight in the NN, $\theta_k$, as,
\begin{equation}
    \label{eqn:num_diff}
    \frac{\partial}{\partial \theta_k}f_\mathrm{NN}(X;\theta_k) \approx \frac{f_\mathrm{NN}(X;\theta_k + h)-f_\mathrm{NN}(X;\theta_k - h)}{2h},
\end{equation}
for a suitably chosen $h$. Compared to AD-based methods such as the backpropagation, numerical differentiation may suffer from numerical issues, i.e., the truncation error resulting from the finite difference approximations used to approximate derivatives.
For one training step updating the weights of the NNs of the architecture from \cite{Han_2018} via gradient descent, 
\begin{equation}
    \label{eqn:gradient_desc}
    \theta_{j+1} = \theta_j - \eta \nabla l(\theta_j),
\end{equation}
where $\eta$ is the learning rate. It is worth pointing out, that due to the non-convex training landscape of the NNs, we cannot make a statement on how many training steps are needed to train the NNs to a satisfactory precision. Let $n_\theta$ be the total number of parameters, where $n_\theta$ is approximately of the order of $d^2$, where $d$ is the input dimension, i.e., the number of neurons in the first layer (assuming the number of layers does not change with $d$). The number of trainable parameters in a feedforward NN is given by \cite{goodfellow2016deep},
\begin{equation}
    \label{eqn:num_params}
    n_\theta = \sum_{l=1}^{L-1} n_l (n_{l+1} + 1)
\end{equation}
where $L$ is the number of layers, $n_l$ is the number of neurons in the $l$th layer and the additional $n_l$ parameters per layer are the biases. For one training step, when using numerical differentiation, we require $O(n_{\theta})$ evaluations of the loss function from \Cref{eqn:loss} to get an estimate of the gradient.
We can thus state the following lemma.

\begin{fact}
    \label{lemma:params_dimension}
     If in a classical feedforward NN of interest, the number of layers is constant in the input dimension $d$, and the number of neurons per layer grows at most linearly in $d$, we have that for the number of trainable parameters $n_\theta$,
    \begin{equation}
        n_\theta = O(d^2).
    \end{equation}

\end{fact}

In contrast to numerical differentiation, with backpropagation the cost for a single training step is $O(1)$ times the runtime for the evaluation of the NN \cite{griewank2008evaluating}. As a consequence, backpropagation is widely used to train NNs. However, backpropagation comes with the downside of memory requirements that are greater and harder to predict compared to methods based on forward mode AD, where the memory requirements are simply twice that of the function evaluation \cite{margossian2019review}.

A possible alternative to backpropagation and numerical differentiation is the forward gradient method presented in \cite{baydin2022gradients}, which estimates the so-called forward gradient,
\begin{equation}
    \label{eqn:fwd_grad_def}
    (\nabla f_\mathrm{NN}\cdot v) v,
\end{equation}
which, for a suitably chosen vector $v$, constitutes an unbiased estimate of the gradient $\nabla f_\mathrm{NN}$.
A key ingredient for the forward gradient method in real world applications is, as for backpropagation, AD. However, using the forward gradient method does not necessitate holding on to intermediate activations, unlike in the case of backpropagation, as it relies on forward mode AD, as opposed to reverse mode AD. Being based on AD, the forward gradient method does not suffer from numerical issues in the way numerical differentiation may do. In terms of the runtime, there is only a constant overhead compared to the evaluation of the function (i.e., the NN) \cite{griewank2008evaluating}. The forward gradient method, by using AD and with a runtime with only a constant overhead over the evaluation of the NN (represented by the function $f_\mathrm{NN}$), returns not only the output of the NN but also $\nabla f_\mathrm{NN}\cdot v$ in a single forward run, without computing $\nabla f_\mathrm{NN}$ itself. Here the gradient is taken with respect to the trainable parameters of the NN and $v$ is a vector along which the gradient is projected. The computation of $\nabla f_\mathrm{NN}$ with the forward gradient method would, as the authors in \cite{baydin2022gradients} point out, require $O(n_\theta)$ evaluations, by choosing $v$ as each basis vector once. In order to be competitive with backpropagation, they need to work with $O(1)$ evaluations, not $O(n_\theta)$. They have to thus choose $v$ such that the overall sensitivity is attributed back to each individual weight parameter in the NN. This feat is achievable by choosing the entries of $v$ to be independently and identically distributed (iid) according to $\mathcal{N}(0,1)$. In \cite{baydin2022gradients}, the authors show that the forward gradient $(\nabla f_\mathrm{NN}\cdot v) v$ is an unbiased estimator of the gradient $\nabla f_\mathrm{NN}$. Furthermore, they provide numerical results indicating that the forward gradient method is competitive with backpropagation with regard to the runtime as well as to number of update iterations to bring the value of the loss function to a certain threshold, for different network architectures. What is more, the forward gradient method requires significantly less memory than backpropagation, as it does not require the storage of all intermediate steps, as pointed out in \cite{baydin2022gradients}. It is worth highlighting, that the forward gradient is by no means the only noisy estimate of a gradient used in machine learning. In stochastic gradient descent (SGD), one estimates the gradient with only a small subset of the available data \cite{robbins1951stochastic,bottou2012stochastic}. While the convergence speed is limited by the noisy approximation of the true gradient, SGD is nevertheless widely used in practice. There exist several techniques for dealing with noisy estimates of gradients in the field of machine learning. Examples of these techniques are gradient clipping \cite{zhang2019clipping}, whereby each entry of the gradient exceeding a certain threshold in absolute value is set back to that threshold, or gradient norm scaling \cite{chen2018gradnorm}, where the norm of the gradient is scaled down to prevent so-called ``gradient explosions'', which may occur otherwise. 

Note that there also exist quantum algorithms which can compute gradients of functions, where the runtime is linear in the runtime of the function itself \cite{gilyen2019optimizing,jordan2005fast}. However, these methods require phase oracles or probability oracles, which is less practical in our case of interest, see \Cref{sct:qmc_pde}. Going forward, we discuss how we may apply the methods discussed above, in particular the forward gradient method in the quantum case.

\subsection{Training Neural Networks with QAMC}
\label{subsct:qamc_nn}
We now describe the application of the forward gradient method from the previous subsection to the problem of training a classical NN in a quantum circuit, where the inputs are given by a distribution encoded in a superposition state. We compare different ways of applying the forward gradient method and also compare them to other methods for training a NN in the classical and quantum setting.

When combining the forward gradient method with QAMC and AD, there are a priori two options regarding how to estimate the forward gradient. In the first option, $v$ is loaded in the form of a (quantum mechanical) superposition state, as in \Cref{def:dist_loading}, according to its distribution $p_v$. In this option, we hope to get the speedup from using QAMC, but expect to get a slowdown from having to resort to estimating each of the $n_\theta$ entries of the forward gradient separately, as QAMC only allows for the estimation of scalar quantities (see the argument in \Cref{subsubsubsct:multivar_mc}). In the second option we sample $v$ classically, which will prevent us from getting the speedup from QAMC for $v$ (but still for $X$). This is motivated by only estimating the scalar $\nabla f_\mathrm{NN} \cdot v$ (as opposed to a $n_\theta$-dimensional vector), such that we hope to avoid the slowdown in $n_\theta$. Furthermore, it would allow us to use classical MVMC (see \Cref{subsubsubsct:multivar_mc}). It is worth pointing out that since the application of forward mode AD (as well as reverse mode AD) results in a runtime with only a constant overhead compared to just evaluating the NN, it is meaningful to compare the query complexities to the unitaries implementing the evaluation of the NN, with or without some form of AD \cite{griewank2008evaluating}.

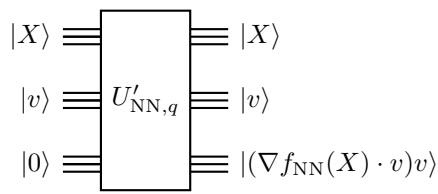
\begin{figure}[H]
\centering
\begin{quantikz}
     \lstick{$|{X}\rangle$} & \gate[3]{U'_{\mathrm{NN},q}}\qwbundle[alternate]{} &   \rstick{$|{X}\rangle$}\qwbundle[alternate]{} 
     \\
    \lstick{$\ket{v}$} & \qwbundle[alternate]{} & \rstick{$\ket{v}$}\qwbundle[alternate]{}
      \\
     \lstick{$\ket{0}$} & \qwbundle[alternate]{} & \rstick{$|(\nabla f_\mathrm{NN}(X) \cdot v)v\rangle$}\qwbundle[alternate]{} 
\end{quantikz}
\caption{Unitary evaluating the forward gradient, with $v$ encoded in a quantum state}
\label{fig:circ_nn_grad_opt2}
\end{figure}

We proceed to discuss the first option, where we prepare $v$ (as well as $X$) as a quantum state according to their respective distribution, as in \Cref{def:dist_loading} and estimate each entry of the forward gradient using QAMC. 
In this case, the gradient we estimate with QAMC is the weighted average of the gradients, weighed by the probabilities of the inputs $X$ of the NNs, as well as those in $v$. 
We name the unitary implementing the computation of the forward gradient for the first option $U'_{\mathrm{NN},q}$, as shown in \Cref{fig:circ_nn_grad_opt2}.

\noindent In this case, the quantum state for a single NN before the unitary is,
\begin{equation}
    \label{eqn:nn_oracle_input_opt2}
    \sum_{i,j} \sqrt{p_{X,i}} \sqrt{p_{v,j}} |X_i\rangle |v_j\rangle |0\rangle  ,
\end{equation}
\noindent where $v$ is distributed according to $p_v$. As the output of the unitary $U'_{\mathrm{NN},q}$ we get,
\begin{equation}
    \label{eqn:nn_oracle_output_opt2}
    \sum_{i,j} \sqrt{p_{X,i}} \sqrt{p_{v,j}} |X_i\rangle  |v_j\rangle |(\nabla f_\mathrm{NN}(X_j) \cdot v_j)v_j\rangle. 
\end{equation}
When estimating the mean of $(\nabla f_\mathrm{NN}(X) \cdot v)v$ using QAMC, we now have to consider the randomness from $X$, quantified by $p_X$, as well as that of $v$, quantified by $p_v$. One issue pops up with this method. In order to estimate the forward gradient, we need to apply QAMC to estimate each entry of the forward gradient, leading to a slowdown of $O(n_\theta)$. Recall that $n_\theta$ is the number of trainable parameters in the NN, which corresponds to the dimension of $(\nabla f_\mathrm{NN}(X) \cdot v)v$. Therefore, while we might get a speedup in the error $\epsilon$, we still expect a slowdown in $n_\theta$ compared to classical backpropagation. In order to derive the query complexity for estimating the forward gradient in the case where $X$ and $v$ are encoded in quantum states according to their distributions $p_X$ and $p_v$, respectively, we proceed to upper bound the variance of each entry of the forward gradient with respect to both sources of randomness. 
Next, we derive an upper bound on the variance with respect to $p_X$ and $p_v$ of the $j$th component of $(\nabla f_\mathrm{NN}(X) \cdot v)v$, i.e., $\mathbb{V}\left[ \left((\nabla f_\mathrm{NN}(X) \cdot v)v\right)^{(j)} \right]$, which we will need to quantify the query complexity. To make our notation more compact, we will refer to $\nabla f_\mathrm{NN}(X)$ as $\nabla f_\mathrm{NN}$. We have that,
\begin{equation}
    \label{eqn:gradient_var_bound}
    \begin{aligned}
    & \mathbb{V}\left[ \left((\nabla f_\mathrm{NN} \cdot v)v\right)^{(j)} \right] = \mathbb{V}\left[ (\nabla f_\mathrm{NN} \cdot v)v^{(j)} \right] = 
    \mathbb{V}\left[ \sum_{k=1}^d (\nabla f_\mathrm{NN})^{(k)} v^{(k)} v^{(j)} \right] \\
    &= \sum_{k=1}^d \mathbb{V}\left[ (\nabla f_\mathrm{NN})^{(k)} v^{(k)} v^{(j)} \right] \\
    & + \sum_{k\neq i} \mathbb{E}\left[ ((\nabla f_\mathrm{NN})^{(k)} v^{(k)} v^{(j)} - \mathbb{E}\left[(\nabla f_\mathrm{NN})^{(k)} v^{(k)} v^{(j)}
\right])((\nabla f_\mathrm{NN})^{(i)} v^{(i)} v^{(j)} - \mathbb{E}\left[ (\nabla f_\mathrm{NN})^{(i)} v^{(i)} v^{(j)}
\right]) \right],
    \end{aligned}
\end{equation}
where we used the formula for the variance of a sum in the last step. The second summand contains the covariance terms, which we have rewritten in the form of the definition in terms of expectation values to allow for further analysis. The covariance terms are,
\begin{equation}
    \label{eqn:cov_zero}
    \begin{aligned}
        & \sum_{k\neq i} \mathbb{E}\left[ ((\nabla f_\mathrm{NN})^{(k)} v^{(k)} v^{(j)} - \mathbb{E}\left[(\nabla f_\mathrm{NN})^{(k)} v^{(k)} v^{(j)}
        \right])((\nabla f_\mathrm{NN})^{(i)} v^{(i)} v^{(j)} - \mathbb{E}\left[ (\nabla f_\mathrm{NN})^{(i)} v^{(i)} v^{(j)}
        \right]) \right]\\
        & = \sum_{k\neq i} \mathbb{E}\left[(\nabla f_\mathrm{NN})^{(k)} v^{(k)} v^{(j)}
        (\nabla f_\mathrm{NN})^{(i)} v^{(i)} v^{(j)} \right] 
        + \mathbb{E}\left[(\nabla f_\mathrm{NN})^{(k)} v^{(k)} v^{(j)} \mathbb{E}\left[(\nabla f_\mathrm{NN})^{(i)} v^{(i)} v^{(j)}\right]\right] \\ & 
        + \mathbb{E}\left[\mathbb{E}\left[(\nabla f_\mathrm{NN})^{(k)} v^{(k)} v^{(j)} \right] (\nabla f_\mathrm{NN})^{(i)} v^{(i)} v^{(j)}\right] 
        + \mathbb{E}\left[\mathbb{E}\left[(\nabla f_\mathrm{NN})^{(k)} v^{(k)} v^{(j)}
        \right]\mathbb{E}\left[ (\nabla f_\mathrm{NN})^{(i)} v^{(i)} v^{(j)}
        \right]
        \right].
    \end{aligned}
\end{equation}
Note that in each of the four terms in the sum in \Cref{eqn:cov_zero}, there is at least one entry of $v$ that appears only once inside an expectation value because in each term $k\neq i$ (it is possible that $j=k$ or $j=i$). Since the individual components of $v$ are iid with mean $0$ each of the four terms in the sum in \Cref{eqn:cov_zero} will vanish in every summand. We continue with our analysis of the variance of the output of the NN we aim to estimate using QAMC, namely,
\begin{equation}
    \label{eqn:grad_var_bound_cont}
    \begin{aligned}
        \mathbb{V}\left[ \left((\nabla f_\mathrm{NN} \cdot v)v\right)^{(j)} \right] &=
\sum_{k=1}^d \mathbb{V}\left[ (\nabla f_\mathrm{NN})^{(k)} v^{(k)} v^{(j)} \right].
    \end{aligned}
\end{equation}
For the summands from \Cref{eqn:grad_var_bound_cont} we make a case distinction to separate the cases where $k\neq j$ from the case where $k=j$. Introducing,
\begin{equation}
    \label{eqn:gmax_def}
    g_\mathrm{max} \eqqcolon \lVert\nabla f_\mathrm{NN}\rVert_\infty^2,
\end{equation}
we have for $k\neq j$,
\begin{equation}
    \label{eqn:kneqj_var_bound}
    \begin{aligned}
    \mathbb{V}\left[ (\nabla f_\mathrm{NN})^{(k)} v^{(k)} v^{(j)} \right] & = \mathbb{E}\left[ ((\nabla f_\mathrm{NN})^{(k)} v^{(k)} v^{(j)})^2 \right] - \mathbb{E}\left[ (\nabla f_\mathrm{NN})^{(k)} v^{(k)} v^{(j)} \right]^2 \\ 
    & = \mathbb{E}\left[ ((\nabla f_\mathrm{NN})^{(k)} v^{(k)} v^{(j)})^2 \right]\\ & = \mathbb{E}\left[ ((\nabla f_\mathrm{NN})^{(k)})^2 \right] \mathbb{E}\left[ (v^{(k)})^2 \right] \mathbb{E}\left[ (v^{(j)})^2 \right] \\
    & =\mathbb{E}\left[ ((\nabla f_\mathrm{NN})^{(k)})^2 \right] \\ 
    & \leq \mathbb{E}\left[ \lVert\nabla f_\mathrm{NN}\rVert_\infty^2 \right] \leq g_\mathrm{max},
    \end{aligned}
\end{equation}
where the second equality is due to the fact that $k\neq j$ and thus the three terms $(\nabla f_\mathrm{NN})^{(k)}$, $v^{(k)}$ and $v^{(j)}$ are independent, with $\mathbb{E}[v^{(k)}]=\mathbb{E}[v^{(j)}]=0$. The third equality again follows from the independence of $(\nabla f_\mathrm{NN})^{(k)}$, $v^{(k)}$ and $v^{(j)}$. The first equality rests on $\mathbb{V}[v^{(k)}]=\mathbb{V}[v^{(j)}]=1$ being true. In the case where $k=j$ we have,
\begin{equation}
    \label{eqn:keqj_var_bound}
    \begin{aligned}
        \mathbb{V}\left[ (\nabla f_\mathrm{NN})^{(k)} (v^{(j)})^2\right] &= \mathbb{E}\left[((\nabla f_\mathrm{NN})^{(k)})^2 (v^{(j)})^4 \right] - \mathbb{E}\left[(\nabla f_\mathrm{NN})^{(k)} v^{(j)} \right]^2 \\
        &= \mathbb{E}\left[((\nabla f_\mathrm{NN})^{(k)})^2\right] \mathbb{E}\left[(v^{(j)})^4 \right] - \mathbb{E}\left[(\nabla f_\mathrm{NN})^{(k)}\right]^2  \mathbb{E}\left[v^{(j)} \right]^2 \\ & = 3\mathbb{E}\left[((\nabla f_\mathrm{NN})^{(k)})^2\right]- \mathbb{E}\left[(\nabla f_\mathrm{NN})^{(k)}\right]^2 \leq 3g_\mathrm{max},
    \end{aligned}
\end{equation}
where we used that $\mathbb{E}\left[ \left(v^{(j)}\right)^4 \right] = 3$. Combining these results, we have that,
\begin{equation}
    \label{eqn:var_bound_final}
    \begin{aligned}
        \mathbb{V}\left[ \left((\nabla f_\mathrm{NN} \cdot v)v\right)^{(j)} \right] =
\sum_{k=1}^d \mathbb{V}\left[ (\nabla f_\mathrm{NN})^{(k)} v^{(k)} v^{(j)} \right] \leq (d + 2)g_\mathrm{max}.
    \end{aligned}
\end{equation}
This result, which also holds in the classical case, we may use for an upper bound of the variance of the forward gradient, which we want to estimate using QAMC. 

\begin{result}[Forward Gradient for NN training with QAMC, First Option]
\label{thm:fwdgrad_option1}
    \, \, \, \, \, Consider a classical NN including its scalar loss function, $f_\mathrm{NN}: \mathbb{R}^d \mapsto \mathbb{R}$, with $n_\theta$ trainable parameters. Let the function computing the quantity $(\nabla f_\mathrm{NN}(X)\cdot v)v$ using forward mode AD be implemented in a quantum circuit in the form of a unitary $U'_{\mathrm{NN},q}$ as seen above, where the gradient is taken with respect to the trainable parameters in the NN. let the input to the NN be given by a $d$-dimensional random variable $X$, governed by a distribution $p_X$, as well as a $n_\theta$-dimensional random variable $v$, whose components are distributed iid according to $\mathcal{N}(0,1)$. Then we can estimate the mean output of $(\nabla f_\mathrm{NN}(X)\cdot v)v$ with respect to the distributions of $X$ and $v$ using QAMC with a sample complexity, with respect to $U'_{\mathrm{NN},q}$ of,
\begin{equation}
    \label{eqn:fwd_grad_opt1_complexity}
    O\left( n_\theta \sqrt{d g_\mathrm{max}} / \epsilon \right),
\end{equation}
    up to error $\epsilon$ in the $l_\infty$ norm with success probability 0.99.
\end{result}
As we expected, we get a slowdown of $n_\theta$ in the query complexity, but a speedup in the error tolerance $\epsilon$ compared to classically using backpropagation. We proceed to discuss the second option of applying the forward gradient method to the setting at hand.

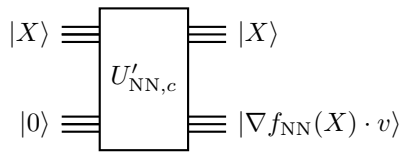
\begin{figure}[H]
\centering
\begin{quantikz}
     \lstick{$|{X}\rangle$} & \gate[2]{U'_{\mathrm{NN},c}}\qwbundle[alternate]{} &   \rstick{$|{X}\rangle$}\qwbundle[alternate]{} 
     \\
     \lstick{$\ket{0}$} & \qwbundle[alternate]{} & \rstick{$|\nabla f_\mathrm{NN}(X) \cdot v\rangle$}\qwbundle[alternate]{} 
\end{quantikz}
\caption{Unitary evaluating the directional derivative, with $v$ classical}
\label{fig:circ_nn_grad_opt1}
\end{figure}

In the second option, we classically sample the vector $v$. We use QAMC to estimate the inner product $\nabla f_\mathrm{NN} \cdot v$ with respect to the distribution of the input $X$, $p_X$, for a fixed classical sample of $v$. To compute the forward gradient, we classically postprocess $\nabla f_\mathrm{NN} \cdot v$ to arrive at $(\nabla f_\mathrm{NN} \cdot v)v$. Since $\nabla f_\mathrm{NN} \cdot v$ is a scalar, we hope to avoid a slowdown in $n_\theta$. 
We proceed to provide some equations and circuits to better illustrate this option. \Cref{fig:circ_nn_grad_opt1} displays a unitary also returning the forward gradient. We name this unitary $U'_{\mathrm{NN},c}$.

\noindent Similarly to the case in \Cref{eqn:nn_oracle_basic_input} we input the following state to $U'_{\mathrm{NN},c}$,
\begin{equation}
    \label{eqn:nn_oracle_input_opt1}
    \sum_{i} \sqrt{p_{X,i}} |X_i\rangle  |0\rangle,
\end{equation}
and get as output,
\begin{equation}
    \label{eqn:nn_oracle_output_opt1}
    \sum_{i} \sqrt{p_{X,i}} |X_i\rangle |\nabla f_\mathrm{NN}(X_i) \cdot v\rangle. 
\end{equation}
\noindent The unitary $U'_{\mathrm{NN},c}$ has access to the fixed sample of $v$ classically and returns the directional derivative using forward mode AD.  
In this option, the quantity we estimate with QAMC is $f_\mathrm{NN}(X) \cdot v$, which is a scalar. Once we have estimated $f_\mathrm{NN}(X) \cdot v$ we multiply it with the same classical sample of $v$ to get the forward gradient. One might argue now, that by carrying out the multiplication of $f_\mathrm{NN}(X) \cdot v$ and $v$ we get an overhead in runtime of $O(n_\theta)$, which is true. However, we do not need to query the unitary implementing the NN $O(n_\theta)$ times to compute the directional derivative. As outlined in \Cref{subsct:neuralnetworks,subsct:ad}, the runtime of evaluating the NN (as well as running AD in forward or reverse mode) is $O(n_\theta)$. However, we only query the NN a constant number of times to obtain a sample of the forward gradient. In the case of, e.g., numerical differentiation or the first option for using the forward gradient with QAMC outlined above, we have to query the unitary $O(n_\theta)$ times, in the case of the second option (as in the classical case with backpropagation) only once, to obtain one sample. 
For now, we proceed to further discuss the second option (where $v$ is classical). 

In order to incorporate the second option of evaluating the forward gradient method into our framework of using NNs together with QAMC and quantify a potential speedup, we need to bound the variance of the quantity $\nabla f_\mathrm{NN}(X) \cdot v$ with respect to $p_X$, if we are to estimate it using AD alongside the evaluation of the NN itself. We have,
\begin{equation}
\begin{aligned}
\label{eqn:fwd_grad_var_bound}    
    \mathbb{V}\left[ \nabla f_\mathrm{NN}(X) \cdot v \right] &= \mathbb{E}\left[ \lvert\nabla f_\mathrm{NN}(X) \cdot v\rvert^2 \right] - \mathbb{E}\left[\nabla f_\mathrm{NN}(X) \cdot v \right]^2 \\
    & \leq \mathbb{E}\left[ \lvert\nabla f_\mathrm{NN}(X) \cdot v\rvert^2 \right] \\
    & \leq \mathbb{E}\left[ \lVert\nabla f_\mathrm{NN}(X)\rVert_\infty^2 \cdot \lVert v\rVert_1^2 \right] \\
    & \leq \mathbb{E}\left[ \lVert\nabla f_\mathrm{NN}(X)\rVert_\infty^2 \cdot n_\theta \lVert v\rVert_2^2 \right] \\
    & \leq g_\mathrm{max} n_\theta \lVert v\rVert_2^2,
\end{aligned}
\end{equation}
where we used H\"older's inequality in the second inequality and the norm inequalities in the third inequality. We point out that due to the norm inequalities, the term $n_\theta$ shows up somewhat unexpectedly. We proceed to formulate the following result.

\begin{result}[Directional Derivative for NN training with QAMC, Second Option]
    \label{thm:fgnnqamc}
     \, \, \,\, \newline Consider a classical NN including its scalar loss function, $f_\mathrm{NN}: \mathbb{R}^d \mapsto \mathbb{R}$, with $n_\theta$ trainable parameters. Let the function computing the quantity $\nabla f_\mathrm{NN}(X)\cdot v$ using forward mode AD be implemented in a quantum circuit in the form of a unitary $U'_{\mathrm{NN},c}$, where the gradient is taken with respect to the trainable parameters in the NN. let the input to the NN be given by a $d$-dimensional random variable $X$, governed by a distribution $p_X$. Furthermore, let the components of $v$ be fixed as classical iid samples from the distribution $\mathcal{N}(0,1)$. Then we can estimate the mean output of $\nabla f_\mathrm{NN}(X)\cdot v$ with respect to the input distribution $p_X$ using QAMC with a query complexity, with respect to $U'_{\mathrm{NN},c}$, of,
    \begin{equation}
        \label{eqn:fwg_thm_sample}
        \Tilde{O}(\sqrt{g_\mathrm{max}n_\theta} \lVert v\rVert_2 / \epsilon)
    \end{equation}
    up to additive error $\epsilon$ with probability 0.99.
\end{result}  

As mentioned above, the entries of $v$ are iid according to $\mathcal{N}(0,1)$. Therefore, $\lVert v\rVert_2^2$ follows a $\chi^2$ distribution with mean $n_\theta$ and variance $2 n_\theta$. Using Chebyshev's inequality, we can upper bound the probability that $\lVert v\rVert_2^2$ deviates from its mean by more than $k$ standard deviations,
\begin{equation}
    \label{eqn:fwg_thm_chebyshev}
    P[\lvert \lVert v\rVert_2^2 - n_\theta\rvert \geq k\sqrt{2n_\theta}] \leq \frac{1}{k^2}.
\end{equation}
Thus, $\lVert v \rVert_2$ grows as $\sqrt{n_\theta}$, since, due to Jensen's inequality,  
\begin{equation}
    \label{eqn:jensen}
    \mathbb{E}\left[\sqrt{\lVert v \rVert_2^2}\right] \leq \sqrt{\mathbb{E}\left[ \lVert v \rVert_2^2 \right]} = \sqrt{n_\theta}.
\end{equation}
Our hope for the second option was that, by estimating only a scalar, $n_\theta$ would not show up in the query complexity, which has been dashed. 

So far in the second option we have treated $v$ as a fixed sample. However, in practice one would also have to sample $v$ multiple times. To upper bound the error when considering the stochasticity originating from $X$ as well as $v$, we proceed as follows. Let $w$ denote our estimate of $\nabla f_\mathrm{NN} \cdot v$ as in \Cref{thm:fgnnqamc}, taking into account the stochasticity of $X$ for a fixed sample of $v$. Let $(\nabla f_\mathrm{NN}\cdot v)v$ denote the forward gradient where the only source of error stems from the stochasticity of $v$, and let $\nabla f_\mathrm{NN}$ be the true gradient. Then we have,
\begin{equation}
    \label{eqn:fwd_grad_total_error}
    \begin{aligned}
    \lVert w\cdot v - \nabla f_\mathrm{NN} \lVert_\infty & = \lVert w - (\nabla f_\mathrm{NN}\cdot v)v + (\nabla f_\mathrm{NN}\cdot v)v - \nabla f_\mathrm{NN} \lVert_\infty \\
    & \leq \lVert w\cdot v - (\nabla f_\mathrm{NN}\cdot v)v \rVert_\infty + \lVert(\nabla f_\mathrm{NN}\cdot v)v - \nabla f_\mathrm{NN} \lVert_\infty \\
    & = \lVert (w - (\nabla f_\mathrm{NN}\cdot v))v \rVert_\infty + \lVert(\nabla f_\mathrm{NN}\cdot v)v - \nabla f_\mathrm{NN} \lVert_\infty \\
    & \leq 3 \lvert w - (\nabla f_\mathrm{NN}\cdot v) \rvert + \lVert(\nabla f_\mathrm{NN}\cdot v)v - \nabla f_\mathrm{NN} \lVert_\infty,
    \end{aligned}
\end{equation}
where we made use of the following observation in the last inequality. Since each entry of $v$ is distributed according to $\mathcal{N}(0,1)$, we may truncate each entry of $v$ at $\pm 3$, as the error contributions from the tails of the Gaussian outside of this range occur with a probability of roughly 0.003. This argument may also be extended to greater constants, e.g, if the Gaussians are truncated at $\pm 7$, which is still a constant, the probability of the Gaussian being outside of this range is around $2.56\cdot 10^{-12}$. In \Cref{thm:fgnnqamc} we outlined the query complexity for estimating $w$ such that the first summand from the last line of \Cref{eqn:fwd_grad_total_error} is below a certain error $\epsilon$. In order to do the same for the second summand, i.e., the query complexity considering samples of $v$ to estimate the gradient up to a certain error tolerance, we can make use of \Cref{lemma:mvmc}, as $v$ is sampled classically. To estimate the gradient $\nabla f_\mathrm{NN}$ up to error $\epsilon$ in the $l_\infty$ norm using the forward gradient $(\nabla f_\mathrm{NN}\cdot v)v$, i.e.,
\begin{equation}
    \label{eqn:classical_fwd_grad_sample}
    \lVert (\nabla f_\mathrm{NN}\cdot v)v - \nabla f_\mathrm{NN} \rVert_\infty < \epsilon,
\end{equation}
we upper bound $\lVert (\nabla f_\mathrm{NN}\cdot v)v\rVert_\infty$ as follows. Making use of the same arguments as in \Cref{eqn:fwd_grad_var_bound}, we have,
\begin{equation}
    \label{eqn:fwd_grad_upper_bound_infty}
    \begin{aligned}
        \lVert (\nabla f_\mathrm{NN}\cdot v)v\rVert_\infty & \leq 3 \lvert \nabla f_\mathrm{NN}(X) \cdot v\rvert \\
        & \leq 3\lVert\nabla f_\mathrm{NN}(X)\rVert_\infty \cdot \sqrt{n_\theta} \lVert v\rVert_2 \\
        & = O\left( \sqrt{g_\mathrm{max}n_\theta} \lVert v\rVert_2 \right) \\
        & = O\left( \sqrt{g_\mathrm{max}} n_\theta \right),
    \end{aligned}
\end{equation}
in expectation, where we use the same argument as in \Cref{eqn:fwd_grad_total_error} for upper bounding the greatest entry of $v$. Consequently, with success probability $\delta$, we need,
\begin{equation}
\label{eqn:classical_fwd_grad_samples}    
O\left( \frac{g_\mathrm{max} n_\theta^2}{\epsilon^2}\log\frac{n_\theta}{\delta} \right),
\end{equation}
many samples in expectation.  
Since for each sample of $v$, we require the sample complexity outlined in \Cref{thm:fgnnqamc}, we formulate the following corollary.

\begin{corollary}[Forward Gradient for NN training with QAMC, Second Option]
    \label{cor:fwd_grad_mixed}
    \,\,\,\,\,\, Consider a classical NN including its scalar loss function, $f_\mathrm{NN}: \mathbb{R}^d \mapsto \mathbb{R}$, with $n_\theta$ trainable parameters. Let the function computing the quantity $\nabla f_\mathrm{NN}(X)\cdot v$ using forward mode AD be implemented in a quantum circuit in the form of a unitary $U'_{\mathrm{NN},c}$, where the gradient is taken with respect to the trainable parameters in the NN. let the input to the NN be given by a $d$-dimensional random variable $X$, governed by a distribution $p_X$. Furthermore, let the components of $v$ be sampled as classical iid samples from the distribution $\mathcal{N}(0,1)$. Then we can estimate the mean of $(\nabla f_\mathrm{NN}(X)\cdot v)v$ with respect to the input distribution $p_X$ and the distribution $p_v$ of $v$ using QAMC to estimate the directional derivative $\nabla f_\mathrm{NN}(X)\cdot v$ for a fixed sample of $v$, followed by classical MVMC estimation in $v$ as in \Cref{lemma:mvmc} with a query complexity, with respect to $U'_{\mathrm{NN},c}$, of,
    \begin{equation}
    \label{eqn:fwd_grad_total_samples}
    \Tilde{O}\left(\frac{(n_\theta^{2.5} g_\mathrm{max}^{1.5}}{\epsilon^3}\right).
    \end{equation}
    up to error $\epsilon$ in the $l_\infty$ norm with probability 0.99.
\end{corollary}

We proceed to summarise the query complexities for estimating the gradient of a NN in different settings and using different methods. We also include the purely classical results. The query complexities are for estimating the gradient $\nabla f_\mathrm{NN}$ up to error $\epsilon$ in the $l_\infty$ norm with success probability 0.99. 

\begin{table}[ht]
\renewcommand{\arraystretch}{1.5}
\begin{tabular}{l|l|l|l}
$X$       & $v$       & Method                    & Query Complexity \\ \hline
C & -         & Backpropagation           &     $\Tilde{O}\left(g_\mathrm{max}/\epsilon^2  \right)$ (\Cref{lemma:mvmc} and \Cref{eqn:gmax_def})           
\\ \hline
C & C & Forward Gradient          &        $\Tilde{O}\left( d^4 g_\mathrm{max}/\epsilon^2 \right)$ (\Cref{lemma:mvmc} and \Cref{eqn:fwd_grad_upper_bound_infty})         
\\ \hline
C & -         & Numerical Differentiation &       $\Tilde{O}\left( d^2 g_\mathrm{max}/\epsilon^2  \right)$ (\Cref{eqn:chebyshev,eqn:num_diff,eqn:gmax_def})          
\\ \hline
Q & -         &  Backpropagation &       $\Tilde{O}\left(d^2\sqrt{g_\mathrm{max}}/\epsilon  \right)$ (\Cref{lemma:qmc} and \Cref{eqn:gmax_def})           
\\ \hline
Q   & C & Forward Gradient          &      $\Tilde{O}\left(d^5 g_\mathrm{max}^{1.5}/\epsilon^3 \right)$ (\Cref{cor:fwd_grad_mixed})         
\\ \hline
Q   & Q   & Forward Gradient          &      $\Tilde{O}\left(  d^{2.5}\sqrt{g_\mathrm{max}} / \epsilon  \right)$ (\Cref{thm:fwdgrad_option1})           
\\ \hline
Q   & -         & Numerical Differentiation &    $\Tilde{O}\left( d^{2.5}\sqrt{g_\mathrm{max}} / \epsilon  \right)$ (\Cref{lemma:qmc} and \Cref{eqn:var_bound_final})             
\end{tabular}
\caption{Query complexities (to the NN) to estimate the gradient of a NN with respect to its trainable parameters for different methods when the input $X$ is sampled from a distribution $p_X$ and $v$ (for the forward gradient) from $p_v$, either classically (C) or quantumly (Q). We make use of \Cref{lemma:params_dimension}, replacing $n_\theta$ by $d^2$ in the complexities and taking the expectation value for the terms containing norms of $v$, as in \Cref{eqn:fwg_thm_chebyshev,eqn:jensen}.}
\label{tab:grad_query_complexities}
\end{table}
Some further comments on are in order. Compared to the classical case of using the forward gradient method to train the NN, we get a speedup in $g_\mathrm{max}$ and $\epsilon$ by using QAMC to estimate $(\nabla f_\mathrm{NN}\cdot v)v$ in the fully quantum case. 
However, in the classical case one would presumably resort to backpropagation to compute the gradient of the NN with respect to the trainable parameters instead of using the forward gradient method (provided memory constraints are not a concern), as no additional stochasticity is introduced. 
When using backpropagation, the randomness with respect to the input $X$ would remain, but there is no need for sampling $v$. Without the stochasticity introduced by $v$, the sample complexity is proportional to $\epsilon^{-2}$, as per \Cref{lemma:mvmc}. It is also worth pointing out that in all the scenarios where $X$ is quantum, we have a slowdown of at least $d^2$ compared to classical backpropagation. This stems from the fact that QAMC, as seen in \Cref{lemma:qmc}, only allows for the estimation of a scalar at a time, unlike in the classical case (see the discussion in \Cref{subsubsubsct:multivar_mc}). Nevertheless, we have shown that with the purely quantum cases, we can achieve a speedup in $\epsilon$, as we had hoped for. The case where $X$ is quantum and $v$ is classical did not turn out to give us the best of both worlds, as we had envisaged. This is due to the fact that, while we can get a speedup in estimating the directional derivative (in an inner loop), we have an outer loop where we sample $v$ classically, resulting in the slowdown in $\epsilon$. This consideration is shown mathematically in \Cref{eqn:fwd_grad_total_error}.

To conclude, as seen in \Cref{thm:nn_qamc_eval}, when using QAMC to estimate the loss function of the NN where the input is sampled from a known distribution, we achieve a quadratic speedup for the query complexity in the error tolerance $\epsilon$ compared to the classical case, as was our goal when applying QAMC. 
When training classical NNs implemented in quantum circuits where the input is given by samples from a known distribution, we compared several classical as well as quantum alternatives. 
As seen in \Cref{tab:grad_query_complexities}, by using QAMC to estimate the gradient of the NN, we get a slowdown in $d$ compared to classical backpropagation, but a speedup in $\epsilon$. This holds when applying QAMC to backpropagation, the forward gradient method (in the fully quantum scenario) as well as numerical differentiation. 
However, compared to when QAMC is applied to backpropagation, an application of QAMC to the forward gradient method and numerical differentiation incurs a slowdown of $\sqrt{d}$.
The forward gradient method, however, being based on AD, is less prone to numerical issues as numerical differentiation, and has the potential for being less memory intensive than backpropagation, which is vital in quantum circuits.

\section{Quantum-Accelerated Monte Carlo Methods and PDEs}
\label{sct:qmc_pde}
In this section we proceed with the algorithm from \cite{Han_2018} to explore the possible application of quantum-accelerated Monte Carlo (QAMC). As outlined in \Cref{subsct:dl_architecture}, the deep learning algorithm from \cite{Han_2018} uses neural networks (NNs) to solve nonlinear partial differential equations (PDEs) by reformulating the nonlinear PDE as a stochastic differential equation (SDE). 
In order to approximate the gradient in each temporal discretisation step of the function described by the nonlinear PDE, a NN is employed to learn the spatial gradient based on samples from the SDE governing the spatial variable. In order to achieve a quantum speedup in the algorithm presented  in \cite{Han_2018}, one possibility consists of using quantum subroutines for estimating the loss function and the solution to the PDE, which depend on the stochastic process governing the spatial variable. A method which may allow us to do so is QAMC, presented in \Cref{subsubsubsct:qmc}, by addressing the bottleneck imposed by Chebyshev's inequality.

\subsection{Applying Quantum-Accelerated Monte Carlo to Loss Function Estimation}
\label{subsct:applying-qmc}

\textbf{Setting} $-$ We now proceed to apply QAMC to the architecture from \cite{Han_2018} with the hope of achieving a quantum speedup. We briefly outline how we aim to do so, and then proceed to elaborate on the individual steps. Our goal is to estimate the payoff function $f_p$ from \Cref{eqn:payoff}, $\left|g(\hat{X}_{t_N}) - \hat{u}_{t_N}\right|^2$, as well as the gradients of the parameters of the NNs, with QAMC.

In order to apply QAMC to the setting from \cite{Han_2018}, we begin by representing the stochastic process governing the stochastic process $X_t$ (see \Cref{eqn:stoch_proc}) in its differential form as follows,
\begin{equation}
    \mathrm{d}X_t = \mu(t, X_t) + \sigma(t, X_t)\mathrm{d}W_t,
    \label{eqn:stoch_proc_diff}
\end{equation}
where $\mu$ and $\sigma$ are real functions and $W_t$ is a standard Brownian motion (as defined in \Cref{subsct:dl_architecture}). Furthermore, we also represent $u_t$ (see \Cref{eqn:bsde}) in its differential form as follows,
\begin{equation}
    \label{eqn:bsde_diff}
    \mathrm{d}u_t = - f(t, X_t, u(t,X_t), \sigma^\top(t,X_t)\nabla u(t,X_t)) + \nabla u(t, X_t)^\top\sigma (t, X_t)\mathrm{d}W_t,
\end{equation}
where $f$ is a real function and in our setting, as seen in \Cref{eqn:ffnn}, $\nabla u(t, X_t)^\top\sigma (t, X_t) = (\sigma^\top\nabla u)(t, X_t)$ is a function represented by the NNs. Next, we need to make a set of assumptions, the use of which we will lay out shortly.

\begin{assumption}
    \label{ass:musig_lipschitz}
    We assume that $\mu$, $\sigma$, $f$ and $\sigma^\top\nabla {u}$ are Lipschitz continuous, as introduced in \Cref{subsct:notation}, on the domain of interest in the squared $l_2$ norm. 
\end{assumption} 
By making \Cref{ass:musig_lipschitz}, we derive the following growth bound on the functions mentioned in \Cref{ass:musig_lipschitz}. For $\mu$ we have,
\begin{equation}
    \label{eqn:nn_growth_bound3}
    \begin{aligned}
    \lVert \mu(t,X) \rVert_2^2 &= \lVert \mu(t,X) - \mu(0,0) + \mu(0,0) \rVert_2^2 \\ 
    &\leq \lVert \mu(t,X) - \mu(0,0)\rVert_2^2 + \lVert \mu(0,0) \rVert_2^2 \\
    & \leq L_{\mu} (t^2 + \lVert X \rVert_2^2) + \lVert \mu(0,0) \rVert_2^2 \\
    & \leq (L_{\mu} + \lVert \mu(0,0) \rVert_2^2) (1 + t^2 + \lVert X \rVert_2^2).
    \end{aligned}
\end{equation}
Similar bounds may be derived analogously for the other functions mentioned in \Cref{ass:musig_lipschitz}. We continue with our next assumptions.
\begin{assumption}
    \label{ass:finite_x0}
    We assume that $\mathbb{E}(\lVert X_{t_0}\rVert_2^m) < \infty$ for $m\geq0$.
\end{assumption}
\Cref{ass:musig_lipschitz,ass:finite_x0} guarantee the existence and uniqueness of a strong solution of the SDEs, meaning that for every Brownian path that $W_t$ may take, the SDEs are guaranteed to have a unique solution \cite{kloeden1992stochastic}. The next assumption we will make use of as this section progresses.

\begin{assumption}
    \label{ass:payoff_lipschitz}
    We assume that the function inside the expectation value in \Cref{eqn:loss}, which we call the payoff function $f_p$ (see \Cref{eqn:payoff}), is Lipschitz continuous on the domain of interest in the sense that
    \begin{equation}
        \lvert f_p(\{\hat{X}_{t_n}\}_{0\leq n\leq N}, \{\Delta W_{t_n}\}_{0\leq n\leq N}) - f_p(\{\Tilde{X}_{t_n}\}_{0\leq n\leq N}, \{\Delta \Tilde{W}_{t_n}\}_{0\leq n\leq N}) \rvert^2 \leq L_{f_p}  \sup_{0\leq n\leq N}\lVert \hat{X}_{t_n} - \Tilde{X}_{t_n} \rVert_2^2.
    \end{equation}
\end{assumption}
Analogously to \Cref{eqn:nn_growth_bound3}, we furthermore derive the following growth bound for $f_p$,
\begin{equation}
    \label{eqn:payoff_growth_bound}
    \begin{aligned}
    f_p(\{\hat{X}_{t_n}\}_{0\leq n\leq N}, \{\Delta W_{t_n}\}_{0\leq n\leq N}) &\leq \left(L_{f_p} + \lvert f_p(\{\hat{Y}_{t_n}\}_{0\leq n\leq N})\rvert^2\right) (1+\sup_{0\leq n\leq N}\lVert\hat{X}_{t_n}\rVert_2^2) \\ & \coloneqq K_{f_p} (1+\sup_{0\leq n\leq N}\lVert\hat{X}_{t_n}\rVert_2^2),
    \end{aligned}
\end{equation}
where $Y_{t_n}$ signifies the path such that $\sup_{0\leq n\leq N}\lVert\hat{X}_{t_n} - Y_{t_n}\rVert_2^2 = \sup_{0\leq n\leq N}\lVert\hat{X}_{t_n}\rVert_2^2$ for all other paths $\hat{X}_{t_n}$.

Note that the payoff function presented in \Cref{eqn:payoff}, $\left|g(\hat{X}_{t_N}) - \hat{u}_{t_N}\right|^2$, is not globally Lipschitz continuous in general, and also involves outputs of the nonlinear function $f$. To avoid violating the Lipschitz condition, we argue that in any practical setting, one would only deal with a bounded interval of interest. Assuming that the loss function does not contain any singularities on the interval of interest, it is Lipschitz continuous on the same interval, with the Lipschitz constant being the maximum of its derivative on the interval. The same argument can be applied to justify the growth bound assumption.

As mentioned above, we make \Cref{ass:musig_lipschitz,ass:finite_x0} so that \Cref{eqn:stoch_proc_diff,eqn:bsde_diff} are guaranteed to have a unique strong solution \cite{kloeden1992stochastic}. For the solution of the SDE it then holds that,
\begin{equation}
    \label{eqn:existence}
    \sup_{t_0 \leq t\leq T} \mathbb{E}\left[\lVert X_t\rVert_2^2 \right] < \infty, 
\end{equation}
as well as,
\begin{equation}
    \label{eqn:stoch_proc_growth_bound}
    \mathbb{E}\left[\sup_{0\leq n\leq N}\lVert{X}_{t_n}\rVert_2^2\right] \leq \mathbb{E}\left[ C(1+\lVert X_{t_0} \rVert_2^2 \right],
\end{equation}
where $C$ may depend on the time interval $T-t_0$ and on the Lipschitz constants of $\mu$ and $\sigma$ \cite{kloeden1992stochastic}. If there is a solution to an SDE of the form in \Cref{eqn:stoch_proc_diff,eqn:bsde_diff} for each given path the Brownian process $W_t$ takes, we say that the SDE has a strong solution \cite{kloeden1992stochastic}. \Cref{ass:musig_lipschitz,ass:finite_x0} are satisfied (at least on a bounded domain) for several important special cases surveyed in \cite{Han_2018}, such as the nonlinear Black-Scholes equation, the Hamilton-Jacobi-Bellman equation as well as the Allen-Cahn equation. In the same three example cases, \Cref{ass:payoff_lipschitz} is satisfied for bounded domains.

In order to present the cost of using QAMC to solve PDEs with the architecture from \cite{Han_2018} we assume that access to $t_0$, $u_0$, $\sigma^\top \nabla u_0$, $X_{t_0}$, $\mu$, $\sigma$ as well as $f$ is given via unitaries, which we denote by $U_{t_0}$, $U_{u_0}$, $U_{\sigma^\top \nabla u_0}$, $U_{X_{t_0}}$, $U_\mu$, $U_\sigma$ and $U_f$, respectively. In addition, we assume that we have access to a unitary $U_{\mathrm{Gauss}}$ which can prepare a state of the form presented in \Cref{def:dist_loading} with a Gaussian distribution. We also assume that evaluating the NNs (i.e., carrying out a single feedforward pass) is done by querying a unitary $U_{\mathrm{NN}}$. Since the architecture of all the NNs in \cite{Han_2018} is the same, we also treat the unitaries for the different NNs in the architecture from \cite{Han_2018} as having the same cost from a query complexity point of view. 

\textbf{Variance bound} $-$ In order to apply \Cref{lemma:qmc} to the architecture outlined in \cite{Han_2018} and quantify a potential quantum speedup, we proceed to find an upper bound on the variance $\lambda^2$ of the quantity whose expectation value we aim to estimate, namely the payoff function from \Cref{eqn:payoff}. We have for $\lambda^2$,
\begin{equation}
    \label{eqn:var_bound}
    \begin{aligned}
    \lambda^2 &= \mathbb{V}[f_p(\{\hat{X}_{t_n}\}_{0\leq n\leq N}, \{\Delta W_{t_n}\}_{0\leq n\leq N})] 
    \\
    & \leq \mathbb{E}[\lvert f_p(\{\hat{X}_{t_n}\}_{0\leq n\leq N}, \{\Delta W_{t_n}\}_{0\leq n\leq N})\rvert^2] 
    \\
    & \leq \mathbb{E}\left[K_{f_p}\left(1 + \sup_{0\leq n\leq N} \lVert\hat{X}_{t_n}\rVert_2^2\right)\right] 
    \\
    & = \mathbb{E}\left[K_{f_p}\left(1 + \sup_{0\leq n\leq N} \lVert\hat{X}_{t_n} - {X}_{t_n} + {X}_{t_n}\rVert_2^2\right)\right] \\
    & \leq \mathbb{E}\left[K_{f_p}\left(1 + \sup_{0\leq n\leq N} \lVert\hat{X}_{t_n} - {X}_{t_n}\rVert_2^2 + \sup_{0\leq n\leq N}\lVert{X}_{t_n}\rVert_2^2\right)\right] 
    \\
    & \leq K_{f_p}\left(1 +  \mathbb{E}\left[\sup_{0\leq n\leq N}\lVert\hat{X}_{t_n} - {X}_{t_n}\rVert_2^2\right] + \mathbb{E}\left[\sup_{0\leq n\leq N}\lVert{X}_{t_n}\rVert_2^2\right]\right) 
    \\
    & \leq K_{f_p}\left(1 +  K_2(\Delta t)^{2r} + C \left(1 + \lVert X_{t_0}\rVert_2^2 \right)\right) \eqqcolon \lambda_\mathrm{max}^2,
    \end{aligned}
\end{equation}
where the first inequality follows from the definition of the variance, the second inequality stems from the growth bound of the payoff function (see \Cref{ass:payoff_lipschitz}), and the last inequality stems from the definition the strong order $r$ in \Cref{eqn:strong_order} and from \Cref{eqn:stoch_proc_growth_bound}, thanks to \Cref{ass:musig_lipschitz,ass:finite_x0,ass:payoff_lipschitz}. Note that the bound in \Cref{eqn:var_bound} also holds in the classical case, as we have not yet employed any quantum subroutines.

\textbf{Error Analysis} $-$ Before going through the application of QAMC to the architecture from \cite{Han_2018}, we comment on the error sources when estimating the payoff function $f_p$, in order to understand what effects different errors have and how they differ in the classical and quantum scenario. Note that the inaccuracy of the NNs when representing the spatial gradient as well as the error stemming from the temporal discretisation of the SDEs are not error sources hindering us at accurately estimating the payoff function, rather they have an effect on the (true) value of the solution. When estimating the payoff function, we consider two kinds of errors.

\begin{enumerate}
    \item{The discretisation error of the Gaussian increments $\Delta W_t$. The estimate of the payoff function considering only the discretisation error of the Gaussian increments $\Delta W_t$ we denote as $I_2$.}
    \item{The estimation error, when using QAMC. The estimate which contains both kinds of errors we denote by $I_3$.}
\end{enumerate}
In order to bound the error between the ideal estimation of the payoff function (with no error whatsoever), which we denote by $I_1$, and our actual estimate containing both errors, $I_3$, we proceed as follows,
\begin{equation}
    \label{eqn:error_bound}
    \lvert I_1 - I_3\rvert = \lvert I_1 - I_2 + I_2 - I_3\rvert \leq \lvert I_1 - I_2\rvert + \lvert I_2 - I_3\rvert.
\end{equation}
In order to upper bound the total error $\lvert I_1 - I_3\rvert$ by $\epsilon$ we proceed to bound the terms on the right hand side of \Cref{eqn:error_bound}. 

We begin by bounding the first error term, $\lvert I_1 - I_2 \rvert$, which results from discretising the Gaussian increments $\Delta W_{t_n}$. The number $N_\mathrm{Gauss} = 2^{n_\mathrm{Gauss}}$ represent how finely we discretise the Gaussian distribution, where $n_\mathrm{Gauss}$ is the number of qubits we use to represent a single univariate Gaussian distribution. The error resulting from this discretisation of the Gaussian increments may be bounded by a Riemann sum. In general, the following holds for a left- or right-rule Riemann sum approximating an integral of a function $f_r$ \cite{hughes2020calculus},
\begin{equation}
    \label{eqn:riemann_sum_error}
    \left| \int_a^b f_r(x) \,dx - \sum_{k=0}^{n-1} \Delta x f_r(a + k\cdot\Delta x) \right| \leq \frac{L_{f_r} (b-a)^2}{2n},
\end{equation}
where $\Delta x = \frac{\lvert a - b \rvert}{N}$, $L_{f_r}$ is the Lipschitz constant of $f_r$ (on the relevant interval) and $n$ represents how finely we discretise the integral. In the case of two integrals being approximated by a left- or right-rule Riemann sum, we have,
\begin{equation}
    \label{eqn:2_riemann}
    \begin{aligned}
    & \left| \int_a^b \int_c^d f_r(x,y) \,dydx - \sum_{k=0}^{n-1} \sum_{l=0}^{m-1} \Delta x \Delta y f_r(a + k\cdot\Delta x, c + l\cdot \Delta y) \right| \\ 
    & \leq \left| \int_a^b \int_c^d f_r(x,y) \,dydx - \int_a^b \sum_{l=0}^{m-1} \Delta y f_r(x, c + l\cdot \Delta y)\,dx \right| \\ 
    & + \left| \int_a^b \sum_{l=0}^{m-1} \Delta y f_r(x, c + l\cdot \Delta y)\,dx - \sum_{k=0}^{n-1} \sum_{l=0}^{m-1} \Delta x \Delta y f_r(a + k\cdot\Delta x, c + l\cdot \Delta y) \right| \\ 
    & \leq \int_a^b \left| \int_c^d f_r(x,y) \,dy - \sum_{l=0}^{m-1} \Delta y f_r(x, c + l\cdot \Delta y) \right|\,dx \\ 
    & + \sum_{l=0}^{m-1} \Delta y \left| \int_a^b f_r(x, c + l\cdot \Delta y)\,dx - \sum_{k=0}^{n-1} \Delta x f_r(a + k\cdot\Delta x, c + l\cdot \Delta y) \right| \\
    & \leq \frac{L_{f_r}(d-c)^2}{2m} \lvert b-a\rvert + \frac{L_{f_r}(b-a)^2}{2n}\lvert d-c\rvert.
    \end{aligned}
\end{equation}
If, as in our case of interest, $a=c$, $b=d$ and $n=m$ we end up with an error bound of,
\begin{equation}
    \label{eqn:2_riemann_error}
    \frac{2 L_{f_r}\lvert b-a\rvert^3}{2n}.
\end{equation}
We generalise the above error to the case of $M$ integrals and we get an error bound of,
\begin{equation}
    \label{eqn:m_riemann_error}
    \frac{M L_{f_r}\lvert b-a\rvert^{M+1}}{2n}.
\end{equation}
We return to our setting where we are analysing the discretisation error for the Gaussian increments. If we set the bounds to be at $\pm 3\sqrt{\Delta t}$ (recall that $\Delta t$ is the variance of our Gaussian increments) away from the mean, respectively, we can consider the error contributions from the tails of the Gaussian which lie outside of our discretised range to be negligible, at roughly $0.003$. We then upper bound the error resulting from the discretisation of the Gaussian increments by using \Cref{eqn:m_riemann_error} as follows,
\begin{equation}
    \label{eqn:error_bound_2}
    \begin{aligned}
    \lvert I_1 - I_2 \rvert \leq \frac{N d L_{f_p}\lvert 6\Delta t\rvert^{Nd+1}}{2N_\mathrm{Gauss}},
    \end{aligned}
\end{equation}
where $N$, $t_0$ and $T$ are as seen in \Cref{fig:dl_arch}, and $L_{f_p}$ is the Lipschitz constant of the payoff function, see \Cref{ass:payoff_lipschitz}. Note that this result also holds for the classical case (i.e. using classical MC methods) and is in line with a similar result in \cite{zeng2021threshold}. In order to upper bound the error from discretising the Gaussian increments by $\epsilon/3$, we choose,
\begin{equation}
    \label{eqn:ngauss}
    N_\mathrm{Gauss} \geq \frac{3Nd L_{f_p}\lvert 6\Delta t\rvert^{Nd+1}}{2\epsilon} = O\left( \epsilon^{-1-1/r} d L_{f_p}\lvert 6\Delta t\rvert^{\epsilon^{-1/r} d +1} \right),
\end{equation}
where we inserted the relation between $N$ and $\epsilon$ from \Cref{eqn:N_eps_bound}. Since $N_\mathrm{Gauss} = 2^{n_\mathrm{Gauss}}$, where $n_\mathrm{Gauss}$ is the number of qubits we require for a discretisation of a one-dimensional univariate Gaussian distribution, we thus have to deploy,
\begin{equation}
    \label{eqn:ngauss_qubits}
    n_\mathrm{Gauss} = \Tilde{O}\left( \epsilon^{-1/r} d \right),
\end{equation}
qubits for the discretisation of the Gaussian increments in the whole architecture from \cite{Han_2018} to achieve the desired precision. In the above calculation we did not make use of any quantum subroutines, and thus the analysis of the error stemming from the discretisation of the Gaussian increments holds in the classical case (for bits) as well as in the quantum case (for qubits).

The second error source when estimating the payoff function stems from estimating $I_2$ with $I_3$. According to \Cref{lemma:powering} and \Cref{lemma:qmc} there exists a quantum algorithm that estimates $I_2$ up to error,
\begin{equation}
    \label{eqn:error_mc}
    \lvert I_2 - I_3 \rvert \leq \epsilon/3,
\end{equation}
with probability at least 0.99 with $\Tilde{O}(\epsilon^{-1})$ estimations of $I_2$. Using classical Monte Carlo (MC) methods, we would require $\Tilde{O}(\epsilon^{-2})$ estimations. This is where the quantum speedup comes into play.

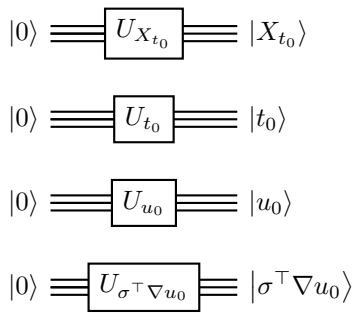
\begin{figure}[H]
\centering
\begin{quantikz}
     \lstick{$\ket{0}$} & \gate{U_{X_{t_0}}}\qwbundle[alternate]{} &   \rstick{$\ket{X_{t_0}}$}\qwbundle[alternate]{} \\
     \lstick{$\ket{0}$} & \gate{U_{t_0}}\qwbundle[alternate]{} & \rstick{$\ket{t_0}$}\qwbundle[alternate]{} \\
     \lstick{$\ket{0}$} & \gate{U_{u_0}}\qwbundle[alternate]{} & \rstick{$\ket{u_0}$}\qwbundle[alternate]{} \\
     \lstick{$\ket{0}$} & \gate{U_{\sigma^\top \nabla u_0}}\qwbundle[alternate]{} & \rstick{$\ket{\sigma^\top \nabla u_0}$}\qwbundle[alternate]{} 
\end{quantikz}

\caption{Quantum circuit preparing the initial values}
\label{fig:circ_init_preparation}
\end{figure}

\textbf{Quantum Circuit} $-$ Having analysed the error sources when estimating the loss function, we now proceed to describe quantum circuit for the proposed algorithm step by step, by writing out the state of the circuit after each operation, and simultaneously keeping count of the number of unitary calls and arithmetic operations we make. We assume that we have a sufficient number of qubits available to us. We indicate the spare qubits, which we may use in the future, by $|0\cdots 0\rangle$.  
We start off by calling the unitaries $U_{X_{t_0}}$, $U_{t_0}$, $U_{\sigma^\top \nabla u_0}$ and $U_{u_0}$ to load the initial values $t_0$, $u_0$, $\sigma^\top \nabla u_0$ and $X_{t_0}$, respectively. Each of these unitaries we call only once for each estimation of the mean in \Cref{eqn:loss}. After calling each of these unitaries, our quantum state is,
\begin{equation}
    \label{eqn:circuit_init}
    |X_{t_0}\rangle
    |t_0\rangle
    |\sigma^\top \nabla u_0\rangle
    |u_0\rangle  
    |0\cdots 0\rangle.
\end{equation}
In our circuit diagrams, the number of wires is generally not representative of the number of qubits. Bearing this consideration in mind, we display the circuit in \Cref{fig:circ_init_preparation} as follows.

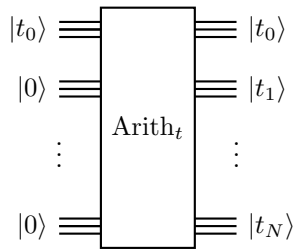
\begin{figure}[H]
\centering
\begin{quantikz}
     \lstick{$\ket{t_0}$} & \gate[4, nwires=3]{\mathrm{Arith}_{t}}\qwbundle[alternate]{} & \rstick{$\ket{t_0}$}\qwbundle[alternate]{} 
     \\
     \lstick{$\ket{0}$} & \qwbundle[alternate]{} & \rstick{$\ket{t_1}$}\qwbundle[alternate]{}
     \\
     \vdots && \vdots \\
     \lstick{$\ket{0}$} & \qwbundle[alternate]{} & \rstick{$\ket{t_N}$}\qwbundle[alternate]{}
\end{quantikz}

\caption{Quantum circuit preparing the time steps $t_n$}
\label{fig:circ_t_preparation}
\end{figure}

\noindent Note that since the initial values featured in \Cref{fig:circ_init_preparation} are classical parameters and will not be in superposition or entangled later on, storing the initial values in qubits is not necessary, they could be introduced into the circuit classically. However, in view of later operations on the quantum circuit, we think it nevertheless makes sense for illustrative reasons to prepare the initial values in qubits. Next, we use $N$ additions to increase $t$,
\begin{equation}
    \label{eqn:circuit_t}
    |X_{t_0}\rangle
    |t_0\rangle
    |\sigma^\top \nabla u_0\rangle
    |u_0\rangle  
    |t_1\rangle 
    \cdots  
    |t_N\rangle
    |0\cdots 0\rangle.
\end{equation}
The circuit adding $\Delta t$ is displayed in \Cref{fig:circ_t_preparation} (we zoom into the states representing the time stamps). The gate labeled $\mathrm{Arith}_{t}$ repeatedly adds $\Delta t$, i.e., it adds $\Delta t$ to $t_0$ to get $t_1$, then adds $\Delta t$ to $t_1$ to get $t_2$ and so on, for a total of $N$ additions.

\noindent For the Gaussian increments we need $N\cdot d$ univariate Gaussian random variables (RVs) to represent the $N$ $d$-dimensional Gaussian RVs, whose entries are identically and independently distributed (iid), loaded in the sense of \Cref{def:dist_loading}. Therefore, we need to call $U_\mathrm{Gauss}$ $N\cdot d$ times. After calling $U_\mathrm{Gauss}$ $N\cdot d$ times, we have,
\begin{equation}
\begin{aligned}
    \label{eqn:circuit_W}
    |X_{t_0}\rangle
    |t_0\rangle
    |\sigma^\top \nabla u_0\rangle
    |u_0\rangle
    \sum_{k_{t_0,1}=1}^{N_\mathrm{Gauss}} 
    \cdots 
    \sum_{k_{t_{N-1},d}=1}^{N_\mathrm{Gauss}} &
    \sqrt{p_{k_{t_0,1}}}
    \cdots
    \sqrt{p_{k_{t_{N-1},d}}}
    |k_{t_0,1}\rangle|\Delta W_{t_0,1}(k_{t_0,1})\rangle
    \cdots \\ &
    |k_{t_{N-1},d}\rangle|\Delta W_{t_{N-1},1}(k_{t_{N-1},d})\rangle \\ &
    |t_1\rangle 
    \cdots  
    |t_N\rangle
    |0 \cdots 0\rangle.
    \end{aligned}
\end{equation}
The circuit for this computation is depicted in \Cref{fig:circ_gr_preparation} below.

\begin{figure}[H]
\centering
\begin{quantikz}
     \lstick{$\ket{0}\ket{0}$} & \gate{U_{\mathrm{Gauss}}}\qwbundle[alternate]{} &   \rstick{$\sum_{k_{t_0,1}=1}^{N_\mathrm{Gauss}}\sqrt{p_{k_{t_0,1}}}|k_{t_0,1}\rangle|\Delta W_{t_0,1}(k_{t_0,1})\rangle$}\qwbundle[alternate]{} \\
     & \cdots \\
     \lstick{$\ket{0}\ket{0}$} & \gate{U_{\mathrm{Gauss}}}\qwbundle[alternate]{} &   \rstick{$\sum_{k_{t_{N-1},d}=1}^{N_\mathrm{Gauss}}\sqrt{p_{k_{t_{N-1},d}}}|k_{t_{N-1},d}\rangle|\Delta W_{t_{N-1},1}(k_{t_{N-1},d})\rangle$}\qwbundle[alternate]{} \\
\end{quantikz}

\caption{Quantum circuit preparing the Gaussian superpositions}
\label{fig:circ_gr_preparation}
\end{figure}
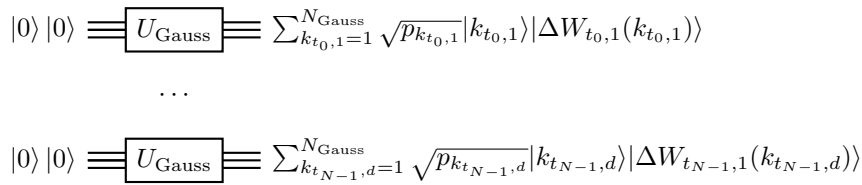
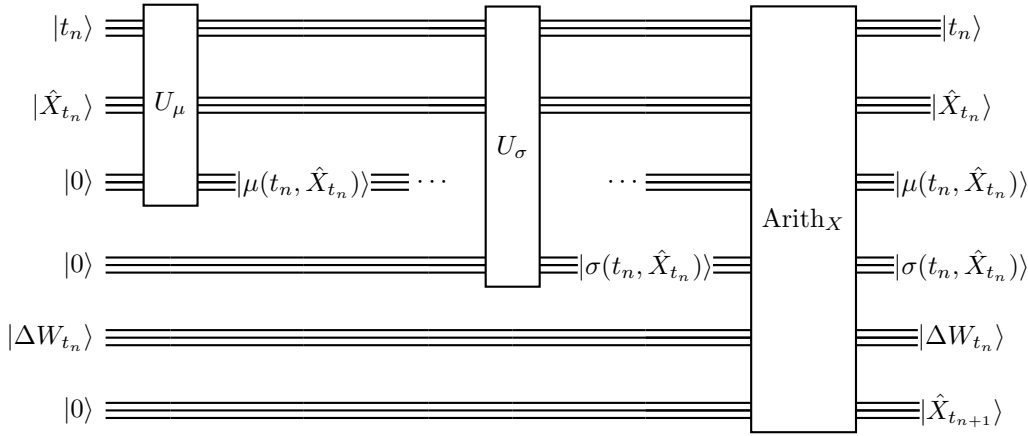
\begin{figure}[H]

\centering
\begin{quantikz}  
\lstick{$|t_n\rangle$} & \gate[wires=3]{U_\mu}\qwbundle[alternate]{} & \qwbundle[alternate]{} & \qwbundle[alternate]{} & \gate[4, nwires=3]{U_\sigma}\qwbundle[alternate]{} & \qwbundle[alternate]{} & \gate[wires=6]{\mathrm{Arith}_X}\qwbundle[alternate]{} & \push{\ket{t_n}}\qwbundle[alternate]{}
\\
\lstick{$|\hat{X}_{t_n}\rangle$}& \qwbundle[alternate]{} & \qwbundle[alternate]{} & \qwbundle[alternate]{} & \qwbundle[alternate]{} & \qwbundle[alternate]{} & \qwbundle[alternate]{} & \push{\altket{\hat{X}_{t_n}}}\qwbundle[alternate]{}
\\
\lstick{$|0\rangle$}& \qwbundle[alternate]{} & \push{\altket{\mu(t_n, \hat{X}_{t_n})}}\qwbundle[alternate]{} & \qwbundle[alternate]{}\hspace{.1cm} \cdots && \hspace{-0.5cm}\cdots\hspace{.1cm} & \qwbundle[alternate]{} & \push{\altket{\mu(t_n, \hat{X}_{t_n})}}\qwbundle[alternate]{}
\\
\lstick{$|0\rangle$} & \qwbundle[alternate]{} & \qwbundle[alternate]{} & \qwbundle[alternate]{} & \qwbundle[alternate]{} & \push{\altket{\sigma(t_n, \hat{X}_{t_n})}}\qwbundle[alternate]{} & \qwbundle[alternate]{} & \push{\altket{\sigma(t_n, \hat{X}_{t_n})}}\qwbundle[alternate]{}
\\
\lstick{$|\Delta W_{t_n}\rangle$} & \qwbundle[alternate]{} & \qwbundle[alternate]{} & \qwbundle[alternate]{} & \qwbundle[alternate]{} & \qwbundle[alternate]{} & \qwbundle[alternate]{} & \push{\altket{\Delta W_{t_n}}}\qwbundle[alternate]{}
\\
\lstick{$|0\rangle$} & \qwbundle[alternate]{} & \qwbundle[alternate]{} & \qwbundle[alternate]{} & \qwbundle[alternate]{} & \qwbundle[alternate]{} & \qwbundle[alternate]{} & \push{\mid\hspace{-0.11cm}\hat{X}_{t_{n+1}}\rangle}\qwbundle[alternate]{} &
\end{quantikz}
\caption{Quantum circuit preparing $\hat{X}_{t_{n+1}}$}
\label{fig:circ_x_preparation}
\end{figure}

\noindent After that, to prepare $\hat{X}_t$ with $t\in[1,\dots ,\, N]$, we make $N$ calls to $U_\mu$ and $U_\sigma$ each, which act as $U_{\mu}|t, X\rangle|0\rangle = |t, X\rangle|\mu(t, X)\rangle$ and $U_{\sigma}|t, X\rangle|0\rangle = |t, X\rangle|\sigma(t, X)\rangle$, respectively. We also require $d\cdot N$ multiplications (multiplying the $\mu$ vectors with the scalars $\Delta t$ at each time step) and $d^2 \cdot N$ multiplications as well as $(d-1)\cdot d\cdot N$ additions for the multiplications of the $\sigma$ matrices with the respective $\Delta W_{t_n}$ vectors. Finally, we require $2\cdot N$ additions to carry out the update steps, as seen in \Cref{eqn:pde_disc_1}. We then arrive at the following state,
\begin{equation}
\begin{aligned}
    \label{eqn:circuit_X}
    |X_{t_0}\rangle
    |t_0\rangle
    |\sigma^\top \nabla u_0\rangle
    |u_0\rangle
    \sum_{k_{t_0,1}=1}^{N_\mathrm{Gauss}} 
    \cdots 
    \sum_{k_{t_{N-1},d}=1}^{N_\mathrm{Gauss}} &
    \sqrt{p_{k_{t_0,1}}}
    \cdots
    \sqrt{p_{k_{t_{N-1},d}}}
    |k_{t_0,1}\rangle|\Delta W_{t_0,1}(k_{t_0,1})\rangle
    \cdots \\ &
    |k_{t_{N-1},d}\rangle|\Delta W_{t_{N-1},1}(k_{t_{N-1},d})\rangle \\ &
    |\hat{X}_{t_1}(k_{t_0,1},\dots,\, k_{t_0,d})\rangle
    \cdots
    |\hat{X}_{t_N}(k_{t_0,1},\dots,\, k_{t_{N-1},d})\rangle \\ &
    |t_1\rangle 
    \cdots  
    |t_N\rangle
    |0 \cdots 0\rangle. 
\end{aligned}
\end{equation}
We now proceed to represent these operations in a quantum circuit. To simplify notation, we will make use the shorthands $|\Delta W_{t_n}\rangle$ and $|\hat{X}_{t_n}\rangle$. We use these kets to represent any \textit{single} term of the superposition of terms, i.e., any \textit{single} Brownian path. More formally, we have,

\begin{equation}
\label{eqn:W_shorthand}
|\Delta W_{t_n}\rangle = 
    |\Delta W_{t_n,1}(k_{t_n,1})\rangle
    \cdots
    |\Delta W_{t_{n},1}(k_{t_{n},d})\rangle,
\end{equation}
for the pointers $k_{t_n,1}$ to $k_{t_n,d}$ representing a Brownian path. And analogously,
\begin{equation}
\label{eqn:X_shorthand}
|\hat{X}_{t_n}\rangle = |\hat{X}_{t_{n}}(k_{t_0,1},\dots,\, k_{n-1,d})\rangle,
\end{equation}
for a random choice of the pointers $k_{t_0,1}$ to $k_{t_n,d}$, which represent a Brownian path. For the following quantum circuits it is worth bearing in mind that the unitaries take terms of the form $|\Delta W_{t_n}\rangle$ and $|\hat{X}_{t_n}\rangle$, i.e., individual paths. We will display the circuit diagrams in this way, for clearer illustration, as seen in \Cref{fig:circ_x_preparation}. However, when running the algorithm, we query these same unitaries in superposition, by feeding as input terms of the form of \Cref{eqn:circuit_X}, i.e., superpositions of terms of the form of $|\Delta W_{t_n}\rangle$ and $|\hat{X}_{t_n}\rangle$. We now present the quantum circuit that computes the discretised values of the stochastic process $X_t$, $\hat{X}_{t_n}$. The unitary $\mathrm{Arith}_X$ computes the step described in \Cref{eqn:pde_disc_1}.

\begin{figure}[H]
\centering
\begin{quantikz}
     \lstick{$|\hat{X}_{t_n}\rangle$} & \gate[2]{U_{\mathrm{NN}_n}}\qwbundle[alternate]{} &   \rstick{$|\hat{X}_{t_n}\rangle$}\qwbundle[alternate]{} 
     \\
     \lstick{$\ket{0}$} & \qwbundle[alternate]{} & \rstick{$|\sigma^\top \nabla u_t (\hat{X}_{t_n}) \rangle$}\qwbundle[alternate]{} 
\end{quantikz}
\caption{Quantum circuit computing the gradient $\sigma^\top \nabla u_{t}$ at time step $t_n$}
\label{fig:circ_nn}
\end{figure}
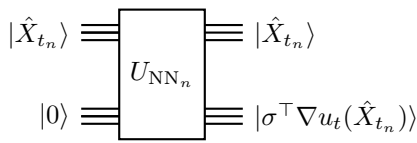

\noindent Note that at the end of the above circuit we are left with the states $|\mu(t_n, \hat{X}_{t_n})\rangle$ and $|\sigma(t_n, \hat{X}_{t_n})\rangle$. These will not used anymore after the $\mathrm{Arith}_X$ unitary. As a consequence, to simplify our notation, we will not display them in the following equations where we write out the quantum state of the circuit. The above circuit in \Cref{fig:circ_x_preparation} is repeated for a total of $N$ times, with the output $\hat{X}_{t_{n}}$ of the $n$th execution of the circuit from \Cref{fig:circ_x_preparation} used as an input for the $n+1$th circuit. The other inputs needed for the $n+1$th execution of the  circuit from \Cref{fig:circ_x_preparation} are $|t_{n}\rangle$ and $|\Delta W_{t_{n}}\rangle$, see \Cref{fig:circ_t_preparation} and the discussion on distribution loading in \Cref{sct:comp_model}. Next, we make $N-1$ calls to the neural network unitaries $U_{\mathrm{NN}}$ to compute the quantities $\sigma^\top \nabla u_t$, in the sense that $U_\mathrm{NN}|\hat{X}_{t_n}\rangle|0\rangle = |\hat{X}_{t_n}\rangle|\sigma^\top \nabla u_t (\hat{X}_{t_n})\rangle$. The state thus becomes the following, where $|\sigma^\top \nabla u_t\rangle$ is only present for $t\in[1,\, N-1]$,
\begin{equation}
\begin{aligned}
    \label{eqn:circuit_nn}
    |u_0\rangle
    |\sigma^\top\nabla u_0\rangle
    |t_0\rangle
    |X_{t_0}\rangle
    \sum_{k_{t_0,1}=1}^{N_\mathrm{Gauss}} 
    \cdots 
    \sum_{k_{t_{N-1},d}=1}^{N_\mathrm{Gauss}} &
    \sqrt{p_{k_{t_0,1}}}
    \cdots
    \sqrt{p_{k_{t_{N-1},d}}}
    |k_{t_0,1}\rangle|\Delta W_{t_0,1}(k_{t_0,1})\rangle
    \cdots \\ &
    |k_{t_{N-1},d}\rangle|\Delta W_{t_{N-1},1}(k_{t_{N-1},d})\rangle \\ &
    |\hat{X}_{t_1}(k_{t_0,1},\dots,\, k_{1,d})\rangle
    \cdots
    |\hat{X}_{t_N}(k_{t_0,1},\dots,\, k_{t_{N-1},d})\rangle \\ &
    |t_1\rangle 
    \cdots  
    |t_N\rangle \\ &
    |\sigma^\top \nabla u_t (\hat{X}_{t_1}) \rangle
    \cdots
    |\sigma^\top \nabla u_t (\hat{X}_{t_{N-1}}) \rangle
|0 \cdots 0\rangle.
\end{aligned}
\end{equation}
It is worth pointing out that, in the above equation, we omitted the pointer arguments $k$ for the states of the form $|\sigma^\top \nabla u_t (\hat{X}_{t_n}) \rangle$ to allow for a more compact notation. We now illustrate the circuit performing this task for a time step $t_n$ in \Cref{fig:circ_nn}. 

\noindent In \Cref{fig:circ_nn} we labeled the unitaries for the NNs with their respective index $n$, since the parameters of the individual NNs, and consequently the unitaries themselves, will in general not be the same. Nevertheless, since their architectures can be expected to be the same, in the sense that they have the same number of layers, the same input and output dimensions and the same number of trainable parameters (as described in \cite{Han_2018}), it makes sense to treat a call to each of the NNs the same from a query complexity point of view, even though the NNs at different time steps do not implement the same function. Thus, apart from \Cref{fig:circ_nn} we drop the index $n$, since, from a query complexity point of view, we assume that their cost will be comparable (if not, we can always just take the most computationally costly NN as our reference). Next, we proceed to compute $\hat{u}_t$ for $t\in [1,\dots ,\, N]$. This computation requires $N$ calls to $U_f$, $N$ additions and multiplications each for multiplying the outcome of $U_f$ with $\Delta t$ and adding it to the previous value of $\hat{u}_t$, and $d\cdot N$ multiplications as well as $d\cdot N$ additions for multiplying the vectors $\sigma^\top \nabla u_t (\hat{X}_{t_n})$ and adding the result to the previous value of $\hat{u}_t$. We then have the following quantum state,
\begin{equation}
    \label{eqn:circuit_u}
    \begin{aligned}
    |u_0\rangle
    |\sigma^\top\nabla u_0\rangle
    |t_0\rangle
    |X_{t_0}\rangle
    \sum_{k_{t_0,1}=1}^{N_\mathrm{Gauss}} 
    \cdots 
    \sum_{k_{t_{N-1},d}=1}^{N_\mathrm{Gauss}} &
    \sqrt{p_{k_{t_0,1}}}
    \cdots
    \sqrt{p_{k_{t_{N-1},d}}}
    |k_{t_0,1}\rangle|\Delta W_{t_0,1}(k_{t_0,1})\rangle 
    \cdots \\ &
    |k_{t_{N-1},d}\rangle|\Delta W_{t_{N-1},1}(k_{t_{N-1},d})\rangle \\ &
    |\hat{X}_{t_1}(k_{t_0,1},\dots,\, k_{1,d})\rangle
    \cdots
    |\hat{X}_{t_N}(k_{t_0,1},\dots,\, k_{t_{N-1},d})\rangle \\ &
    |t_1\rangle 
    \cdots  
    |t_N\rangle
    |\sigma^\top \nabla u_t (\hat{X}_{t_1}) \rangle
    \cdots
    |\sigma^\top \nabla u_t (\hat{X}_{t_{N-1}}) \rangle \\ &
    |\hat{u}_{t_1}\rangle
    \cdots
    |\hat{u}_{t_N}\rangle
    |0 \cdots 0\rangle.
\end{aligned}
\end{equation}
Again, we omitted the pointer arguments $k$ for the states $|\hat{u}_{t_n}\rangle$ to simplify our notation. The pointer arguments for $|\hat{u}_{t_n}\rangle$ would be the same as for $|\hat{X}_{t_n}\rangle$, i.e., $(k_{t_0,1},\dots,\, k_{t_{n-1},d})$. The circuit implementing these operations is displayed in \Cref{fig:circ_u_preparation} below, where the unitary $\mathrm{Arith}_u$ carries out the arithmetic operations outlined in \Cref{eqn:pde_disc_2}.

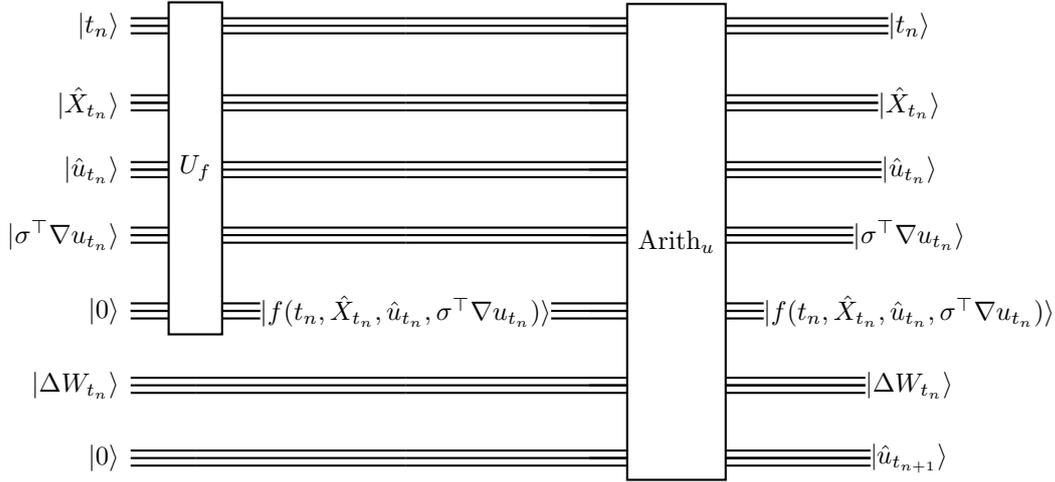
\begin{figure}[H]

\centering
\begin{quantikz}  
\lstick{$|t_n\rangle$} & \gate[wires=5]{U_f}\qwbundle[alternate]{} & \qwbundle[alternate]{} & \qwbundle[alternate]{} & \gate[wires=7]{\mathrm{Arith}_u}\qwbundle[alternate]{} & \push{\altket{t_n}}\qwbundle[alternate]{}
\\
\lstick{$|\hat{X}_{t_n}\rangle$}& \qwbundle[alternate]{} & \qwbundle[alternate]{} & \qwbundle[alternate]{} & \qwbundle[alternate]{} & \push{\altket{\hat{X}_{t_n}}}\qwbundle[alternate]{}
\\
\lstick{$|\hat{u}_{t_n}\rangle$} & \qwbundle[alternate]{} & \qwbundle[alternate]{} & \qwbundle[alternate]{} & \qwbundle[alternate]{} & \push{\altket{\hat{u}_{t_n}}}\qwbundle[alternate]{}
\\
\lstick{$|\sigma^\top \nabla u_{t_n}\rangle$} & \qwbundle[alternate]{} & \qwbundle[alternate]{} & \qwbundle[alternate]{} & \qwbundle[alternate]{} & \push{\altket{\sigma^\top \nabla {u}_{t_n}}}\qwbundle[alternate]{}
\\
\lstick{$|0\rangle$} & \qwbundle[alternate]{} & \push{\altket{f(t_n, \hat{X}_{t_n}, \hat{u}_{t_n}, \sigma^\top \nabla u_{t_n})}}\qwbundle[alternate]{} & \qwbundle[alternate]{} & \qwbundle[alternate]{} & \push{\altket{f(t_n, \hat{X}_{t_n}, \hat{u}_{t_n}, \sigma^\top \nabla u_{t_n})}}\qwbundle[alternate]{} 
\\
\lstick{$|\Delta W_{t_n}\rangle$} & \qwbundle[alternate]{} & \qwbundle[alternate]{} & \qwbundle[alternate]{} & \qwbundle[alternate]{} & \push{\altket{\Delta W_{t_n}}}\qwbundle[alternate]{}
\\
\lstick{$|0\rangle$} & \qwbundle[alternate]{} & \qwbundle[alternate]{} & \qwbundle[alternate]{} & \qwbundle[alternate]{} & \push{\mid\hspace{-0.11cm}\hat{u}_{t_{n+1}}\rangle}\qwbundle[alternate]{} &
\end{quantikz}
\caption{Quantum circuit preparing $\hat{u}_{t_{n+1}}$}
\label{fig:circ_u_preparation}
\end{figure}
\begin{figure}[H]
\centering
\begin{quantikz}
     \lstick{$|\hat{X}_{t_N}\rangle$} & \gate[3]{U_\mathrm{loss}}\qwbundle[alternate]{} &   \rstick{$|\hat{X}_{t_N}\rangle$}\qwbundle[alternate]{} 
     \\
     \lstick{$|\hat{u}_{t_N}\rangle$} & \qwbundle[alternate]{} &   \rstick{$|\hat{u}_{t_N}\rangle$}\qwbundle[alternate]{} 
     \\
     \lstick{$\ket{0}$} & \qwbundle[alternate]{} & \rstick{$|f_\mathrm{p}(\hat{X}_{t_N},\hat{u}_{t_N})\rangle$}\qwbundle[alternate]{} 
\end{quantikz}
\caption{Quantum circuit computing the payoff function $f_\mathrm{p}$}
\label{fig:circ_rotation}
\end{figure}
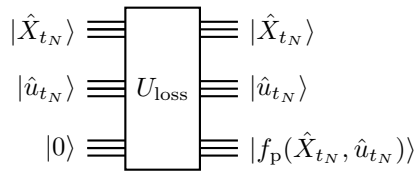

\noindent In order to compute $\hat{u}_{t_n}$ for $n\in [1,\ldots,N]$ we have to execute the circuit from \Cref{fig:circ_u_preparation} $N$ times.
Finally, we require one call to the unitary $U_\mathrm{loss}$ to compute the loss function. The quantum state then becomes,
\begin{equation}
    \label{eqn:circuit_r}
    \begin{aligned}
    |u_0\rangle
    |\sigma^\top\nabla u_0\rangle
    |t_0\rangle
    |X_{t_0}\rangle
    \sum_{k_{t_0,1}=1}^{N_\mathrm{Gauss}} 
    \cdots 
    \sum_{k_{t_{N-1},d}=1}^{N_\mathrm{Gauss}} &
    \sqrt{p_{k_{t_0,1}}}
    \cdots
    \sqrt{p_{k_{t_{N-1},d}}}
    |k_{t_0,1}\rangle|\Delta W_{t_0,1}(k_{t_0,1})\rangle
    \cdots \\ &
    |k_{t_{N-1},d}\rangle|\Delta W_{t_{N-1},1}(k_{t_{N-1},d})\rangle \\ &
    |\hat{X}_{t_1}(k_{t_0,1},\dots,\, k_{1,d})\rangle
    \cdots
    |\hat{X}_{t_N}(k_{t_0,1},\dots,\, k_{t_{N-1},d})\rangle \\ &
    |t_1\rangle 
    \cdots  
    |t_N\rangle \\ &
    |\sigma^\top \nabla u_t (\hat{X}_{t_1}) \rangle
    \cdots
    |\sigma^\top \nabla u_t (\hat{X}_{t_{N-1}}) \rangle \\ &
    |\hat{u}_{t_1}\rangle
    \cdots
    |\hat{u}_{t_N}\rangle \\ &
    |f_p(\{\hat{X}_{t_n}\}_{0\leq n\leq N}, \{\Delta W_{t_n}\}_{0\leq n\leq N})\rangle
    |0 \cdots 0\rangle.
\end{aligned}
\end{equation}
We illustrate this step with the following circuit in \Cref{fig:circ_rotation}.

\noindent Now that we have $|f_\mathrm{p}(\hat{X}_{t_N},\hat{u}_{t_N})\rangle$ in the superposition, weighted by the distributions of the Gaussian increments, we can apply QAMC from \Cref{subsubsubsct:qmc} to estimate the mean of $f_\mathrm{p}(\hat{X}_{t_N},\hat{u}_{t_N})$ with respect to the distributions given by the Gaussian increments $\Delta W_{t_n}$. We proceed to summarise the query complexities for the individual unitaries.

\textbf{Query Complexities} $-$ Based on the above, estimating the payoff function from \Cref{eqn:payoff} using QAMC, thus requires the asymptotic number of queries to the following unitaries and arithmetic operations,

\begin{itemize}
    \item{$\Tilde{O}(\lambda_\mathrm{max}/\epsilon)$ queries to $U_{X_{t_0}}$, $U_{u_0}$, $U_{\sigma^\top \nabla u_0}$, $U_{t_0}$ and $U_\mathrm{loss}$ each}
    \item{$\Tilde{O}(N\cdot \lambda_\mathrm{max}/\epsilon)$ queries to $U_\mu$, $U_\sigma$, $U_f$ and $U_{\mathrm{NN}}$ each}
    \item{$\Tilde{O}(d\cdot N\cdot \lambda_\mathrm{max}/\epsilon)$ queries to $U_\mathrm{Gauss}$}
    \item{$\Tilde{O}(d^2\cdot N\cdot \lambda_\mathrm{max}/\epsilon)$ arithmetic operations},
\end{itemize}
How does this compare to the classical case? We do not get the speedup from QAMC in the classical case. Therefore, we have,

\begin{itemize}
    \item{${O}(\lambda_\mathrm{max}^2/\epsilon^2)$ queries to $U_{X_{t_0}}$, $U_{u_0}$, $U_{\sigma^\top \nabla u_0}$, $U_{t_0}$ and $U_\mathrm{loss}$ each}
    \item{${O}(N\cdot \lambda_\mathrm{max}^2/\epsilon^2)$ queries to $U_\mu$, $U_\sigma$, $U_f$ and $U_{\mathrm{NN}}$ each}
    \item{${O}(d\cdot N\cdot \lambda_\mathrm{max}^2/\epsilon^2)$ queries to $U_\mathrm{Gauss}$}
    \item{${O}(d^2\cdot N\cdot \lambda_\mathrm{max}^2/\epsilon^2)$ arithmetic operations},
\end{itemize}
from which we can see the slowdown in $\lambda_\mathrm{max}$ and $\epsilon$ compared to the quantum enhanced version. We conclude the analysis with the following result.

\begin{result}[QAMC Evaluation]
\label{thm:qamc_training}
Using the QAMC method from \cite{montanaro2015quantum} and described in \Cref{subsubsubsct:qmc}, there is the potential for achieving a quantum speedup from a query complexity point of view for estimating the loss function in the algorithm for solving nonlinear PDEs described in \cite{Han_2018} with the number of calls to unitaries as well as arithmetic operations as outlined above.
\end{result}

Subsequently, we discuss the problem of training the NNs, referencing the findings from \Cref{sct:training_qmc_nn}.

\subsection{Training the Neural Networks}
\label{subsct:training_nn}

\begin{figure}[t]
    \centering
    \includegraphics[width=0.8\textwidth]{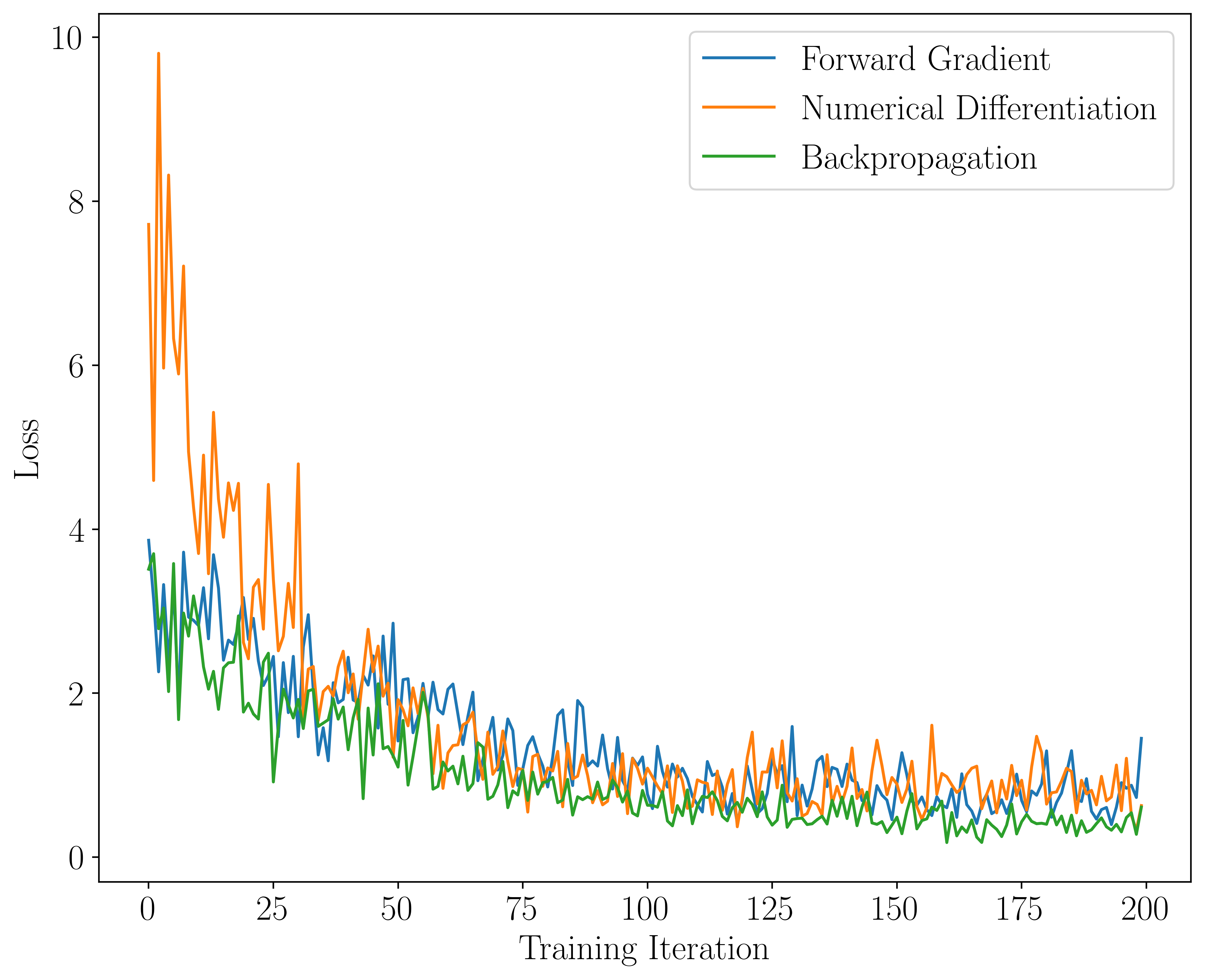}
    \caption{Loss from \Cref{eqn:loss} for the HJB PDE against the number of training iterations when the forward gradient method, numerical differentiation and backpropagation are employed to update the weights. In each case, we have $N=20$ temporal discretisation steps, each of the 20 neural networks has 225 trainable parameters, and the terminal time is $T=1$. We take 20 training samples per iteration and the learning rate is 0.01. In the case when the forward gradient method is employed, we take 100 samples of $v$ per iteration. For numerical differentiation, $h$ from \Cref{eqn:num_diff} is 0.001.}
    \label{fig:fwd_grad}
\end{figure}

As described in \Cref{sct:training_qmc_nn}, using the forward gradient method allows us to estimate the gradients of the parameters in the NNs using QAMC. Since the payoff function $f_p$ (see \Cref{eqn:payoff}) is assumed to be Lipschitz continuous (see \Cref{ass:payoff_lipschitz}) and has a scalar output we may apply the methods surveyed in \Cref{tab:grad_query_complexities}, where the gradient is taken with respect to the trainable parameters in the architecture from \cite{Han_2018}, i.e., the weights in the NNs and the initial values of $u$ and $\sigma^\top \nabla u$. To numerically verify that the forward gradient method as well as numerical differentiation work in the deep learning architecture from \cite{Han_2018}, we implement these methods in the purely classical setting to update the weights when solving the Hamilton-Jacobi-Bellman (HJB) PDE. The HJB PDE is a special case of \Cref{eqn:pde} where $\sigma(t,X)=2\mathbb{I}$, $\mu(t,X)=0$ and $f(t,X,u(t,X),\sigma^\top(t,X)\nabla u(t,X)) = \lVert \nabla u(t,X)\rVert_2^2$, and, to evaluate the loss function, $g(X) = \log((1+\lVert X \rVert_2^2)/2)$. We take $N=20$ temporal discretisation steps, where we approximate the spatial gradient at each step with a NN with 225 trainable parameters, and the terminal time is $T=1$. We present the values of the loss function at each training iteration in \Cref{fig:fwd_grad}.

As we can see in \Cref{fig:fwd_grad}, and as experimentally shown for other problems in \cite{baydin2022gradients}, the forward gradient method is indeed competitive with backpropagation, although the loss function in the latter case appears to converge slightly below the value in the forward gradient case, possibly due to the additional stochasticity introduced by the forward gradient, as mentioned in \cite{baydin2022gradients}. The same observation holds for numerical differentiation, where the slightly higher loss at convergence may stem from the truncation error. 

\subsection{Multi-Level Monte Carlo Methods and PDEs}
\label{subsct:mlmc}

We have so far discussed estimating the loss function in the deep learning architecture from \cite{Han_2018}. We now discuss the estimation of the solution to the PDE itself, once the loss function has been reduced to a satisfactory level. Once the loss function is below a certain threshold or has ceased to improve, the circuits outlined above (in \Cref{fig:circ_init_preparation,fig:circ_t_preparation,fig:circ_gr_preparation,fig:circ_x_preparation,fig:circ_nn,fig:circ_u_preparation}) may simply be truncated after the $n$th time step and used to estimate $u_{t_n}$. Since the NNs as well as $\mu$ and $\sigma$ are Lipschitz continuous, one may estimate ${u}_{t_n}$ with the given assumptions (\Cref{ass:musig_lipschitz,ass:finite_x0,ass:payoff_lipschitz}). As $u_{t_n}$ are scalars, one may again make use of QAMC to estimate $u_{t_n}$ with a speedup in $\epsilon$ compared to the classical case. 


When considering the precision with which we estimate the solution at final time, $u_{t_N}$, we now also have to consider the error resulting from the discretisation of the SDEs. It is worth mentioning that the error stemming from the inaccuracies of the NNs when representing the spatial gradients are also an error source contributing to the total error when estimating the solution. However, the highly non-convex training landscapes of the NNs prevent us from bounding this contribution to the error. We proceed to analyse the contribution from the discretisation of the SDEs, neglecting the contribution from the inaccuracies of the NNs.

As in the case where we estimate the loss function, the error originating from the discretisation of the Gaussian increments, as well as the estimation error when using MC methods, are relevant. Let the ideal estimation of the solution be $I_0$ and the estimate of the solution considering only discretisation the error of the SDE be $I_1$. Furthermore, analogously to the discussion in \Cref{subsct:applying-qmc}, let $I_2$ be the estimate of the solution ${u}_{t_N}$ when also considering the error from the discretisation of the Gaussian increments and let $I_3$ be the estimate where all three kinds of errors are taken into account. Then we have,
\begin{equation}
    \label{eqn:error_bound_solution}
    \lvert I_0 - I_3\rvert = \lvert I_0 - I_1 + I_1 - I_2 + I_2 - I_3\rvert \leq \lvert I_0 - I_1\rvert + \lvert I_1 - I_2\rvert + \lvert I_2 - I_3\rvert.
\end{equation}
In order to estimate the solution $u_{t_N}$ up to error $\epsilon$ we need to upper bound the term on the left hand side in \Cref{eqn:error_bound_solution}, which we can achieve by upper bounding the right hand side of the same equation. In \Cref{subsct:applying-qmc} we showed that upper bounding the term $\lvert I_1 - I_2\rvert$ gives us a requirement on how many bits or qubits are needed to represent the Gaussian increments. Furthermore, upper bounding $\lvert I_2 - I_3\rvert$ gave us a requirement on how many samples we need to take with classical MC or QAMC. We thus still need to upper bound the first term, $\lvert I_0 - I_1\rvert$,
\begin{equation}
    \label{eqn:error_bound_1}
    \begin{aligned}
    \lvert I_0 - I_1\rvert &= \lvert \mathbb{E}[u_{t_N}(\{{X}_{t_n}\}_{0\leq n\leq N})] - \mathbb{E}[u_{t_N}(\{\hat{X}_{t_n}\}_{0\leq n\leq N})]\rvert \\ & \leq \mathbb{E}[\lvert u_{t_N}(\{{X}_{t_n}\}_{0\leq n\leq N}) - u_{t_N}(\{\hat{X}_{t_n}\}_{0\leq n\leq N})\rvert] \\ & \leq L_{u_{t_N}} \mathbb{E}\left[\sup_{0\leq n\leq N} \lVert X_{t_n} - \hat{X}_{t_n} \rVert_2 \right] = O((\Delta t)^r),
    \end{aligned}
\end{equation}
where the first equality is the definition of $I_0$ and $I_1$ (we dropped the Gaussian increments in the arguments to simplify the notation), the first inequality follows from applying Jensen's inequality (recall that the absolute value is a convex function). The second inequality follows from \Cref{ass:payoff_lipschitz}, since the local Lipschitz continuity of $f_p$ implies local Lipschitz continuity of ${u_{t_N}}$ (we name its Lipschitz constant $L_{u_{t_N}}$). The last equation follows from the definition of the strong order $r$, see \Cref{eqn:strong_order}. In order to bound $\lvert I_0 - I_1 \rvert$ by $\epsilon/3$, we need to choose $\Delta t = O(\epsilon^{1/r})$ and consequently,
\begin{equation}
    \label{eqn:N_eps_bound}
    N = O(\epsilon^{-1/r}).
\end{equation}
It is worth pointing out that this relation between $N$ and $\epsilon$ holds as much for the quantum case as for the classical case.

By inserting the dependency of $N$ on $\epsilon$ from \Cref{eqn:N_eps_bound} into the query complexities from \Cref{subsct:applying-qmc}, we arrive at the following complexities,

\begin{itemize}
    \item{$\Tilde{O}(\lambda_\mathrm{max}/\epsilon)$ queries to $U_{X_{t_0}}$, $U_{u_0}$, $U_{\sigma^\top \nabla u_0}$, $U_{t_0}$ and $U_\mathrm{loss}$ each}
    \item{$\Tilde{O}(\lambda_\mathrm{max}/\epsilon^{1+1/r})$ queries to $U_\mu$, $U_\sigma$, $U_f$ and $U_{\mathrm{NN}}$ each}
    \item{$\Tilde{O}(d\cdot \lambda_\mathrm{max}/\epsilon^{1+1/r})$ queries to $U_\mathrm{Gauss}$}
    \item{$\Tilde{O}(d^2\cdot \lambda_\mathrm{max}/\epsilon^{1+1/r})$ arithmetic operations}.
\end{itemize}
and by inserting $r=1/2$ for the Euler-Maruyama scheme which the architecture from \cite{Han_2018} uses, we arrive at,

\begin{itemize}
    \item{$\Tilde{O}(\lambda_\mathrm{max}/\epsilon)$ queries to $U_{X_{t_0}}$, $U_{u_0}$, $U_{\sigma^\top \nabla u_0}$, $U_{t_0}$ and $U_\mathrm{loss}$ each}
    \item{$\Tilde{O}(\lambda_\mathrm{max}/\epsilon^{3})$ queries to $U_\mu$, $U_\sigma$, $U_f$ and $U_{\mathrm{NN}}$ each}
    \item{$\Tilde{O}(d\cdot \lambda_\mathrm{max}/\epsilon^{3})$ queries to $U_\mathrm{Gauss}$}
    \item{$\Tilde{O}(d^2\cdot \lambda_\mathrm{max}/\epsilon^{3})$ arithmetic operations}.
\end{itemize}
How does this compare to the classical case? The relation between $N$ and $\epsilon$ from \Cref{eqn:N_eps_bound} holds in the classical case as well. However, we do not get the speedup from QAMC in the classical case. Therefore, we have,

\begin{itemize}
    \item{${O}(\lambda_\mathrm{max}^2/\epsilon^2)$ queries to $U_{X_{t_0}}$, $U_{u_0}$, $U_{\sigma^\top \nabla u_0}$, $U_{t_0}$ and $U_\mathrm{loss}$ each}
    \item{${O}(N\cdot \lambda_\mathrm{max}^2/\epsilon^2)$ queries to $U_\mu$, $U_\sigma$, $U_f$ and $U_{\mathrm{NN}}$ each}
    \item{${O}(d\cdot N\cdot \lambda_\mathrm{max}^2/\epsilon^2)$ queries to $U_\mathrm{Gauss}$}
    \item{${O}(d^2\cdot N\cdot \lambda_\mathrm{max}^2/\epsilon^2)$ arithmetic operations},
\end{itemize}
which, after inserting the expressions for $N$ and $r$ leaves us with,

\begin{itemize}
    \item{${O}(\lambda_\mathrm{max}^2/\epsilon^2)$ queries to $U_{X_{t_0}}$, $U_{u_0}$, $U_{\sigma^\top \nabla u_0}$, $U_{t_0}$ and $U_\mathrm{loss}$ each}
    \item{${O}(\lambda_\mathrm{max}^2/\epsilon^4)$ queries to $U_\mu$, $U_\sigma$, $U_f$ and $U_{\mathrm{NN}}$ each}
    \item{${O}(d\cdot \lambda_\mathrm{max}^2/\epsilon^4)$ queries to $U_\mathrm{Gauss}$}
    \item{${O}(d^2\cdot \lambda_\mathrm{max}^2/\epsilon^4)$ arithmetic operations},
\end{itemize}
As in the case of estimating the loss function (see \Cref{subsct:applying-qmc}), QAMC thus offers the potential for a quantum speedup in $\epsilon$ when estimating the solution ${u_{t_N}}$ of the PDE.

As described in \Cref{subsubsubsct:mlmc}, multi-level MC (MLMC) methods offer the potential to improve the sample complexity when estimating the mean value of functions depending on the discretised solutions of SDEs. We thus discuss the application of MLMC as well as quantum-accelerated MLMC (QAMLMC) to the problem of estimating the solution ${u_{t_N}}$ in the deep learning architecture from \cite{Han_2018}.

We begin by checking that \Cref{ass:mlmc_payoff,ass:mlmc_lipschitz,ass:mlmc_scheme} are satisfied in the setting of this section so far. \Cref{ass:mlmc_lipschitz} is satisfied since we make \Cref{ass:musig_lipschitz}. The numerical scheme employed in \cite{Han_2018} is the Euler-Maruyama scheme which is of the form \Cref{eqn:taylor-ito} and satisfies \Cref{ass:mlmc_scheme} \cite{kloeden1992stochastic}. Finally, \Cref{ass:mlmc_payoff} is satisfied since we make \Cref{ass:payoff_lipschitz}, the latter being a stronger assumption. Thus, the assumptions made in \cite{an2021quantum} are satisfied and we can in principle apply \Cref{lemma:classical_mlmc,lemma:qamlmc}.

We continue with considerations on how one might use QAMLMC (or classical MLMC) in combination with the architecture from \cite{Han_2018}. It is worth pointing out, that estimating each of the terms $Y_k$ from \Cref{eqn:mlmc_estimator} which approximate the solution at final time $u_{t_N}$ with increasing precision, having $N_k$ discretisation steps in the architecture from \cite{Han_2018}, will also have $N_k$ NNs to train. In total, with $K$ estimators $Y_k$ with $2^k$ NNs each, where $K=O(\log(2\epsilon^{-1}))$, we would have $O(2^{\log(2\epsilon^{-1})})=O(\epsilon^{-1})$ NNs to train. Since in each estimator there are different NNs (i.e., we do not assume that any weight sharing across different estimators takes place, although that might be possible), MLMC or QAMLMC may only offer a speedup to the estimation of the solution ${u_{t_N}}$ and would not be of help in estimating the gradients for each estimator. The gradients would need to be computed as discussed in \Cref{subsct:qamc_nn}. 

When using classical or quantum-accelerated MLMC methods, observe that the cutoff strong order for which the $1/r$ term vanishes in the exponent of $\epsilon$ is not the same for the classical and the quantum case, as seen in \Cref{lemma:classical_mlmc,lemma:qamlmc}. In the classical case, it is $r=1/2$ and in the quantum case it is $r=1$. This difference stems from the different base case MC sample complexities, $O(\lambda^2/\epsilon^2)$ in the classical case and $\Tilde{O}(\lambda/\epsilon)$ in the quantum case, where $\lambda^2$ is the variance of the term in question \cite{an2021quantum}. Recall that in the architecture from \cite{Han_2018} we have $r=1/2$ as the Euler-Maruyama scheme is used. We continue to analyse potential speedups when using classical and quantum-accelerated MLMC methods to estimate the solution ${u_{t_N}}$ from the deep learning architecture.

Employing MLMC methods in the deep learning architecture in the classical case to estimate the solution ${u_{t_N}}$, with $r=1/2$ we have as per \Cref{lemma:classical_mlmc},

\begin{itemize}
    \item{$\Tilde{O}(\epsilon^{-2})$ queries to $U_{X_{t_0}}$, $U_{u_0}$, $U_{\sigma^\top \nabla u_0}$, $U_{t_0}$, $U_\mathrm{loss}$, $U_\mu$, $U_\sigma$, $U_f$ and $U_{\mathrm{NN}}$ each}

    \item{$\Tilde{O}(d\cdot \epsilon^{-2})$ queries to $U_\mathrm{Gauss}$}

    \item{$\Tilde{O}(d^2 \cdot \epsilon^{-2})$ arithmetic operations},
\end{itemize}
where $d$ is the dimension of the stochastic process $X_t$ and where we treated the individual NNs as equivalent from a query complexity point of view. In the case of QAMLMC we have in the same setting,
\begin{itemize}
    \item{$\Tilde{O}(\epsilon^{-2})$ queries to $U_{X_{t_0}}$, $U_{u_0}$, $U_{\sigma^\top \nabla u_0}$, $U_{t_0}$, $U_\mathrm{loss}$, $U_\mu$, $U_\sigma$, $U_f$ and $U_{\mathrm{NN}}$ each}
    \item{$\Tilde{O}(d\cdot \epsilon^{-2})$ queries to $U_\mathrm{Gauss}$}
    \item{$\Tilde{O}(d^2 \cdot \epsilon^{-2})$ arithmetic operations},
\end{itemize}
which is the same as for the classical case (the lower threshold in the strong order $r$ for the classical case compensates for the speedup in $\epsilon$ in the quantum case). However, both methods improve on the complexities outlined above in this subsection. It is worth highlighting again, that MLMC methods would only speed up the estimation of the solution but not the estimation of the gradients with respect to the parameters of the NNs. Thus, the results of \Cref{sct:training_qmc_nn} in \Cref{tab:grad_query_complexities} are still relevant for training the NNs in the sense that they offer the potential for speedups in $\epsilon$ when estimating the gradient, even in the scenario where MLMC methods are applied to estimating the solution.

Improvements in the query complexities for QAMLMC (which would allow for QAMLMC to outperform classical MLMC) may be brought about by implementing a numerical scheme with $r>1/2$, e.g., the Milstein scheme \cite{milstein1974approximate} with $r=1$. However, this step comes with its own difficulties, particularly for the architecture from \cite{Han_2018}. For implementing higher order schemes such as the Milstein scheme, derivatives of $\sigma$ and $\sigma^\top \nabla u$ come into play. A possibility to compute what amounts to the Hessian of $u$ might be to add a second NN per discretisation step. However, we leave this investigation to future research.

Summarising our findings from this section, we saw that by applying QAMC to the deep learning architecture for solving nonlinear PDEs from \cite{Han_2018} there exists the potential for a speedup in the error tolerance $\epsilon$ when estimating the loss function. In \Cref{subsct:mlmc} we found that applying classical MLMC methods offers the potential for an even greater speedup in $\epsilon$ compared to applying QAMC when estimating the solution of the PDE at final time. Interestingly, with the numerical scheme employed in the deep learning architecture, QAMLMC does not offer a speedup (nor suffers from a slowdown) compared to (classical) MLMC. Future research may, however, be able to find the possibility for a quantum speedup by increasing the strong order $r$ of the numerical scheme for approximating the SDE. This might allow QAMLMC to provide a speedup over classical MLMC again. As mentioned in \Cref{subsct:mlmc}, increasing the strong order $r$ in the deep learning architecture comes with its own set of challenges, which we leave for future research to address. It is worth pointing out that applying MLMC methods only offers the potential of speeding up the estimation of the loss function, but not the estimation of the gradient, thus our findings from \Cref{sct:training_qmc_nn} on estimating the gradient with QAMC methods is still relevant.

\section{Quantum Algorithm for Accelerating Neural Network Training}
\label{sct:quantum_feedforward}
In this section we discuss the application of a quantum algorithm from the literature (see \cite{allcock2020quantum}) which offers the potential for accelerating the training and evaluating of classical NNs.
This algorithm constitutes a fault-tolerant quantum algorithm that we aim to incorporate into the architecture from \cite{Han_2018} for solving nonlinear partial differential equations (PDEs). After having reviewed the algorithm, we discuss how it can be applied in the setting from \cite{Han_2018}, and what advantages and disadvantages are associated with doing so.

It is worth pointing out that while in previous sections we mainly discussed the query complexity to the unitaries implementing the NNs, this section (and the algorithm we introduce therein) are not about the query complexity to the unitaries, but rather the runtime of the NN, i.e., the runtime of the unitaries. A key component of this algorithm is the robust inner product estimation (RIPE) quantum subroutine, introduced in \Cref{subsct:ripe}. 
At the heart of many machine learning architectures lie NNs that are trained in a supervised setting, meaning that the NN is trained with labelled examples. Applications of supervised NNs are found in manufacturing \cite{kim2020wafer}, drug discovery \cite{baskin2016renaissance}, and solving differential equations, as in the algorithm from \cite{Han_2018}, which we outlined in \Cref{subsct:dl_architecture}.

When classically training feedforward NNs, the total runtime of the training algorithm is \allowbreak ${O}(T_{\mathrm{iter}}ME)$, where $T_{\mathrm{iter}}$ is the number of training steps, $M$ is the number of data points trained on per training step, and $E$ is the number of edges in the network. This linear dependence on the number of edges renders the training of large fully connected feedforward NNs expensive. To alleviate this issue, the authors in \cite{allcock2020quantum} present a quantum-enhanced algorithm for training and evaluating feedforward NNs, where the linear dependence on the number of edges $E$ may be exchanged for a linear dependence in the number of neurons $n_\mathrm{nodes}$, at the cost of the dependence on $T_{\mathrm{iter}}M$ becoming $(T_{\mathrm{iter}}M)^{3/2}$, i.e., ${O}((T_{\mathrm{iter}}M)^{3/2} n_\mathrm{nodes})$. 
Recall the NN training via backpropagation outlined in \Cref{subsct:neuralnetworks}. In particular notice that we can reconsider \Cref{eqn:forward_prop,eqn:backprop} by emphasising the inner product in each of the two equations,
\begin{equation}
    \label{eqn:forward_prop_ip}
    z^{(j)}_l = (W^{(j)}_l \cdot a_{l-1}) + b^{(j)}_l,
\end{equation}
and
\begin{equation}
    \label{eqn:backprop_ip}
    \delta^{(j)}_l = f_\mathrm{nl}'(z^{(j)}_l) ((W_{l+1}^\top)^{(j)}\cdot \delta_{l+1}).
\end{equation}
The authors in \cite{allcock2020quantum} point out that when training and evaluating NNs using \Cref{eqn:forward_prop,eqn:backprop} in the classical setting, the runtime is $O(T_{\mathrm{iter}}ME)$. This complexity with its linear dependence on $E$ arises for each training iteration and for each data point for the following reason. One has to, in each layer, evaluate one activation function per neuron, as well as compute the inner product from \Cref{eqn:forward_prop_ip}, which depends on the previous layer. Thus, we have a time complexity of $O\left( \sum_{l=2}^L n_{l} n_{l-1} = E \right)$ per data point per training iteration. 

The authors then propose to use the RIPE algorithm (see \Cref{lemma:ripe}) to estimate the inner products in \Cref{eqn:forward_prop_ip,eqn:backprop_ip}. In addition to achieving potential speedups, the authors argue that introducing noise into the inner product estimation may help regularise NNs, i.e., prevent them from overfitting. Overfitting refers to the phenomenon whereby NNs fail to identify patterns underlying the data and instead memorise the data on which they are trained and consequently fail to generalise well to new data.
Next, the authors point out that merely initialising and storing the weight matrices $W_l$ takes time at least $O(E)$. To circumvent this issue, the authors in \cite{allcock2020quantum} come up with two solutions. Firstly, they use low-rank initialisation for the weight matrices. Observe, that due to \Cref{eqn:weight_update} and setting $\eta_{0,1} = -1$ we can write the weights as,
\begin{equation}
    \label{eqn:weight_sum}
    W^{(j,k)}_{t,l} = \sum_{\tau = 0}^{t-1} \sum_{m=1}^M \frac{-\eta_{\tau,l}}{M} \delta^{(j)}_{\tau,m,l} a^{(k)}_{\tau,m,l-1}.
\end{equation}
With low-rank initialisation, only a fraction of the $M$ summands $a^{(k)}_{0,m,l}\delta^{(j)}_{0,m,l}$ are set to be initially nonzero. They justify this, by pointing out that other (classical) algorithms for training NNs have made use of this approach as well to speed up the training of NNs \cite{denton2014exploiting,yu2017compressing,sainath2013low}. Thus, the authors avoid writing out $O(E)$ weight values for the initial weights. Secondly, they make use of an implicit weight storage scheme using QRAM, see \Cref{def:qram}. Recall that QRAM is a quantum mechanical analogue of classical RAM, allowing for classical data to be queried in superposition. By storing matrices $X_{t,l,j}\in\mathbb{R}^{t\times M}$ with elements $X_{t,l,j}^{(\tau, m)} = \frac{-\eta_{\tau,l}}{M}\delta^{(j)}_{\tau,m,l}\lVert a_{\tau,m,l-1}\rVert_2$ in an $l_2$-binary search tree in QRAM, the states $|W^{(j)}_{t,l}\rangle$ can be computed efficiently on the fly. An $l_2$ binary search tree is a tree (data structure) where the nodes have two children, unless they are leaves, and where the root stores $\lVert x \rVert_2^2$ and the leaves store the components $(x^{(k)})^2$ along with the sign of $x^{(k)}$. The authors in \cite{kerenidis2016quantum} showed that storing a classical vector $x\in\mathbb{R}^K$ in an $l_2$ binary search tree in QRAM takes time $\Tilde{O}(K)$ and retrieving it polylogarithmic time in $K$. For a given training iteration $t$, there are $n_\mathrm{nodes} - n_1$ such matrices stored in QRAM. The key takeaway is that in order to avoid the cost of $O(E)$, the authors in \cite{allcock2020quantum} store the matrices $X_{t,l,j}$ implicitly in QRAM, from which they can compute the weights $W^{(j)}_l$ efficiently on the fly. For more details on the implicit storage of the weight matrices, we refer to \cite{allcock2020quantum}.
The cost of performing the forward propagation in an NN then becomes $O(n_\mathrm{nodes} T_\mathrm{RIPE})$, where $T_\mathrm{RIPE}$ is the mean time to carry out the RIPE algorithm from \Cref{lemma:ripe}, which is applied to estimate the inner products in \Cref{eqn:forward_prop_ip,eqn:backprop_ip}. Using the expression from \Cref{eqn:ripe_result}, the overall running time for the forward propagation is
\begin{equation}
    \label{eqn:forward_time}
    {O}\left( \sqrt{T_{\mathrm{iter}}M}n_\mathrm{nodes} \frac{\log 1/\gamma}{\epsilon} R_{t,m,a} \right),
\end{equation}
where $\gamma$ and $\epsilon$ are as in \Cref{lemma:ripe}, $R_{t,m,a}$ depends on the matrices $X_{t,l,j}$ and $W^{(j)}_{t,l}$ as well as the terms $a_{t,m,l-1}$. The authors provide experimental evidence, that $R^{t,m,a}$ does, however, in practice not impact the running time significantly. Similarly, by again using low-rank initialisation and the implicit weight storage scheme, the authors show that one backpropagation pass can be carried out in time
\begin{equation}
    \label{eqn:backward_time}
    {O}\left( \sqrt{T_{\mathrm{iter}}M}N \frac{\log 1/\gamma}{\epsilon} R_{t,m,\delta} \right),
\end{equation}
where $R_{t,m,\delta}$ depends on the matrices $X_{t,l,j}$ and $W^{(j)}_{t,l}$ and on the terms $\delta_{t,m,l-1}$. Again, the authors provide evidence indicating that in practical settings $R_{t,m,\delta}$ does not significantly impact the runtime. When training the NN for $T_\mathrm{iter}$ iterations with $M$ data points each, the runtime gets multiplied by a factor of $T_{\mathrm{iter}}M$.

It is worth pointing out again, that the authors in \cite{allcock2020quantum} managed to overcome the dependence on $E$ in the runtime of training and evaluating NNs, which comes at the cost of $\sqrt{M}n_\mathrm{nodes}$ showing up in the runtime. Replacing $E$ by $n_\mathrm{nodes}$ in the runtime approximately amounts to a quadratic speedup (in the case of a fully connected feedforward NN), since each neuron, in that scenario, is connected with every neuron in the preceding and succeeding layer,  see \Cref{lemma:params_dimension}. Furthermore, in practice, for large NNs, $n_\mathrm{nodes} \gg \sqrt{T_{\mathrm{iter}}M}$ \cite{allcock2020quantum}. We summarise the key findings from \cite{allcock2020quantum} (Algorithm 3 therein) in \Cref{lemma:kerenidis}.

\begin{lemma}[RIPE-Accelerated NN Training and Evaluation \cite{allcock2020quantum}]
    \label{lemma:kerenidis}
    There exist quantum algorithms for training and evaluating a feedforward NN using the RIPE algorithm (see \Cref{lemma:ripe}). The time complexity for the training procedure is ${O}\left((T_{\mathrm{iter}}M)^{3/2}n_\mathrm{nodes} \frac{\log 1/\gamma}{\epsilon} R_b \right)$ and ${O}\left( \sqrt{T_{\mathrm{iter}}M}n_\mathrm{nodes} \frac{\log 1/\gamma}{\epsilon} R_f \right)$ for the evaluation of the NN. Here, $T_\mathrm{iter}$ is the number of training iterations, $M$ the number of data points trained on per iteration, $n_\mathrm{nodes}$ the number of neurons in the NN, $\gamma$ and $\epsilon$ are from the RIPE algorithm (see \Cref{lemma:ripe}), and $R_f$ and $R_b$ are factors depending on the NN and the training samples. In practice, the last two parameters are expected to not significantly impact the runtime.
\end{lemma}

\subsection{Application to Deep Learning Approach}
\label{subsct:quantum_feedforward_application}
We proceed to discuss how the results from \cite{allcock2020quantum} that we introduced above may be applied to the architecture from \cite{Han_2018}, and discuss the implications. To begin with, it is worth pointing out that the NNs in the deep learning architecture are in fact trained in a supervised manner. For the sampled Brownian paths $\{ X_{t_n} \}_n$, which constitute the training data, the labels are given by $g(X_{t_N})$, as outlined in \Cref{subsct:dl_architecture}. Thus, when starting from the classical algorithm outlined in \cite{Han_2018}, one may train and evaluate the NNs using the results from \cite{allcock2020quantum}, with the caveat that QRAM is required, unlike in the algorithms discussed so far. As outlined above, this approximately results in a quadratic speedup in $d$ for the training and evaluation of the NNs, where $d$ is the dimension of the stochastic process $X_t$ from \cite{Han_2018}, (see \Cref{lemma:params_dimension}), whereas the number of neurons per NN is of the order of $d$.

A question that naturally emerges at this point, is whether the RIPE-accelerated training and evaluation may be combined with other methods, with which we have aimed to enhance the deep learning architecture. In \Cref{subsct:mlmc} we outlined how the estimation of the loss function in the deep learning may be accelerated using classical multi-level Monte Carlo (MLMC) methods. Since employing the classical MLMC method does not change the nature of any of the NNs involved in the deep learning architecture from \cite{Han_2018}, the application of the RIPE-accelerated methods appears straightforward. However, the combination of the RIPE-accelerated method and quantum-accelerated Monte Carlo (QAMC) methods does not. In the RIPE-accelerated methods, the quantum subroutine that is employed is the RIPE subroutine (\Cref{lemma:ripe}). The other operations, such as applying the activation functions, are not carried out on a quantum processor, as described in \cite{Han_2018}. If we were to combine the RIPE-accelerated methods with QAMC, the whole deep learning architecture, from preparing the stochastic process $X_t$ to computing the payoff function $f_p$ that we aim to estimate, has to be carried out in a quantum circuit, lest the superpositions get collapsed, on which QAMC and the underlying amplitude estimation subroutine crucially rely. Therefore, the implicit weight storage scheme, which is an essential part of the RIPE-accelerated methods, would have to be adapted. Recall that QRAM merely allows for the loading of classical data in superposition, but does not allow for the storage of states in superposition. Thus, the terms that are stored in QRAM in \cite{allcock2020quantum} would instead be stored in qubits in the quantum circuit, in superposition. This procedure would require some kind of random access memory in a quantum circuit, as outlined in, e.g., \cite{ambainis2007quantum}. We point out that this procedure would most likely be very expensive regarding the number of qubits required. We leave the further exploration of combining RIPE-accelerated NN methods with QAMC to future research. 

\section{Discussion}
\label{sct:discussion}

In this section we discuss the findings from throughout this work. In \Cref{sct:variational} we investigated introducing parametrised quantum circuits (PQCs) into the neural networks (NNs) in the deep learning architecture for solving nonlinear partial differential equations (PDE). The motivation for doing so was the hope that by exploiting a Hilbert space whose dimension grows exponentially in the number of qubits, the PQC could explore a large feature space and potentially find solutions, like other (classical) feature maps. We followed an approach from the literature for avoiding the barren plateau issue in our PQCs. Compared to other variational quantum algorithms for solving differential equations, our approach does not aim to encode the solution in a quantum mechanical formulation, followed by a single variational circuit, but instead uses the more intricate structure from the deep learning approach. The evidence suggested that introducing a PQC into the NNs did not improve the performance. When completely replacing the NNs with a PQC, the performance even worsened. These findings suggest that, as the nature of the problem is of a classical nature, i.e., approximating the gradient of a classical function, the classical NNs are also better at this task. Nevertheless, we cannot rule out the possibility, that in a very high-dimensional scenario, which we cannot simulate classically, there may be an advantage when introducing parametrised quantum circuits into the deep learning model. In a hypothetical problem where, on the contrary, we expect the problem to be of a quantum mechanical nature, such as learning some unitary evolution, we might expect a PQC to perform better. 

In \Cref{sct:training_qmc_nn} we outlined the problem of estimating the (scalar) loss function and the gradient with respect to the trainable parameter of a classical NN, when the inputs are sampled from known distributions. When using classical Monte Carlo (MC) methods to approach this problem, the sample complexity to achieve a certain error threshold is given by Chebyshev's inequality. Using a quantum subroutine termed quantum-accelerated MC (QAMC) which relies on amplitude estimation, there is, however, the potential for speeding up this process. By employing QAMC to estimate the loss function of a classical NN implemented in a quantum circuit, where the input is sampled from a known distribution, we showed the potential for a speedup in the error tolerance. Furthermore, we discussed different methods for training the NNs in this scenario, differing in their memory requirements and susceptibility to numerical issues, and compared them to each other and to their classical counterparts. In the quantum case, we generally expect a slowdown in the input dimension of the NN, but a possible speedup in the error tolerance when estimating the gradient of the NN, again by making use of QAMC. We discussed these methods with the deep learning architecture in mind. We believe it may be relevant elsewhere too, wherever NNs are trained on data points sampled from known distributions. 

The tools from \Cref{sct:training_qmc_nn} we applied to the deep learning architecture in \Cref{sct:qmc_pde}. Compared to the classical case, this allows us to potentially achieve a speedup in the error dependency of the query complexity compared to the classical. In contrast to many other fault-tolerant quantum algorithms for solving differential equations, our algorithm does not rely on the quantum linear systems algorithm for the speedup. On one hand, we do therefore not get the potential for an exponential speedup in the dimension of the problem. On the other hand, we also avoid the use of quantum random access memory (QRAM), and our solution is not amplitude encoded. Due to our algorithm relying on deep learning techniques, we inherit the general lack of a convergence guarantee from the classical method. To clarify, we can get guarantees on estimating the loss function and the gradient, but that does not guarantee us a convergence to the solution of the PDE, as the optimisation landscape of NNs is highly non-convex. We also investigated the usage of multi-level MC (MLMC) techniques (and their quantum-accelerated versions) for estimating the solution of the PDE, as the latter depends on the approximation of stochastic differential equations (SDEs). We found that using classical MLMC improves on the sample complexity compared to using QAMC. As the requirements on the numerical scheme for approximating the SDE with MLMC methods are different in the classical and quantum case, quantum-accelerated MLMC (QAMLMC) offers no speedup in the deep learning architecture, as originally introduced. However, future research may be able to increase the strong order of the numerical scheme, paving the way for a quantum speedup by using QAMLMC to estimate the solution of the PDE. It is worth pointing out that the deep learning architecture, as well as MLMC methods do not interfere with the training process of the NNs. Thus, the discussion on estimating the gradient using QAMC from the previous section still applies in this context.

In \Cref{sct:quantum_feedforward} we discussed a quantum algorithm from the relevant literature for accelerating the evaluation and training of classical NNs using the robust inner product estimation (RIPE) quantum subroutine and how it may be applied to the deep learning architecture to speed up the training of the NNs. It does, however, crucially rely on QRAM. RIPE-accelerated methods also may straightforwardly be combined with classical MLMC methods in the context of the deep learning architecture. For now, Since QAMC and MLMC methods offer the potential for a speedup in the error tolerance $\epsilon$ when estimating the loss function and the gradient, and RIPE-accelerated methods for evaluating and training classical NNs offers a possible speedup in the input dimension $d$, we can outline two regimes. In the case where $d\gg1/\epsilon$, the speedup in $d$ when training the NNs with the RIPE-accelerated method would be more valuable. On the contrary, when $d\ll1/\epsilon$, QAMC and MLMC offer a more significant advantage. The task of combining RIPE-accelerated methods for NNs with QAMC we leave to future research.

With the methods discussed in this work we have shown the potential to achieve speedups, using classical and quantum methods, when solving nonlinear PDEs relevant in, e.g., finance, game theory and physics.

\section{Acknowledgments}
We acknowledge valuable discussions with Jia-Yang Gao. 
This research is supported by the National Research Foundation, Singapore, and A*STAR under its CQT Bridging Grant and its Quantum Engineering Programme under grant NRF2021-QEP2-02-P05.
LM also acknowledges funding from the Swiss-European Mobility Programme (SEMP).
FR acknowledges financial support by the Swiss National Science Foundation (Ambizione grant no. PZ00P2$\_$186040).

\bibliography{apssamp}
\appendix
\section{Quantum Computational Model}
\label{sct:comp_model}

Before we look at specific quantum algorithms, we review our computational model and other assumptions and definitions. The computational model of quantum computing we work in is the standard quantum circuit model \cite{nielsen2002quantum}. In a quantum circuit with $n$ qubits, operations on quantum states are represented as unitary matrices (termed quantum gates) acting on at least one qubit. Since the operations are unitary, they are reversible. We assume that the whole evolution of the quantum state is completely under our control \cite{schumacher2010quantum}. We illustrate a quantum circuit by a set of horizontal lines (wires) \cite{nielsen2002quantum}. The unitary operations are represented as boxes on these wires, being executed from left to right. The quantum circuit model is useful for showing how a complicated operation can be constructed from relatively simple operations. At the end of the circuit a measurement, typically in the computational basis, takes place. As mentioned in \cite{nielsen2002quantum}, an arbitrary classical circuit can be simulated by an equivalent (reversible) quantum circuit. Indeed, the class of problems that can be solved in polynomial time with a probabilistic classical computer is a subset of the class of problems that can be solved in polynomial time with a quantum computer \cite{bernstein1993quantum}. 
For a probabilistic classical circuit with runtime $T$, there exists a corresponding quantum circuit with runtime $O(T^{\log_2(3)})$ \cite{buhrman2001time} (formulation from \cite{bausch2021quantum}).

In our work, we carry out arithmetic computations by employing a fixed point representation of real numbers. We make the assumption that there are enough qubits available to us, such that we can store numbers with enough precision, such that numerical errors become negligible. We use the fixed point encoding for real numbers as in \cite{haner2018optimizing} with the formulation from \cite{rebentrost2021alphatron}. 

\begin{definition}[Fixed-Point Encoding of Real Numbers]
\label{def:fixed_point_arith}
Let $c_1$ and $c_2$ be positive integers and $a\in \{0,1\}^{c_1}$, $b\in\{0,1\}^{c_2}$ as well as $s\in\{0,1\}$, then we define a rational number as,
\begin{equation}
    {Q}(a,b,s) = (-1)^s \left( 2^{c_1 -1}a^{(c_1 -1)} + \dots + 2a^{(1)} + a^{(0)} + 2^{-1}b^{(0)} + \dots + 2^{-c_2}b^{(c_2 -1)}\right) \in [-R, R],
    \label{eqn:fixed_point}
\end{equation}
where $R=2^{c_1} - 2^{-c_2}$. 
\end{definition}
Using the fixed-point encoding of real numbers as in \Cref{def:fixed_point_arith}, we define our arithmetic model.

\begin{definition}[Quantum Arithmetic Model]
\label{def:arith_model}
Given $c_1$, $c_2 \in\mathbb{N}$, we say that we use a quantum arithmetic computing model if the four arithmetic operations can be performed in constant time on a quantum computer. 
\end{definition}
More elaborate discussions on how one can perform arithmetic operations on a quantum computer using the fixed-point representation of real numbers can be found in Appendix C of \cite{zeng2021threshold}, as well as in \cite{haner2018optimizing}.

Controlled rotation operations are central to quantum computation and the cost of carrying out controlled rotations is dependent the number of bits needed to specify the rotation angle \cite{wocjan2009rotation}. In our computational model we associate a controlled rotation with constant cost, with the formulation from \cite{doriguello2021quantum}.

\begin{definition}[Controlled Rotation]
\label{def:controlled_rot}
We say we carry out a controlled rotation with an operation $\mathcal{R}$ if, with constant time and for all rational numbers $x\in[0,1]$ defined by a $(1+c_2)$-bit string as defined above in the fixed point arithmetic, 
\begin{equation}
    \mathcal{R}|x\rangle|0\rangle = |x\rangle\left(\sqrt{1-x}|0\rangle + \sqrt{x}|1\rangle\right).
    \label{eqn:rotation}
\end{equation}
\end{definition}

In order to access (classical) functions in a quantum circuit we make use of oracles and unitaries. If we know how to implement the function $f$ under consideration, we speak of a unitary $U_f$. Typically, the term oracle is used for functions where we do not know how to implement the circuit, e.g., in Grover's search algorithm \cite{grover}. The unitary $U_f$ for a classical function $f$ is defined as follows \cite{nielsen2002quantum}.

\begin{definition}[Access to Function]
\label{def:unitary}
We say that we have quantum access to a classical function $f : \{0,1\}^n \to \{0,1\}^m$, which can be represented via a classical circuit, via a unitary $U_f$, if we can perform the quantum operation,
\begin{equation}
     U_f|x\rangle|y\rangle = |x\rangle|y \oplus f(x)\rangle, \hspace{1em}\mathrm{for\,\,} x\in \{0,1\}^n ,\mathrm{\,\,} y\in\{0,1\}^m.
    \label{eqn:oracle_def}
\end{equation}
in constant time, where $\oplus$ represents the bitwise exclusive OR (XOR) operation. 
\end{definition}
The oracle $O_f$ acts in the same way, for the case when we do not know how to implement the function $f$. Importantly, $U_f$ allows for access to function evaluations in superposition, i.e., in a linear combination of states. For complex normalised coefficients $\{c_i\}_{i=1}^N$ and $\{x_i\}_{i=1}^N$ a set of points, we have,
\begin{equation}
    U_f \sum_{i=1}^N c_i|x_i\rangle|0\rangle = \sum_{i=1}^N c_i|x_i\rangle|f(x_i)\rangle.
    \label{eqn:oracle_superpos}
\end{equation}
Quantum distribution loading refers to the preparation of quantum states corresponding to probability distributions, i.e.,
\begin{equation}
    |\psi\rangle = \sum_i \sqrt{p_i}|i\rangle,
    \label{eqn:grover_rudolph}
\end{equation}
where $\{p_i\}_i$ constitutes a probability distribution. Grover and Rudolph have presented a method to load a discrete approximation of a log-concave probability distribution (e.g., a univariate Gaussian distribution) \cite{grover2002creating}. The inductive Grover-Rudolph method starts from a state of the form,
\begin{equation}
    \label{eqn:gr_start}
    |\psi_m\rangle = \sum_{i=0}^{2^m-1} \sqrt{p_{i,m}}|i\rangle,
\end{equation}
and then proceeds to further divide the $2^m$ regions into $2^{m+1}$. To do so, the following function is computed,
\begin{equation}
    \label{eqn:gr_func}
    f_\mathrm{GR}(i) = \frac{\int_{x_{L,i}}^{(x_{R,i} - x_{L,i})/2} p(x) \,dx}{\int_{x_{L,i}}^{x_{R,i}} p(x) \,dx},
\end{equation}
where $x_{L,i}$ and $x_{R,i}$ are the left and right boundaries of the region $i$. Next, $\theta_i = \arccos (\sqrt{f_\mathrm{GR}(i)})$ is loaded into a quantum circuit such that the following controlled rotation can be carried out,
\begin{equation}
    \label{eqn:gr_rot}
    \sqrt{p_{i,m}}|i\rangle |\theta_i\rangle |0\rangle \mapsto \sqrt{p_{i,m}}|i\rangle |\theta_i\rangle (\cos(\theta_i)|0\rangle + \sin(\theta_i)|1\rangle),
\end{equation}
which, after uncomputing $|\theta_i\rangle$, leaves us in $|\psi_{m+1}\rangle$.

There are, however, issues with the Grover-Rudolph method, as pointed out in \cite{herbert2021no,zeng2021threshold}. Having to compute the integrals in \Cref{eqn:gr_func} may undo any quantum speedup brought about by other quantum subroutines. Nevertheless, in \cite{doriguello2021quantum}, the authors point out that this argument only applies when one needs to sample from the distribution $p(x)$ in order to compute $f_\mathrm{GR}$, which is not always the case. Another way to avoid the slowdown, in the case where the same distribution is to be loaded multiple times, is to compute the quantities $\theta_i$ up to the desired $m$ once and store them in a quantum random access memory (see \Cref{def:qram}), from which they can be loaded multiple times. In \cite{zoufal2019quantum}, the authors present a different method for distribution loading, namely the quantum Generative Adversarial Networks (QGAN). The QGAN can first learn a distribution (from samples) and later, once trained, repeatedly load the distribution into the quantum circuit efficiently. We conclude this discussion with the following definition for quantum distribution loading.

\begin{definition}[Quantum Distribution Loading]
\label{def:dist_loading}
We say we have access to distribution loading for the distribution $\{p_i\}_i$ if we have access to a unitary $U_p$, such that,
\begin{equation}
    U_p|0\cdots 0\rangle = \sum_{i=0}^{N_p} \sqrt{p_i}|i\rangle.
    \label{eqn:grover_rudolph_x}
\end{equation}
    
\end{definition}
Apart from potentially alleviating the issue around quantum distribution loading, quantum random access memory (QRAM) is a key component of quantum algorithms for, e.g., machine learning \cite{gao2017efficient}, evaluating general NAND trees \cite{childs2007every}, and enforcing privacy in database searches \cite{giovannetti2008quantum_privacy}. We conclude this section with the introduction of QRAM. 
\begin{definition}[Quantum Random Access Memory (QRAM) \cite{giovannetti2008qram}]
\label{def:qram}
We refer to a data structure that allows for the loading of classical data in superposition into a quantum circuit, as in,
\begin{equation}
    \label{eqn:qram}
    \sum_i c_i |i\rangle|0\cdots 0\rangle \mapsto \sum_i c_i |i\rangle |D_i\rangle,
\end{equation}
where $|i\rangle$ represent the pointer states (in superposition) and $|D_i\rangle$ the data stored at index $i$, and $c_i$ are normalised complex coefficients, as Quantum Random Access Memory.

\end{definition}

\section{Classical and Quantum Methods}
\label{sct:classical_quantum_methods}
In this section we introduce a range of algorithms and subroutines, classical and quantum, which will be relevant later on in this work. We give an introduction to neural networks (NNs) in \Cref{subsct:neuralnetworks} and automatic differentiation (AD) in \Cref{subsct:ad}. 
We introduce variational quantum methods in \Cref{subsct:vqa}, which are proposed to be well suited for the noisy intermediate-scale quantum (NISQ) era, where circuit depths are kept short to mitigate the effects of noise. Thereafter, we introduce classical Monte Carlo (MC) methods and their quantum-accelerated counterparts in \Cref{subsct:monte_carlo}. Among the fault-tolerant quantum algorithms we introduce, are quantum-accelerated MC (QAMC) methods in \Cref{subsubsubsct:qmc} and the robust inner product estimation (RIPE) algorithm in \Cref{subsct:ripe}.

\subsection{Neural Networks}
\label{subsct:neuralnetworks}
Deep learning is a subfield of machine learning which involves leveraging large NNs and has a wide range of applications, such as web search \cite{huang2013websearch,deerwester1990indexing}, computer vision \cite{voulodimos2018computer_vision} and natural language processing \cite{young2018nlp}. The algorithm from \cite{Han_2018}, which serves as the starting point of our work, applies deep learning to the problem of solving nonlinear partial differential equations (PDEs). We here give a brief overview of feedforward NNs and their training and evaluation algorithms, following \cite{allcock2020quantum}. References for further reading include \cite{goodfellow2016deep,nielsen2015neural}.

A feedforward NN consists of a collection of units, organised into $L$ layers, each layer $l$ having $n_l$ neurons. The connections between the neurons of two adjacent layers, e.g. between layers $l-1$ and $l$, can be described by a weight matrix $W_{l}\in\mathbb{R}^{n_l\times n_{l-1}}$. Furthermore, each layer has its own bias vector $b_l \in \mathbb{R}^{n_l}$. Given a nonlinear function $f_\mathrm{nl}$ (termed the activation function), data is propagated through the NN, in what is called forward propagation, by computing $a^{(j)}_l = f_{\mathrm{nl},l}(z^{(j)}_l)$ and,
\begin{equation}
    \label{eqn:forward_prop}
    z^{(j)}_l = \sum_{k=1}^{n_l} W^{(j,k)}_l a^{(k)}_{l-1} + b^{(j)}_l.
\end{equation}
The goal of a feedforward NN is generally to make correct predictions from data that it has not been trained on. Note that the nonlinearity of $f_\mathrm{nl}$ is crucial, otherwise the whole NN would just amount to one linear transformation. Consider a data set of the form $\{ (x_1,\, y_1),\ldots , (x_m,\, y_m) \}$ where $x_i \in\mathbb{R}^{n_1}$ are data vectors and $y_i \in\mathbb{R}^{n_L}$ are the corresponding labels. We make use of such a data set to train the NN, meaning that its weights $W_l$ and biases $b_l$ are tuned such that the NN minimises a cost function $C: \mathbb{R}^{n_L}\mapsto \mathbb{R}$. Ideally, the NN is trained long enough such that it makes correct predictions on unseen data points with high accuracy. The weights and biases of the NN are typically trained using a method called backpropagation, which works as follows. Given $z_L$ and $a_L$ at the end of the network, after having applied forward propagation to a data point from the training set, we introduce a vector $\delta_L$ with entries $\delta_L^{(j)} = f_\mathrm{nl}'(z_L^{(j)})\frac{\partial C}{\partial a_L^{(j)}}$ and proceed to compute the derivatives of the weights and biases backward through the network using the chain rule, at each step calculating and storing the vectors $\delta_l$ with entries,
\begin{equation}
    \label{eqn:backprop}
    \delta_l^{(j)} = f_\mathrm{nl}'(z_l^{(j)} ) \sum_{k=1}^{n_{l+1}} (W_{l+1}^\top)^{(j,k)}\, \delta_{l+1}^{(k)}.
\end{equation}
The weights and biases are then typically updated via gradient descent,
\begin{equation}
    \label{eqn:weight_update}
    {W}^{(j,k)}_{t+1,l} = {W}^{(j,k)}_{t,l} - \eta_{t,l} \frac{1}{M} \sum_{m=1}^M a_{t,m,l-1}^{(k)} \, \delta_{t,m,l}^{(j)},
\end{equation}
\begin{equation}
    \label{eqn:bias_update}
    b_{t+1,l}^{(j)} = b_{t,l}^{(j)} - \eta_{t,l}\frac{1}{M} \sum_{m=1}^M \delta_{t,m,l}^{(j)},
\end{equation}
where the index $t$ indicates the update step, $\eta$ is the learning rate and $M$ is the size of the training data (or a batch thereof). We refer to one step of the form of \Cref{eqn:weight_update,eqn:bias_update} as an iteration step. 
In the next subsection, we describe the methods used to efficiently evaluate the derivatives that appear in backpropagation, which may also be applied in other scenarios.

\subsection{Automatic Differentiation}
\label{subsct:ad}
\begin{figure}[ht]
\centering
\begin{tikzpicture}[x=1.5cm, y=1.5cm, >=stealth]

\node[circle, draw=black, fill=white, minimum size=0.8cm] (i1) at (0,0) {$a_0^{(1)}$};
\node[circle, draw=black, fill=white, minimum size=0.8cm] (i2) at (0,-1) {$a_0^{(2)}$};

\node[circle, draw=black, fill=white, minimum size=0.8cm] (h1) at (2,0) {$a_1^{(1)}$};
\node[circle, draw=black, fill=white, minimum size=0.8cm] (h2) at (2,-1) {$a_1^{(2)}$};

\node[circle, draw=black, fill=white, minimum size=0.8cm] (o) at (4,-0.5) {$a_2^{(1)}$};

\draw[->] (i1) -- (h1) node[midway, above] {$w^{(1,1)}_{1}$};
\draw[->] (i1) -- (h2) node[above left=0.37cm and 0.6cm of h2] {$w^{(2,1)}_{1}$};
\draw[->] (i2) -- (h1) node[below left=0.35cm and 0.7cm of h1] {$w^{(1,2)}_{1}$};
\draw[->] (i2) -- (h2) node[midway, below] {$w^{(2,2)}_{1}$};
\draw[->] (h1) -- (o) node[midway, above] {$w^{(1,1)}_{2}$};
\draw[->] (h2) -- (o) node[midway, below] {$w^{(1,2)}_{2}$};
\end{tikzpicture}
\caption{Toy Example of a NN, where the values in the neurons represent the pre-activation values, in the cases where an activation is applied, and the $w$ values represent the weights associated with the respective edges. For simplicity, we omit biases.}
\label{fig:toy_nn}
\end{figure}
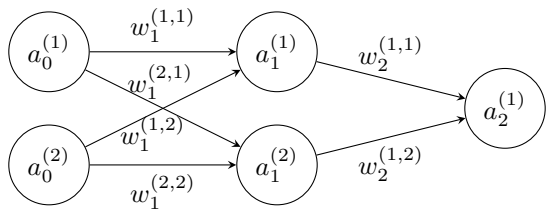

We next introduce automatic differentiation (AD), which was first introduced in \cite{wengert1964simple} and is widely used to numerically evaluate derivatives. AD computes numerical values of derivatives of functions, without developing algebraic expressions for the derivatives, by decomposing the function into a sequence of elementary functional steps. Using predeveloped subroutines (e.g. for the derivatives of certain known functions, as well as for the product rule and the chain rule), the derivative is computed alongside the actual function. The number of arithmetic operations when using AD is increased only by a constant factor compared to the number of arithmetic operations needed to evaluate a given function \cite{griewank2008evaluating}. There exist two modes of AD, forward and reverse mode \cite{baydin2018automatic}. For a function with $n$ inputs and $m$ outputs, forward mode AD computes a column of the Jacobian, and reverse mode computes a row of the Jacobian. We illustrate each mode using the toy example of a NN in \Cref{fig:toy_nn}, where we omit the biases $b_l$ for simplicity. In the context of NNs, reverse mode AD corresponds to the backpropagation algorithm.

We begin by applying reverse mode AD to the toy example from \Cref{fig:toy_nn}. Reverse mode AD in the context of NNs, i.e., backpropagation, consists of a forward and backward pass. In the forward pass, the input values are propagated forwards through the NN and the intermediate values are stored. In the backward pass, the derivatives are computed. For our toy NN from \Cref{fig:toy_nn}, the forward pass looks as follows, where we represent the values we compute at each step in the forward pass in a vector. We begin by loading the initial values, i.e., the input data,
\begin{equation}
\begin{pmatrix}
    a_0^{(1)}  \\
    a_0^{(2)}
\end{pmatrix}
=
\begin{pmatrix}
    x^{(1)}  \\
    x^{(2)}
\end{pmatrix},
\end{equation}
propagating the data through the first linear transformation,
\begin{equation}
\begin{pmatrix}
    a_1^{(1)}  \\
    a_1^{(2)}
\end{pmatrix}
=
\begin{pmatrix}
    x^{(1)}w^{(1,1)}_{1} + x^{(2)}w^{(1,2)}_{1}  \\
    x^{(1)}w^{(2,1)}_{1} + x^{(2)}w^{(2,2)}_{1}
\end{pmatrix},
\end{equation}
applying the nonlinear activation function, 
\begin{equation}
\begin{pmatrix}
    z_1^{(1)}  \\
    z_1^{(2)}
\end{pmatrix}
=
\begin{pmatrix}
    f_\mathrm{nl}(a_1^{(1)})  \\
    f_\mathrm{nl}(a_1^{(2)})
\end{pmatrix},
\end{equation}
followed by the second linear transformation,
\begin{equation}
\begin{pmatrix}
    a_2^{(1)}
\end{pmatrix}
=
\begin{pmatrix}
    z_1^{(1)} w^{(1,1)}_{2} + z_1^{(2)} w^{(1,2)}_{2}
\end{pmatrix},
\end{equation}
which leaves us with the final output of the toy NN, to which one typically applies a loss function, $f_\mathrm{loss}$,
\begin{equation}
    C 
=
    f_\mathrm{loss}(a_2^{(1)}).
\end{equation}
Next, we go through the backward pass of reverse mode AD. At each step in the backward pass, we display the newly computed values, which correspond to the derivatives of the final output of the NN with respect to the weights of a given layer. Note that in practice one only computes the vectors on the right hand side, and not the symbolic derivatives, as doing so may be computationally expensive. We start with the cost of the NN, $C$, and compute the derivatives with respect to the weights in the final linear transformation,
\begin{equation}
\begin{pmatrix}
    \frac{C}{\partial w^{(1,1)}_{2}} \\[6pt]
    \frac{C}{\partial w^{(1,2)}_{2}}
\end{pmatrix}
=
\begin{pmatrix}
    \frac{C}{\partial a_2^{(1)}} \frac{\partial a_2^{(1)}}{\partial w^{(1,1)}_{2}}\\[6pt]
    \frac{C}{\partial a_2^{(1)}}\frac{\partial a_2^{(1)}}{\partial w^{(1,2)}_{2}}
\end{pmatrix}
=
\begin{pmatrix}
    z_1^{(1)} \\[6pt]
    z_1^{(2)}
\end{pmatrix},
\end{equation}
and proceed backwards, where the nonlinear activation function comes into play,
\begin{equation}
\begin{pmatrix}
    \frac{C}{\partial w^{(1,1)}_{1}}
    \\[6pt]
    \frac{C}{\partial w^{(2,1)}_{1}} 
    \\[6pt]
    \frac{C}{\partial w^{(1,2)}_{1}}
    \\[6pt]
    \frac{C}{\partial w^{(2,2)}_{1}}
\end{pmatrix}
=
\begin{pmatrix}
    \frac{C}{\partial a_2^{(1)}} \frac{\partial a_2^{(1)}}{\partial  z_1^{(1)}} \frac{\partial z_1^{(1)}}{\partial a_1^{(1)}} \frac{\partial a_1^{(1)}}{\partial w^{(1,1)}_{1}}
    \\[6pt]
    \frac{C}{\partial a_2^{(1)}} \frac{\partial a_2^{(1)}}{\partial z_1^{(1)}} \frac{\partial z_1^{(1)}}{\partial a_1^{(1)}} \frac{\partial a_1^{(1)}}{\partial w^{(2,1)}_{1}}
    \\[6pt]
    \frac{C}{\partial a_2^{(1)}} \frac{\partial a_2^{(1)}}{\partial z_1^{(2)}} \frac{\partial z_1^{(2)}}{\partial a_1^{(2)}}\frac{\partial a_1^{(2)}}{\partial w^{(1,2)}_{1}}
    \\[6pt]
    \frac{C}{\partial a_2^{(1)}} \frac{\partial a_2^{(1)}}{\partial z_1^{(2)}} \frac{\partial z_1^{(2)}}{\partial a_1^{(2)}} \frac{\partial a_1^{(2)}}{\partial w^{(2,2)}_{1}}
\end{pmatrix}
=
\begin{pmatrix}
    w^{(1,1)}_{2} f_\mathrm{nl}'(a_1^{(1)}) a_0^{(1)}
    \\[6pt]
    w^{(1,1)}_{2} f_\mathrm{nl}'(a_1^{(2)}) a_0^{(1)}
    \\[6pt]
    w^{(1,2)}_{2} f_\mathrm{nl}'(a_1^{(1)}) a_0^{(2)}
    \\[6pt]
    w^{(1,2)}_{2} f_\mathrm{nl}'(a_1^{(2)}) a_0^{(2)}
\end{pmatrix}.
\end{equation}
It is worth pointing out, that we have now computed the entries of the gradient of the output of the NN with respect to the trainable parameters. Since we know the structure of the NN, we do not need to symbolically compute the derivatives (in the sense that we would obtain an algebraic representation of the derivative), but we can merely use the right hand side of the equations in the last two vectors to calculate the numerical values of the derivatives. This allows us to compute the gradient of the NN with respect to the weights with only a constant overhead compared to the evaluation of the NN (i.e., the forward pass) \cite{griewank2008evaluating}, as for the forward as well as the backward pass we need $O(E)$ arithmetic operations and $O(n_\mathrm{nodes})$ evaluations of the activation function and its derivative, where $E$ is the number of trainable parameters and $n_\mathrm{nodes}$ is the number of nodes in the NN. A key point allowing us to do so is the reusability of terms that we computed and stored in the forward pass for the backward pass.

Next, we outline the process of applying forward mode AD to the toy NN. In forward mode, we evaluate the NN alongside computing the derivatives. Again, with AD we do not compute the symbolic expressions for the derivatives (these are only shown for illustration purposes), but only numerical values. In contrast to the forward pass in reverse mode AD, no intermediate values get stored for later retrieval during forward mode AD. We will represent the evaluations and the derivatives as a pair of vectors for each stage. The entries of the derivative vector (termed the tangents, with $\partial$) contain the derivatives of the corresponding entries in the vector containing the function evaluation values (the primals). The reader may ask ``The derivative with respect to what?''. The answer to this question lies in the initialisation of the derivative values, as will become clearer by the end of our computation. We initialise the derivatives of the weight parameters $w$ to be given by the parameters $v$ with the labels and indices. We again begin by loading the input values, and in the case of forward mode AD we have, 
\begin{equation}
\begin{pmatrix}
    a_0^{(1)}  \\
    a_0^{(2)}
\end{pmatrix}
=
\begin{pmatrix}
    x^{(1)}  \\
    x^{(2)}
\end{pmatrix},
\begin{pmatrix}
    \partial a_0^{(1)} \\
    \partial a_0^{(2)}
\end{pmatrix}
=
\begin{pmatrix}
    0  \\
    0
\end{pmatrix},
\end{equation}
and after the linear transformation we have,
\begin{equation}
\begin{pmatrix}
    a_1^{(1)} \\
    a_1^{(2)}
\end{pmatrix}
=
\begin{pmatrix}
    x^{(1)}w^{(1,1)}_{1} + x^{(2)}w^{(1,2)}_{1}  \\
    x^{(1)}w^{(2,1)}_{1} + x^{(2)}w^{(2,2)}_{1}
\end{pmatrix},
\begin{pmatrix}
    \partial a_1^{(1)} \\
    \partial a_1^{(2)}
\end{pmatrix}
=
\begin{pmatrix}
    x^{(1)}v^{(1,1)}_{1} + x^{(2)}v^{(1,2)}_{1} \\
    x^{(1)}v^{(2,1)}_{1} + x^{(2)}v^{(2,2)}_{1}
\end{pmatrix},
\end{equation}
and following the nonlinear activation function we have,
\begin{equation}
\begin{pmatrix}
    z_1^{(1)}   \\
    z_1^{(2)}
\end{pmatrix}
=
\begin{pmatrix}
    f_\mathrm{nl}(a_1^{(1)})  \\
    f_\mathrm{nl}(a_1^{(2)})
\end{pmatrix},
\begin{pmatrix}
    \partial z_1^{(1)} \\
    \partial z_1^{(2)}
\end{pmatrix}
=
\begin{pmatrix}
    f_\mathrm{nl}'(a_1^{(1)})\partial a_1^{(1)} \\
    f_\mathrm{nl}'(a_1^{(2)})\partial a_1^{(2)}
\end{pmatrix},
\end{equation}
and after the final linear transformation we get,
\begin{equation}
\begin{gathered}
\begin{pmatrix}
    a_2^{(1)}
\end{pmatrix}
=
\begin{pmatrix}
    f_\mathrm{nl}(a_1^{(1)})w^{(1,1)}_{2} + f_\mathrm{nl}(a_1^{(2)})w^{(1,2)}_{2}
\end{pmatrix}
,
\\
\begin{pmatrix}
     \partial a_2^{(1)}
\end{pmatrix}
=
\begin{pmatrix}
     f_\mathrm{nl}'(a_1^{(1)})\partial a_1^{(1)} w^{(1,1)}_{2} + f_\mathrm{nl}(a_1^{(1)})v^{(1,1)}_{2} + 
    f_\mathrm{nl}'(a_1^{(2)})\partial a_1^{(2)} w^{(1,2)}_{2} + f_\mathrm{nl}(a_1^{(2)})v^{(1,2)}_{2}
\end{pmatrix}
,
\end{gathered}
\end{equation}
where we applied the product rule. Finally, we compute the loss,
\begin{equation}
    C
=
    f_\mathrm{loss}(a_2^{(1)})
\, , \,
     \partial C 
=
     f_\mathrm{loss}'(a_2^{(1)}) \partial a_2^{(1)}
,
\end{equation}
The quantity $\partial C$, which forward mode AD gives us, corresponds to $\nabla C \cdot v$, whereby $\nabla C$ is the gradient of $C$ with respect to the trainable parameters, and $v$ is the vector containing the initialisations of the tangents in the same order as they appear in the gradient \cite{baydin2018automatic}. To see this in our example, let us expand $\partial C$,
\begin{equation}
    \label{eqn:partial_c}
    \begin{aligned}
        \partial C & = f_\mathrm{loss}'(a_2^{(1)}) \Bigl(f_\mathrm{nl}'(a_1^{(1)})(x^{(1)}v^{(1,1)}_{1} + x^{(2)}v^{(1,2)}_{1}) w^{(1,1)}_{2} + f_\mathrm{nl}(a_1^{(1)})v^{(1,1)}_{2} 
        \\ 
        & + f_\mathrm{nl}'(a_1^{(2)})(x^{(1)}v^{(2,1)}_{1} + x^{(2)}v^{(2,2)}_{1}) w^{(1,2)}_{2} + f_\mathrm{nl}(a_1^{(2)})v^{(1,2)}_{2} \Bigr)
    \end{aligned}
\end{equation}
where this claim is more clearly visible. By choosing the initialisation values for the derivatives, we can thus choose with respect to which weights we want to differentiate, by computing the directional derivative with respect to the initial values of the tangents. If we, e.g., set $v$ to only have one nonzero entry, forward mode AD computes the derivative of $C$ with respect to the corresponding parameter. Similarly to reverse mode AD, forward mode AD allows us to compute the directional derivative of the NN with only a constant overhead in runtime compared to the evaluation of the NN \cite{griewank2008evaluating}, with $O(E)$ arithmetic operations and $O(n_\mathrm{nodes})$ evaluations of the activation function and its derivative.

To conclude, we can numerically compute the gradient of the NN with respect to the trainable parameters with reverse mode AD (backpropagation, in the context of NNs) and we can compute the inner product of the gradient with respect to the initial tangent values with forward mode AD. Both modes allow for the computation of the respective quantity (the gradient and the directional derivative) with a runtime with only a constant overhead (in practice, a factor of 2 to 3 \cite{baydin2018automatic}), compared to just the evaluation of the NN. 

After having reviewed a range of relevant classical algorithms and subroutines, we proceed in the following subsections to give an overview over a range of quantum algorithms and subroutines that we will reference later on in this work.

\subsection{Variational Quantum Algorithms}
\label{subsct:vqa}

We begin by introducing variational quantum algorithms (VQAs) together with their strengths and weaknesses. VQAs have been designed with the aim of being particularly well suited for the noisy intermediate-scale quantum (NISQ) era, by keeping the circuit depths short. Nevertheless, VQAs form a class of quantum algorithms that have been proposed for a variety of problems, such as machine learning \cite{schuld2019quantum}, optimisation \cite{farhi2014qaoa} and chemistry \cite{VQA_cerezo}. In the context of VQAs, one defines a cost function $\mathcal{C}$, which encodes the solution to the problem, as well as a quantum circuit (or ansatz) $|\psi (\theta)\rangle$ parametrised by a set of parameters $\theta$ \cite{VQA_cerezo}. The quantum processor then evaluates the cost function (or its gradients), and a classical optimiser tries to find the optimal parameters $\theta^\star$ for the ansatz,
\begin{equation}
    \label{eqn:variational}
    \theta^\star = \argmin_{\theta}\,\mathcal{C}(|\psi(\theta)\rangle).
\end{equation}
The parameter shift rule allows for the evaluation of the gradient of a parametrised quantum circuit (PQC) with respect to a single parameter \cite{mitarai2018quantum,schuld2019evaluating}. 

\begin{lemma}[Parameter Shift Rule \cite{mitarai2018quantum}]
    \label{lemma:parameter_shift}
    Let $\mathcal{B}$ be an observable and let,
    \begin{equation}
        \langle\mathcal{B}\rangle = \langle0|^{\otimes n}\,\mathcal{U}^\dag (\theta)\,\mathcal{B}\,\mathcal{U}(\theta)\,|0\rangle^{\otimes n}
    \end{equation} 
    be the expectation value of $\mathcal{B}$ with respect to a parametrised $n$-qubit quantum circuit $\mathcal{U}(\theta)|0\rangle^{\otimes n}$ where $\theta$ denotes the set of parameters. If a given parameter $\theta_j$ appears in $\mathcal{U}(\theta)$ only once in the form of a gate $\mathcal{G}(\theta_j) = e^{-i\theta_j G}$ we can write $\mathcal{U}(\theta)|0\rangle^{\otimes n} = \mathcal{U}_1 \mathcal{G}(\theta_j)\mathcal{U}_2|0\rangle^{\otimes n}$. Furthermore, if $G$ is a Hermitian operator with two distinct eigenvalues (such as for any single qubit gate), which we can shift without loss of generality to $\pm r$, it holds that,
   \begin{equation}
    \begin{aligned}
        \partial_{\theta_j}\langle\mathcal{B}\rangle /r = &\, \langle0|^{\otimes n}\,\mathcal{U}_2^\dag \, \mathcal{G}^\dag\left(\theta_j + s\right)\,\mathcal{U}_1^\dag \,\mathcal{B}\,\mathcal{U}_1\, \mathcal{G}\left(\theta_j + s\right) \,\mathcal{U}_2\, |0\rangle^{\otimes n}\\ &- \langle 0|^{\otimes n}\,\mathcal{U}_2^\dag \, \mathcal{G}^\dag\left(\theta_j - s\right)\,\mathcal{U}_1^\dag \,\mathcal{B}\,\mathcal{U}_1\, \mathcal{G}\left(\theta_j - s\right) \,\mathcal{U}_2\, |0\rangle^{\otimes n},
   \end{aligned}
    \end{equation}
    where $s = \frac{\pi}{4r}$.
\end{lemma}
In \cite{crooks2019gradients} the author extends the parameter shift rule to general two-qubit gates, with a constant overhead in runtime.

What makes VQAs uniquely suitable for NISQ devices is their typically short circuit depth, which prevents too many errors from accumulating throughout the computation on the quantum processor. Furthermore, the optimisation is outsourced to a classical processor. At the same time, VQAs make use of the Hilbert space whose size increases exponentially in the number of qubits to encode features and find solutions.

A downside of VQAs is, however, the widespread presence of barren plateaus \cite{mcclean2018barren}. This phenomenon refers to the vanishing of the gradient, which happens exponentially fast in the number of qubits. The gradient is needed to minimise the cost function. When computing an expectation of an observable $\mathcal{C}$ of a parametrised $n$-qubit quantum circuit we have,
\begin{equation}
   \mathcal{U}(\theta) = \prod_k \mathcal{U}_k(\theta_k)\mathcal{W}_k,
   \label{eqn:var_ansatz}
\end{equation}
where $\mathcal{U}_k(\theta_k) = e^{-i\theta_k\mathcal{V}_k}$ and where $\mathcal{V}_k$ are Hermitian and $\mathcal{W}_k$ are not parametrised unitaries. When tuning the parameters $\theta$, we do so by using a classical optimiser which typically relies on calculating the derivative with respect to a parameter $\theta_j$ of an expectation value we want to minimise, i.e.,
\begin{equation}
    \label{eqn:var_expectation_derivative}
    \partial_{\theta_j}\langle\mathcal{C}\rangle = i\langle 0|^{\otimes n}\, \mathcal{U}_-^\dag \, [\mathcal{V}_j,\,\mathcal{U}_+^\dag \mathcal{C}\,\mathcal{U}_+]\,\mathcal{U}_- |0\rangle^{\otimes n},
\end{equation}
where $\mathcal{U}_-$ and $\mathcal{U}_+$ refer to the products of the factors from \Cref{eqn:var_ansatz} with $k<j$ and $k>j$, respectively, and the square brackets denote the commutator. Whenever the training ansatz $\mathcal{U}(\theta)$ is sufficiently random, $\mathcal{U}_-$ or $\mathcal{U}_+$ or both match the Haar random distribution for unitary matrices up to the second moment \cite{brandao2016local_haar}. 
In \cite{mcclean2018barren}, the authors show that if a circuit is sufficiently deep such that $\mathcal{U}_-$ or $\mathcal{U}_+$ or both form a 2-design (matching the Haar random distribution up to the second moment), then with high probability the ansatz state will be on a barren plateau, i.e., the size of the gradient vanishes exponentially fast in the number of qubits and the optimiser will not be able to find a direction along which the parameters can be optimised. Randomly parametrised quantum circuits (PQC) are often used as initial guesses in variational quantum algorithms as a starting point for exploring the space of quantum states \cite{mcclean2018barren}.

Several potential remedies to combat the emergence of barren plateaus have been put forward, often proposing to reduce the entanglement between qubits, or groups of qubits \cite{grant2019barren,patti2021barren,skolik2021barren,volkoff2021barren}. In \cite{kim2021universal,larocca2021theory}, the authors show that overparametrising variational quantum circuits also improves the trainability.  Furthermore, quantum convolutional neural networks \cite{qcnn} have been shown to not exhibit barren plateaus \cite{qcnn_nobarren}.

As touched upon in \Cref{sct:intro}, we refer to quantum algorithms as fault-tolerant, when their design is not concerned about limitations of the underlying hardware. 
The next subsections will feature fault-tolerant quantum algorithms (as well as some classical methods they improve upon) that will be relevant later on in this work.

\subsection{Classical and Quantum-Accelerated Monte Carlo Methods}
\label{subsct:monte_carlo}
Monte Carlo (MC) methods use randomness to estimate numerical properties of systems which would prove intractable for an analytical analysis and are often employed in, e.g., physics \cite{newman1999monte}, finance \cite{glasserman2004monte}, and machine learning \cite{andrieu2003introduction}. Next, we outline classical and quantum MC methods, starting with the univariate case in \Cref{subsubsubsct:qmc} and proceeding to the multivariate case in \Cref{subsubsubsct:multivar_mc} and multi-level MC methods in \Cref{subsubsubsct:mlmc}. We will apply these methods to the deep learning architecture from \cite{Han_2018} later on in this work.

\subsubsection{Univariate Monte Carlo Methods}
\label{subsubsubsct:qmc}

MC methods are often used to estimate the expected output of a randomised algorithm, we begin by focusing on the case where the quantity we want to estimate is a scalar, with the formulation from \cite{montanaro2015quantum} for the general setting. Let $v(\mathcal{A})$ denote the scalar (hence univariate) random variable which returns the outcome of the randomised algorithm $\mathcal{A}$ processed by the function $v$. MC methods then aim to estimate the expectation value of $w$ of $v(\mathcal{A})$ in the following way. They produce $k$ samples by independently running $\mathcal{A}$ $k$ times and taking the average of the samples to produce $\Tilde{w}$ which is used as an estimator of the true value $w$. Assuming the variance of $v(\mathcal{A})$ is bounded by $\sigma^2$, it holds, by Chebyshev's inequality, that,
\begin{equation}
    \label{eqn:chebyshev}
    P[|w - \Tilde{w}| \geq\epsilon] \leq \frac{\sigma^2}{k\epsilon^2},
\end{equation}
for $\epsilon>0$. We can thus conclude, that by taking $k$ samples where
\begin{equation}
 k = O(\sigma^2 /\epsilon^2),
 \label{eqn:classical_mc}
\end{equation}
we can estimate $w$ up to error $\epsilon$ with a success probability of, e.g., 0.99.

By employing a quantum algorithm, it is possible to achieve a quadratic speedup in $\sigma/\epsilon$ for the MC method, as first shown in \cite{montanaro2015quantum}. We start by formulating the setting in terms of a quantum circuit, as done in \cite{rebentrost2018quantum_xanadu}. Assume we have an algorithm $\mathcal{A}$ on $n$ qubits, which, upon measurement, produces the $n$-bit result $x$ with probability $\left|a_x\right|^2$. Furthermore, let $v(x) : \{0,\, 1\}^n \mapsto [0,\, 1]$. The goal is then to estimate the expectation value,
\begin{equation}
    \label{eqn:mc_expectation}
    \mathbb{E}[v(\mathcal{A})] = \sum_{x=0}^{2^n-1} \left|a_x\right|^2 v(x).
\end{equation}
We may obtain this expectation value by combining $\mathcal{A}$,
\begin{equation}
    \label{eqn:mathcal_A}
    \mathcal{A}|0\rangle^{\otimes n} = \sum_{x=0}^{2^n-1} a_x |x\rangle,
\end{equation}
and the rotation operation seen in \Cref{eqn:rotation} to arrive at,
\begin{equation}
    \label{eqn:rotation_A}
    \mathcal{R}(\mathcal{A}\otimes \mathcal{I}_2)|0\rangle^{\otimes n+1} = \sum_{x=0}^{2^n-1} a_x |x\rangle \left(\sqrt{1-v(x)}|0\rangle + \sqrt{v(x)}|1\rangle\right) \eqqcolon |\chi\rangle,
\end{equation}
where $\mathcal{I}_d$ denotes the $d$-dimensional identity operation. Measuring the ancilla qubit in the state $|1\rangle$ has as its success probability the sought expectation value,
\begin{equation}
    \label{eqn:expectation_success}
    \Bar{v}\coloneqq\langle\chi|(\mathcal{I}_{2^n}\otimes |1\rangle\langle1|)|\chi\rangle = \mathbb{E}[v(\mathcal{A})].
\end{equation}
However, since the variance of this (Bernoulli) distribution is $\mathbb{V}[v(\mathcal{A})] = \Bar{v}(1-\Bar{v})/k$, where $k$ is the number of samples, we still have to sample $k$ times where 
\begin{equation}
    k = O\left(\frac{\Bar{v}(1-\Bar{v})}{\epsilon^2}\right),
\end{equation}
to obtain a given accuracy $\epsilon$, as in the classical case. The quadratic speedup in \cite{montanaro2015quantum} comes from employing amplitude estimation \cite{brassard2002quantum}. The speedup with amplitude estimation is attained by encoding the desired expectation value to an eigenfrequency of an oscillating quantum system, and using additional qubits to extract the eigenfrequency. Consider the following unitary,
\begin{equation}
    \label{eqn:ampest_v}
    \mathcal{V} = \mathcal{I}_{2^{n+1}} - 2\mathcal{I}_{2^n}\otimes |1\rangle\langle 1|,
\end{equation}
such that $\langle\chi|\mathcal{V}|\chi\rangle = 1 - 2\Bar{v}$. Because we can write any quantum state in the $(n+1)$-qubit Hilbert space as,
\begin{equation}
    \mathcal{V}|\chi\rangle = \cos(\theta/2)|\chi\rangle + e^{i\phi}\sin(\theta/2)|\chi^\perp\rangle,
    \label{eqn:v_orth}
\end{equation}
where $|\chi^\perp\rangle$ is a specific orthogonal complement of $|\chi\rangle$. Now,
\begin{equation}
    1-2\Bar{v} = \cos(\theta/2),
    \label{eqn:ampest_angle}
\end{equation}
so the goal is to estimate $\theta$. Next, we define the unitaries,
\begin{equation}
    \label{eqn:ampest_U}
    \mathcal{U} = \mathcal{I}_{2^{n+1}} - 2|\chi\rangle\langle\chi|,
\end{equation}
and,
\begin{equation}
    \label{eqn:ampest_Q}
    \mathcal{Q} = \mathcal{UVUV},
\end{equation}
where $\mathcal{Q}$ performs a rotation by an angle of $2\theta$ in the two-dimensional Hilbert space spanned by $|\chi\rangle$ and $\mathcal{V}|\chi\rangle$. The unitary $\mathcal{U}$ can be implemented via,
\begin{equation}
    \label{eqn:prepare_U}
    \mathcal{U} = \mathcal{R}(\mathcal{A}\otimes \mathcal{I}_2) (\mathcal{I}_{2^{n+1}} - 2(|0\rangle\langle0|)^{\otimes n+1}) (\mathcal{R}(\mathcal{A}\otimes \mathcal{I}_2))^\dag .
\end{equation}
The task now becomes to estimate $\theta$ by estimating the eigenvalues $e^{\pm i\theta}$ of $\mathcal{Q}$ (recall that $\mathcal{Q}$ is unitary). This computation can be achieved by using the quantum phase estimation algorithm \cite{nielsen2002quantum}. We present the amplitude estimation and the quantum-accelerated MC (QAMC) method in the following lemmas.

\begin{lemma}[Amplitude Estimation \cite{brassard2002quantum}]
\label{lemma:ampest}

The quantum algorithm termed amplitude estimation takes as input a single copy of a quantum state $|\chi\rangle$, unitary transformations $\mathcal{U} = \mathcal{I} - 2|\psi\rangle\langle\psi|$ and $\mathcal{V} = \mathcal{I} - 2\mathcal{P}$ where $\mathcal{P}$ is a projector, and an integer $k$. Amplitude estimation outputs $\Tilde{a}$, an estimate of $a = \langle\chi|\mathcal{P}|\chi\rangle$ such that,
\begin{equation}
\centering
\left| a - \Tilde{a} \right| \leq 2\pi\frac{\sqrt{a(1-a)}}{k} + \frac{\pi^2}{k^2},
\end{equation}
with probability at least $8/\pi^2$, using $\mathcal{U}$ and $\mathcal{V}$ $t$ times each.
\end{lemma}

\begin{lemma}[Powering Lemma \cite{jerrum1986random}]
\label{lemma:powering}
Let $\mathcal{A}$ be a classical or quantum algorithm which aims to estimate some quantity $\mu$ and whose output $\Tilde{\mu}$ satisfies $|\mu - \Tilde{\mu}| \leq \epsilon$ except with fixed probability $\gamma < 1/2$. Then, for any $\delta > 0$, it suffices to repeat $\mathcal{A}$ $O(\mathrm{log\,}1/\delta)$ times and take the median to obtain an estimate which is accurate within $\epsilon$ with probability at least $1-\delta$.

\end{lemma}
Amplitude estimation, together with the powering lemma, are used in \cite{montanaro2015quantum} to achieve a quantum speedup for Monte Carlo mean estimation. We start with the result for $0\leq v(\mathcal{A})\leq 1$, which is a direct application of \Cref{lemma:ampest,lemma:powering} to the setting outlined so far.

\begin{lemma}[{Mean estimation for [0, 1] bounded functions \cite{montanaro2015quantum}}]
    \label{lemma:qmc_01}
    Let $\mathcal{A}$ be a quantum circuit on n qubits and let $v(\mathcal{A})$ be the random variable that takes on the value $v(x) \in [0,\, 1]$ when $\mathcal{A}$ outputs x. Let $\mathcal{R}$ be defined such that,
    \begin{equation}
    \label{eqn:qmc_controlled_rot}
        \mathcal{R}|x\rangle|0\rangle = |x\rangle(\sqrt{1-v(x)}|0\rangle + \sqrt{v(x)}|1\rangle).
    \end{equation}
    Furthermore, let $|\chi\rangle = \mathcal{R}(\mathcal{A}\otimes\mathcal{I}_2)|0\rangle^{\otimes n+1}$ and let $\mathcal{U} = \mathcal{I}_{2^{n+1}} - 2|\chi\rangle\langle\chi|$. Then there exists a quantum algorithm that uses $O(\log\, 1/\delta)$ copies of $|\chi\rangle$ and uses $\mathcal{U}$ $O(t\,\log\, 1/\delta)$ times to output an estimate $\hat{\Bar{v}}$ such that,
    \begin{equation}
        \left|\hat{\Bar{v}} - \mathbb{E}[v(\mathcal{A})]\right| \leq C \left(\frac{\sqrt{\mathbb{E}[v(\mathcal{A})]}}{t} + \frac{1}{t^2} \right),
    \end{equation}
    with probability at least $1 - \delta$ where $C$ is a universal constant. In particular, for any fixed $\delta > 0$ and and $\epsilon$ such that $0<\epsilon\leq 1$, to produce an estimate $\hat{\Bar{v}}$ such that $\left|\hat{\Bar{v}} - \mathbb{E}[v(\mathcal{A})]\right| \leq \epsilon\mathbb{E}[v(\mathcal{A})]$ it suffices to take $t = O\left(1/(\epsilon\sqrt{\mathbb{E}[v(\mathcal{A})]})\right)$. To achieve $\left|\hat{\Bar{v}} - \mathbb{E}[v(\mathcal{A})]\right| \leq \epsilon$ with probability at least $1-\delta$ it suffices to take $t = O\left(1/\epsilon\right)$.
\end{lemma}
The result in \Cref{lemma:qmc_01} is improved from $v(\mathcal{A}) \in [0,\, 1]$ to $v(\mathcal{A})$ having a bounded variance.

\begin{lemma}[Mean estimation for functions with bounded variance \cite{montanaro2015quantum}]
\label{lemma:qmc}
Let $\mathcal{A}\,$ be a quantum circuit on $n$ qubits and let $v(\mathcal{A})$ be the random variable that takes on the value $v(x)$ when $\mathcal{A}$ outputs x such that $\mathbb{V}[v(\mathcal{A})] < \lambda^2$. Let the accuracy be $\epsilon < 4\lambda$. Let $\mathcal{U}$ and $|\chi\rangle$ be as in \Cref{lemma:qmc_01}. Then there exists a quantum algorithm that uses $O(\log\,\lambda/\epsilon\, (\log\, \log\,\lambda/\epsilon))$ copies of $|\chi\rangle$ and uses $\mathcal{U}$ for a number of times $O(\lambda/\epsilon\,(\log\,\lambda/\epsilon)^{3/2}(\log\,\log\,\lambda/\epsilon))$ and estimates $\mathbb{E}[v(\mathcal{A})]$ up to additive error $\epsilon$ with success probability at least $2/3$. 
\end{lemma}
In \cite{montanaro2015quantum}, \Cref{lemma:qmc} is derived from \Cref{lemma:qmc_01} by decomposing $v(\mathcal{A})$ into a set of random variables whose outputs lie in $[0,\, 1]$ and by estimating the mean of each of these random variables. The success probability can be improved to $1-\delta$ for $\delta>0$ at a multiplicative cost of $O(\log1/\delta)$ using \Cref{lemma:powering}. Therefore, we can, e.g., improve the success probability to 0.99, without an extra cost being reflected if we formulate it using the $\Tilde{O}$ notation, as $\Tilde{O}$ hides polylogarithmic factors. It is worth pointing out that QAMC (see \Cref{lemma:qmc}) offers a (nearly) quadratic speedup in terms of $\epsilon$ compared to the classical method (see \Cref{eqn:classical_mc}). 

We next outline the generalisation of MC methods from the univariate case to the multivariate case.

\subsubsection{Multivariate Monte Carlo Methods}
\label{subsubsubsct:multivar_mc}

In multivariate MC estimation, the goal is to estimate the mean in the case where $v(\mathcal{A})\in\mathbb{R}^d$. In the classical scenario, estimating all entries $\mathbb{E}[v_1(\mathcal{A})],\ldots,\mathbb{E}[v_d(\mathcal{A})]$ can be done simultaneously with the same executions of $\mathcal{A}$ with an overhead of $\log(d)$ in the sample complexity due to Hoeffding's inequality, as shown in Appendix A of \cite{cornelissen_jerbi2021quantum}. We restate the result in \Cref{lemma:mvmc} below.

\begin{lemma}[Classical Multivariate Monte Carlo Estimation (Appendix A in \cite{cornelissen_jerbi2021quantum})]
\label{lemma:mvmc}
Let an algorithm $\mathcal{A}$ generate a $d$-dimensional random variable $v(\mathcal{A})$, bounded in $l_\infty$ norm $\lVert v(\mathcal{A})\rVert_\infty \leq B$, then we can estimate $\mathbb{E}[v(\mathcal{A})]$ up to error $\epsilon$ in the $l_\infty$ norm with success probability $1-\delta$ with a sample complexity of,
\begin{equation}
    O\left(\frac{B^2}{\epsilon^2}\log\frac{d}{\delta}\right).
\end{equation}
\end{lemma}

In the quantum case, however, this simultaneous estimation is impeded as one relies on amplitude estimation to estimate the mean, which is encoded in the relative phase. Due to the periodicity and boundedness of the relative phase, one cannot use the same executions of $\mathcal{A}$ to simultaneously estimate all entries of $v(\mathcal{A})$. This inconvenience generally results in a linear overhead of $d$ in the sample complexity in the quantum case over the univariate (quantum) scenario. As shown in \cite{cornelissen_jerbi2021quantum}, there are special cases where this linear dependence can be slightly improved, however, not down to the logarithmic dependence of the classical case.

\subsubsection{Multi-Level Monte Carlo Methods}
\label{subsubsubsct:mlmc}

MC methods are often used to estimate an expectation value of a random variable determined by the solution of an SDE \cite{an2021quantum}, as is the case in \cite{Han_2018}, see \Cref{eqn:loss}. Here, we outline multi-level MC (MLMC) methods, which have the potential to improve the sample complexity of MC methods for estimating the mean value of a function depending on the (discretised) solution of an SDE, in the classical and quantum case. We follow \cite{an2021quantum} and begin with the setting and then outline the classical and quantum methods.

Let $X_t\in\mathbb{R}^d$ be a stochastic process defined by the SDE,
\begin{equation}
    \mathrm{d}X_t = \mu(t, X_t) + \sigma(t, X_t)\mathrm{d}W_t,
    \label{eqn:stoch_proc_diff_qamlmc}
\end{equation}
where $\sigma(t,X)\in\mathbb{R}^{d\times d}$, $\mu(t,X)\in\mathbb{R}^d$ and $W_t$ is a standard Brownian motion (as introduced in \Cref{subsct:dl_architecture}). 
Consider the problem of estimating an expectation value at time $T$ of the form (Problem 1 in \cite{an2021quantum}),
\begin{equation}
    \mathbb{E}(\mathcal{P}(X_T)|X_{t_0}) \in \mathbb{R},
    \label{eqn:problem_exp}
\end{equation}
where $\mathcal{P}: \mathbb{R}^d \mapsto \mathbb{R}$ is some payoff function and $X_{t_0}$ is the value of $X_t$ at the initial time.
In the case of a general SDE without an explicit solution, one first has to discretise the SDE on an interval $[t_0, T]$ with step size $\Delta t$ to produce an approximate solution using a numerical scheme. 

A numerical scheme for approximating the solution of an SDE $\hat{X}_{t_n}$ with time step size $\Delta t = T/N$ is of strong order $r$ if there exists a constant $K_m >0$, for any $m\in\mathbb{N}$, such that,
\begin{equation}
    \mathbb{E}\left[ \sup_{0\leq n\leq N} \lVert\hat{X}_{t_n} - X_{t_n}\rVert_2^m\right] \leq K_m (\Delta t)^{rm}.
    \label{eqn:strong_order}
\end{equation}
When using a numerical scheme of strong order $r$, the error for estimating \Cref{eqn:problem_exp} scales as $\epsilon = O((\Delta t)^r)$ \cite{kloeden1992stochastic}, resulting in an $\epsilon$-dependence of $O(1/\epsilon^{2 + 1/r})$ in the sample complexity for the classical case \cite{an2021quantum}.
A way to improve the error dependence in the sample complexity is to make use of the MLMC method \cite{heinrich2001multilevel,giles2008multilevel,giles2015multilevel}. We follow the introduction of the MLMC method from \cite{an2021quantum}.

The MLMC method aims to estimate the expectation value of some random variable $P$, $\mathbb{E}[P]$, by means of a sequence of random variables $P_0,\, P_1,\ldots,\, P_K$ where each element of the sequence approximates $P$ with greater accuracy and $K=O(\log(2\epsilon^{-1})$. In the setting of a discretised SDE, we can think of $P$ as the payoff function $\mathcal{P}$ evaluated at the terminal time $\mathcal{P}(X_T)$ which we want to estimate, and the index of the $P_k$ relating to how many approximation steps $N$ we take via $N_k = T/(\Delta t_k) = 2^k T$. MLMC methods then estimate $\mathbb{E}[P]$ by observing that the following telescoping sum holds, due to the linearity of the expectation value,
\begin{equation}
    \label{eqn:telescoping_sum}
    \mathbb{E}[P_K] = \mathbb{E}[P_0] + \sum_{k=0}^K \mathbb{E}[P_k - P_{k-1}],
\end{equation}
where $P_{-1} = 0$. The MLMC method estimates each of the summands in a way that minimises the cost. To estimate $\mathbb{E}[P]$ we introduce the estimator $Y$,
\begin{equation}
    \label{eqn:mlmc_estimator}
    Y = \sum_{k=0}^K Y_k,
\end{equation}
where we have for $Y_k$ the following expression,
\begin{equation}
    \label{eqn:mlmc_estimator2}
    Y_k = \frac{1}{N_k} \sum_{i=0}^{N_k} \left( P_k^{(i)} -P_{k-1}^{(i)} \right),
\end{equation}
where $i$ indicates the sample. Each $Y_k$ is then approximated via MC methods, where $P_k^{(i)} -P_{k-1}^{(i)}$ comes from one Brownian path, but with a different number of discretisation steps for $P_k$ and $P_{k-1}$. 
Further following \cite{an2021quantum}, we bound the error, 
\begin{equation}
    \label{eqn:mlmc_error}
    \mathbb{E}\left[Y - \mathbb{E}[P]\right]^2 \leq \mathbb{V}[Y] + \mathbb{E}[P_K - P]^2 \leq \epsilon^2,
\end{equation}
where we need to consider the cost $C_k$ and the variance $V_k$ of $P_k - P_{k-1}$. The total cost and variance of $Y$ are given by $\sum_{k=0}^K N_k C_k$ and $\sum_{k=0}^K V_k/N_k$, respectively. Assuming that $C_k = O(2^{\gamma k})$ and $V_k = O(2^{-\beta k})$ where $\beta \geq \gamma$, it can be shown that the number of samples needed in the classical case to estimate $\mathbb{E}[P]$ within error $\epsilon$ is $\Tilde{O}(\epsilon^{-2})$, removing the $1/r$ dependence.

When applying MLMC to the problem of estimating $\mathbb{E}(\mathcal{P}(X_T)|X_{t_0})$ from \Cref{eqn:problem_exp} with $X_t$ governed by the SDE from \Cref{eqn:stoch_proc_diff_qamlmc}, the authors in \cite{an2021quantum} make a set of assumptions, beginning with assumptions on quantities appearing in the SDE. 
\begin{assumption}[Assumption 1 in \cite{an2021quantum}]
    \label{ass:mlmc_lipschitz}
    The functions $\mu$ and $\sigma$ are globally Lipschitz continuous. Furthermore, they assume that for the initial value $X_{t_0}$ it holds that $\mathbb{E}[X_{t_0}^m] \leq C_m$ for constants $C_m \geq 0$.
\end{assumption}
Next, the authors make an assumption on the numerical scheme employed (recall the definition o the strong order $r$ in \Cref{eqn:strong_order}). The so-called Tayler-Itô schemes constitute a general class of high order schemes (for solving SDEs) of the following form \cite{kloeden1992stochastic},
\begin{equation}
    {X}_{k+1} = \sum_{\alpha}f_{\alpha}(kh, {X}_k)I_{\alpha},
    \label{eqn:taylor-ito}
\end{equation}
where the $f_{\alpha}$'s are coefficient functions and the $I_{\alpha}$'s are integrals over the time interval $[kh, (k+1)h]$. 

\begin{assumption}[Assumption 2 in \cite{an2021quantum}]
    \label{ass:mlmc_scheme}
    The coefficient functions $f_{\alpha}$'s as in \Cref{eqn:taylor-ito} are globally Lipschitz continuous with respect to $X$.
\end{assumption} 
Finally, the authors in \cite{an2021quantum} make an assumption on the payoff function.
\begin{assumption}[Assumption 3 in \cite{an2021quantum}]
    \label{ass:mlmc_payoff}
    The payoff function $\mathcal{P}$ from \Cref{eqn:problem_exp}, is globally Lipschitz continuous.
\end{assumption}
We summarise the classical result for the application of MLMC to SDEs as follows.
\begin{lemma}[Classical MLMC for SDE Payoff Estimation (Proposition 3 in \cite{an2021quantum})\,]
    Consider the problem of estimating a payoff function as in \Cref{eqn:problem_exp} of a discretised SDE of the form of \Cref{eqn:stoch_proc_diff_qamlmc} under \Cref{ass:mlmc_lipschitz,ass:mlmc_scheme,ass:mlmc_payoff}. Then MLMC with a scheme of strong order $r$ estimates $\mathbb{E}[\mathcal{P}(X_T)|X_{t_0}]$ up to mean-squared error $\epsilon^2$ with probability at least 0.99 with a sample complexity of,
    \begin{equation}
   \centering
   \begin{cases}
      O(\epsilon^{-2}) & \text{$r>1/2$}\\
      O(\epsilon^{-2}(\mathrm{log\ }1/\epsilon)^{2}) & \text{$r=1/2$}\\
      O(\epsilon^{-1/r}) & \text{$r<1/2$}.
    \end{cases}
    \end{equation}
    \label{lemma:classical_mlmc}
\end{lemma}

In \cite{an2021quantum}, the authors present a quantum-accelerated version of MLMC, QAMLMC. The speedup is derived from making use of QAMC from \cite{montanaro2015quantum} (see \Cref{subsubsubsct:qmc}). In QAMLMC, QAMC is used to estimate the expectation values $\mathbb{E}[P_k - P_{k-1}]$. 
The result for applying QAMLMC to the problem of estimating quantities of the form \Cref{eqn:problem_exp} is summarised below.

\begin{lemma}[QAMLMC (Theorem 3 in \cite{an2021quantum})]
\label{lemma:qamlmc}
    Consider the problem of estimating a payoff function as in \Cref{eqn:problem_exp} of a discretised SDE of the form of \Cref{eqn:stoch_proc_diff_qamlmc} under \Cref{ass:mlmc_lipschitz,ass:mlmc_scheme,ass:mlmc_payoff}. Then QAMLMC with a scheme of strong order $r$ estimates $\mathbb{E}[\mathcal{P}(X_T)|X_{t_0}]$ up to additive error $\epsilon$ with probability at least 0.99 with a sample complexity of,
   \begin{equation}
   \centering
   \begin{cases}
      O(\epsilon^{-1}(\mathrm{log\ }1/\epsilon)^{3/2}(\mathrm
      {log\ log\ }1/\epsilon)^2) & \text{$r>1$}\\
      O(\epsilon^{-1}(\mathrm{log\ }1/\epsilon)^{7/2}(\mathrm
      {log\ log\ }1/\epsilon)^2) & \text{$r=1$}\\
      O(\epsilon^{-1/r}(\mathrm{log\ }1/\epsilon)^{3/2}(\mathrm
      {log\ log\ }1/\epsilon)^2) & \text{$r<1$}.
    \end{cases}
    \end{equation}
    \label{thm:qa-mlmc}
\end{lemma}

\subsection{Robust Inner Product Estimation}
\label{subsct:ripe}

Next, we outline a fault-tolerant quantum algorithm for estimating inner products. It is a vital component of the algorithm (presented in the same paper, \cite{allcock2020quantum}) for addressing the bottleneck of training the NNs in the deep learning architecture which we will discuss in \Cref{sct:quantum_feedforward}. The authors in \cite{allcock2020quantum} introduce a quantum algorithm termed robust inner product estimation (RIPE) which is a generalisation of the inner product estimation algorithm from \cite{kerenidis2019qmeans}. It allows for estimating the inner product between two states $v$ and $c$ by using their (amplitude encoded) quantum states, i.e.,
\begin{equation}
    \label{eqn:quantum_vector_encoding}
    |v\rangle = \frac{1}{\lVert v \rVert_2}\sum_j v^{(j)} |j\rangle,
\end{equation}
and analogously for $|c\rangle$. The inner product estimation algorithm (Lemma 4.2 from \cite{kerenidis2019qmeans}) allows for the estimation of the inner product between two vectors $v$ and $c$, with known norms, up to error $\epsilon$ with probability at least $1-\gamma$ and in time $\Tilde{O}(\lVert v\rVert_2 \lVert c\rVert_2 T \log(2/\gamma)/\epsilon)$ where $T$ is the time needed to prepare $|v\rangle$ and $|c\rangle$. We here sketch the proof.

We begin in the following state,
\begin{equation}
    \label{eqn:ipe_1}
    \frac{1}{\sqrt{2}}(|0\rangle + |1\rangle) |0\rangle.
\end{equation}
Next, we load the vectors $|v\rangle$ and $|c\rangle$ with the operations $|0\rangle |0\rangle \mapsto |0\rangle |v\rangle$ and similarly $|1\rangle |0\rangle \mapsto |1\rangle |c\rangle$. This takes us in the state,
\begin{equation}
    \label{eqn:ipe_2}
    \frac{1}{\sqrt{2}}(|0\rangle |v\rangle + |1\rangle|c\rangle).
\end{equation}
After applying a Hadamard gate on the first qubit, we obtain,
\begin{equation}
    \label{eqn:ipe_3}
    \frac{1}{{2}}(|0\rangle (|v\rangle + |c\rangle) + |1\rangle(|v\rangle - |c\rangle)).
\end{equation}
Given the state in \Cref{eqn:ipe_3}, the probability of measuring 1 in the first qubit, $p_1$, is,
\begin{equation}
    \label{eqn:ipe_4}
    p_1 = \frac{1}{4}(2 - 2 \langle v|c\rangle) = \frac{1 - \langle v|c\rangle}{2},
\end{equation}
from which one can calculate $\langle v|c\rangle$, given $\lVert v\rVert_2$ and $\lVert c\rVert_2$. By rewriting $|1\rangle(|v\rangle - |c\rangle)$ as $|y,\,1\rangle$ (swapping the qubits), we now have, 
\begin{equation}
    \label{eqn:ipe_5}
    \sqrt{p_1}|y,\,1\rangle + \sqrt{1-p_1}|G,\,0\rangle,
\end{equation}
where $G$ signifies a garbage state. Next, using amplitude estimation (see \Cref{lemma:ampest}), we arrive at the state,
\begin{equation}
    \label{eqn:ipe_6}
    \sqrt{\alpha}|\hat{p}_1,\, G',\, 1\rangle + \sqrt{1-\alpha}|G'',\, 1\rangle,
\end{equation}
where $\alpha > 8/\pi^2$, $\lvert\hat{p}_1 - p_1 \rvert\leq\epsilon$ and $G'$ as well as $G''$ are further garbage states. To extract $\hat{p}_1$ the authors from \cite{kerenidis2019qmeans} make use of a result from \cite{wiebe2014quantum_median}.

\begin{lemma}[Quantum Median Estimation (Lemma 8 in \cite{wiebe2014quantum_median})]
    \label{lemma:median_est}
    Let $\mathcal{Y}$ be a unitary that maps
    \begin{equation}
        \label{eqn:median_est}
        \mathcal{Y} : |0\rangle^{\otimes n} \mapsto \sqrt{a}|x,\, 1\rangle + \sqrt{1-a}|G,\,0\rangle,
    \end{equation}
    for some $1/2<a\leq 1$ in time T. Then there exists a quantum algorithm which, for any $\gamma>0$ and $1/2 < a_0 \leq a$, prepares a state $|\psi\rangle$ such that $\lVert |\psi\rangle - |0\rangle^{\otimes n L}|x\rangle\rVert_2 \leq \sqrt{\gamma}$ for some integer $L$, in time
    \begin{equation}
        \label{eqn:median_est_time}
        2T\left\lceil \frac{\log 2/\gamma}{2(\lvert a_0\rvert -1/2)^2}\right\rceil .
    \end{equation}
\end{lemma}
\Cref{lemma:median_est} allows us to extract a quantum state $|\psi\rangle$ such that $\lVert |\psi\rangle - |0\rangle^{\otimes L}|\hat{p}_1,\, G'\rangle\rVert_2 \leq \sqrt{\gamma}$. From $\hat{p}_1$ we can then compute $\langle v|c\rangle$ via the relation from \Cref{eqn:ipe_4}.  In \cite{kerenidis2019qmeans}, the authors generalise the inner product estimation algorithm. We present their result in form of \Cref{lemma:ripe}.

\begin{lemma}[Robust Inner Product Estimation (RIPE) \cite{allcock2020quantum}]
    \label{lemma:ripe}
    If quantum states $|v\rangle$ and $|c\rangle$ can each be prepared in time $T$, and if the norms $\lVert v\rVert_2$ and $\lVert c\rVert_2$ are known within multiplicative error $\epsilon/3$, then the mapping $|v\rangle |c\rangle |0\rangle \mapsto |v\rangle |c\rangle |s\rangle$ where, with probability at least $1-\gamma$,
    \begin{equation}
    \label{eqn:ripe_result}
        \lvert s - v\cdot c\rvert \leq 
        \begin{cases}
            \epsilon |v\cdot c| & \text{in time } \Tilde{O}\left(\frac{T (\log1/\gamma) \lVert v\rVert_2 \lVert c\rVert_2}{\epsilon \lvert v\cdot c\rvert}\right)
            \\
            \epsilon & \text{in time } \Tilde{O}\left(\frac{T (\log1/\gamma \lVert v\rVert_2 \lVert c\rVert_2}{\epsilon}\right).
        \end{cases}
    \end{equation}
\end{lemma}

\end{document}